\documentclass[aps,prx,twocolumn,superscriptaddress,longbibliography]{revtex4-2}
\usepackage[utf8]{inputenc}
\usepackage{newtxtext,newtxmath}
\allowdisplaybreaks
\usepackage{newfloat,algcompatible}
\usepackage{etoolbox}
\AtBeginEnvironment{algorithm}{\noindent\hrulefill\par\nobreak\vskip-5pt}
\DeclareFloatingEnvironment[
    fileext=loa,
    listname=List of Algorithms,
    name=ALGORITHM,
    placement=tbhp,
]{algorithm}
\usepackage{multirow}
\usepackage{graphicx,subfigure,bbm}

\usepackage[english]{babel}
\usepackage{mathtools}
\usepackage{array,bm}
\usepackage{fancyhdr}
\usepackage{color}
\usepackage{tabularx}
\usepackage{extarrows}
\usepackage{enumerate}\usepackage{epstopdf}
\usepackage[colorlinks,
            linkcolor=black,
            citecolor=black,
            urlcolor=blue
            ]{hyperref}
\usepackage{natbib}
\usepackage{placeins}
\usepackage{tikz}
\usetikzlibrary{shapes,arrows}
\usepackage{booktabs}
\usepackage{subfigure}
\usepackage{lineno}
\usepackage{lipsum}
\usepackage{float}
\usepackage{xr-hyper}
\usepackage{hyperref}

\definecolor{lightyellow}{RGB}{255,255,224}
\newcolumntype{Y}{>{\raggedright\arraybackslash}X}

\tikzstyle{decision} = [diamond, draw, fill=blue!20,
    text width=4.5em, text badly centered, node distance=3cm, inner sep=0pt]
\tikzstyle{block} = [rectangle, draw, fill=blue!20,
    text width=10em, text centered, rounded corners, minimum height=4em]
\tikzstyle{line} = [draw, -latex']
\tikzstyle{cloud} = [rectangle, draw,fill=red!20, node distance=7cm,
    minimum height=4em]

\def\(({\left(}
\def\)){\right)}
\def\[[{\left[}
\def\]]{\right]}

\newcommand{\be}{\begin{equation}}
\newcommand{\ee}{\end{equation}}
\newcommand{\bea}{\begin{eqnarray}}
\newcommand{\eea}{\end{eqnarray}}

\begin{document}

\title{Tracking large chemical reaction networks and rare events by neural networks}

\author{Jiayu Weng}
\email[These authors contributed equally]{}
\affiliation{Institute of Data Science, University of Hong Kong, Hong Kong SAR, China}
\affiliation{School of Computing and Data Science, University of Hong Kong, Hong Kong SAR, China}
\affiliation{Department of Systems Science, Faculty of Arts and Sciences, Beijing Normal University, Zhuhai 519087, China}

\author{Xinyi Zhu}
\email[These authors contributed equally]{}
\affiliation{School of Physics, University of Electronic Science and Technology of China, Chengdu 611731, China}
\affiliation{Department of Systems Science, Faculty of Arts and Sciences, Beijing Normal University, Zhuhai 519087, China}

\author{Jing Liu}
\affiliation{School of Physical Science and Technology, Beijing University of Posts and Telecommunications, Beijing 102206, China}
\affiliation{Institute of Theoretical Physics, Chinese Academy of Sciences, Beijing 100190, China}

\author{Linyuan Lü}
\affiliation{School of Cyber Science and Technology, University of Science and Technology of China, Hefei 230027, China}

\author{Pan Zhang}
\email[Corresponding authors: ]{panzhang@itp.ac.cn}
\affiliation{Institute of Theoretical Physics, Chinese Academy of Sciences, Beijing 100190, China}
\affiliation{School of Fundamental Physics and Mathematical Sciences, Hangzhou Institute for Advanced Study, UCAS, Hangzhou 310024, China}

\author{Ying Tang}
\email[Corresponding authors: ]{jamestang23@gmail.com}
\affiliation{Institute of Fundamental and Frontier Sciences, University of Electronic Science and Technology of China, Chengdu 611731, China}
\affiliation{School of Physics, University of Electronic Science and Technology of China, Chengdu 611731, China}
\affiliation{Key Laboratory of Quantum Physics and Photonic Quantum Information, Ministry of Education, University of Electronic Science and Technology of China, Chengdu 611731, China}
\affiliation{Non-classical Information Science Basic Discipline Research Center of Sichuan Province, University of Electronic Science and Technology of China, Chengdu 611731, China}

\begin{abstract}
Chemical reaction networks are widely used to model stochastic dynamics in biophysical processes. Their exact probabilistic description via the chemical master equation becomes intractable for high-dimensional systems due to the exponential growth of the state space. This difficulty is further exacerbated when metastable transitions and rare events dominate the dynamics. 
Here, we present an advanced neural-network approach that effectively tackles large reaction networks and rare-event dynamics within a unified framework. 
Building on autoregressive representations of the full joint probability distribution, we achieve efficient training through second-order optimization methods, including natural gradient descent and the time-dependent variational principle.
We leverage enhanced-sampling strategies that enable accurate resolution of low-probability regions governing rare transitions.
We demonstrate the method on a multistable genetic toggle switch and a mitogen-activated protein kinase (MAPK) cascade consisting of 16 species and 35 reactions with dense coupling. We further quantify rare events in spatially extended Schl\"ogl models in both one- and two-dimensional lattices, reaching system sizes with extremely large state spaces up to $3\times10^{15}$. These results highlight the potential of neural modeling as a flexible and scalable framework for studying high-dimensional chemical reaction networks and rare events.
\end{abstract}

\maketitle

\section{Introduction}

The chemical master equation (CME) provides a fundamental probabilistic framework for describing chemical reaction networks~\cite{Gillespie_stochastic_2007,van2007stochastic}, underpinning a wide range of applications in physics~\cite{weber2017master}, chemistry~\cite{qian_chemical_2010}, and biology~\cite{zhang_markovian_2019}. 
While the CME is exact in principle, its direct solution quickly becomes infeasible for realistic systems due to the exponential growth of the state space. The challenge becomes particularly severe in high-dimensional reaction networks~\cite{cao_accurate_2016, cai2025revival}, and the situation is further complicated in spatially extended systems, such as reaction-diffusion lattices~\cite{schlogl_chemical_1972, kim_stochastic_2017}. 
To mitigate this curse of dimensionality, a number of approaches have been introduced, including systematic truncation schemes such as finite state projection (FSP)~\cite{munsky_finite_2006, FSP_cellular_sciadv}, buffer-based state partitioning in the accurate chemical master equation (ACME)~\cite{cao_accurate_2016}, and decompositions of the CME into lower-dimensional subsystems~\cite{fang_advanced_2024}. While effective for specific systems, these methods often require system-dependent designs, and their scalability is difficult to maintain in densely coupled networks.

In addition to high dimensionality, accurately characterizing rare events~\cite{e_applied_2019} poses another major challenge.
Many stochastic reaction systems often reside in metastable states, with infrequent transitions between them governing transition rates between emergent macroscopic states. Such phenomena are closely related to dynamical phase transitions in nonequilibrium networks~\cite{liu_dynamical_2025}, and underlie many processes such as biochemical switching~\cite{wei_non-kramers_2025}, and noise-induced pattern formation~\cite{earlylife2015PRL,halatek2018patternformation}. The stochastic simulation algorithm (SSA)~\cite{gillespie_general_1976,  pagare_stochastic_2024, 10.1063/5.0217316} 
provides a widely used trajectory-based method for CME. However, it becomes inefficient when the dynamics is dominated by rare transitions between metastable states~\cite{nicholson_quantifying_2023}. In such regimes, SSA together with mean-field approximations~\cite{grima_study_2012, pineda_beyond_2017}, often fails to capture the relevant rare configurations. This difficulty leads to large variance or biased estimates of transition rates, and the number of required trajectories increases exponentially with system size. The challenge is amplified in spatially extended reaction-diffusion systems, where the coupling between neighboring sites induces correlated transition and makes rare events harder to sample.

\begin{figure*}[ht!]
    \includegraphics[width=1\textwidth]{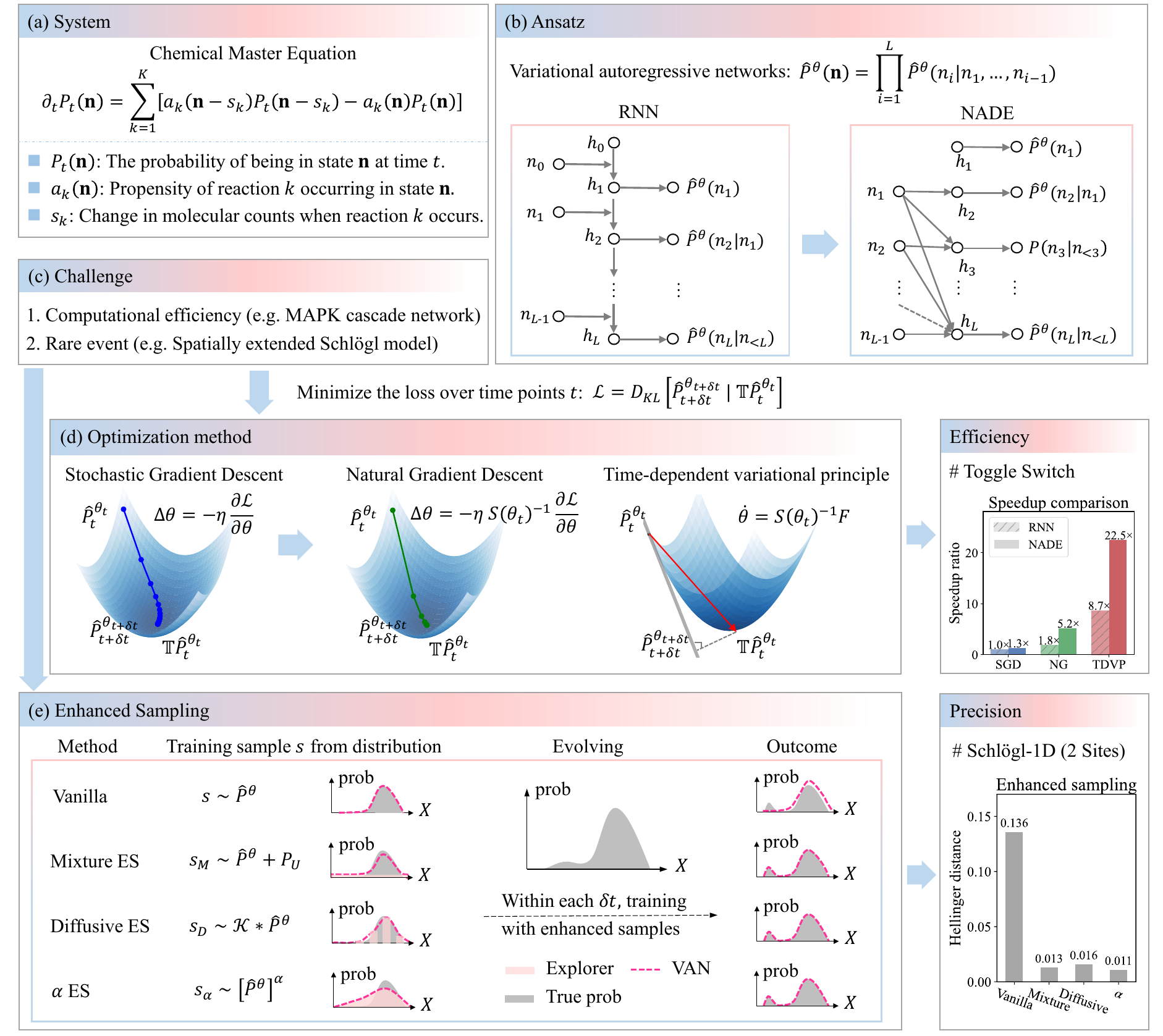}
    \caption{The present work efficiently tracks large reaction networks and rare events.
    (a) The chemical reaction system follows the CME, whose exponentially large state space makes the exact probability distribution intractable.
    (b) The variational autoregressive network factorizes the joint distribution into conditional probabilities. We use NADE-based architectures to improve training efficiency over sequential RNNs.
    (c) Our advances include both reducing computational cost to deal with large networks and quantifying rare-event statistics.
    (d) The VAN is trained by minimizing the KL-divergence between time steps. Moving from stochastic gradient descent (SGD), we now use natural gradient (NG), which improves convergence by rescaling updates with the inverse Fisher information matrix and the time-dependent variational principle (TDVP) with the projected temporal evolution. In the genetic toggle switch model~\cite{gardner_construction_2000,terebus_discrete_2019}, NG and TDVP achieve significant speedup over SGD.
    (e) To capture rare events, we augment the training procedure with enhanced-sampling strategies, including mixture sampling with configurations drawn from a uniform distribution $P_U$, diffusive sampling based on a kernel $\mathcal{K}$, and $\alpha$ sampling using an exponent-based modification.
    In the Schl\"ogl model, these sampling schemes markedly reduce the Hellinger distance to the baseline from expensive Gillespie simulations.}
    \label{Fig1}
\end{figure*}

Recent advancements in machine learning, such as deep neural networks, have introduced new possibilities for tackling high-dimensional stochastic problems~\cite{PhysRevLett.122.250502,mehta_high-bias_2019,carleo_machine_2019, jiang_neural_2021, sukys_approximating_2022, ZHOU2025121574, tang_quantifying_2022, anton_computational_2025}. 
In particular, variational autoregressive networks (VANs) provide a flexible representation of high-dimensional probability distributions~\cite{wu_solving_2019, PhysRevLett.124.020503, barrett2022autoregressive, PhysRevLett.128.090501, shin2021protein}. Such representations are shown to accurately track the time-evolving joint probability distribution for a range of reaction networks in our previous work~\cite{tang_neural-network_2023}.
However, its application to large reaction networks and spatially extended systems remains computationally prohibitive, highlighting the need to extend the approach to more general systems~\cite{alon2006introduction}.

Addressing the second challenge of capturing rare events also requires further development of the neural-network approach. Meanwhile, the recent tensor-network approaches~\cite{nicholson_quantifying_2023} have shown promise in modeling stochastic reaction-diffusion systems.
By factorizing the high-dimensional probability distribution into a sequence of low-rank tensors, they capture correlations between neighboring sites while maintaining manageable computational complexity.
Such methods achieve remarkable accuracy, as demonstrated by a spatially-extended system on one-dimensional lattices. 
Nevertheless, many reaction-diffusion systems of practical interest, such as surface catalytic and membrane-bound reactions, take place beyond one-dimensional lattices~\cite{PhysRevE.81.056110,2010stochastic_reaction-diffusion,2013CRDME}. 
Extending tensor-network techniques to such systems is challenging, for example, in two dimensions the required bond dimension of the conventional projected entangled pair states (PEPS)~\cite{RevModPhys.93.045003} increases rapidly with lattice connectivity, and both tensor contraction and time evolution scale quasi-exponentially with this bond dimension~\cite{PhysRevLett.98.140506, PhysRevB.109.235102, 2014TensorRenormalization}. 
These challenges motivate further developments of tensor-network approaches to high-dimensional reaction-diffusion systems and highlight the need for more flexible and scalable approaches.

In this paper, we present an advanced neural-network framework that addresses the computational challenges of large reaction networks together with the characterization of rare events.
To represent high-dimensional probability distributions, we employ neural autoregressive distribution estimator (NADE)~\cite{larochelle_neural_2011,uria_neural_2016}, which efficiently characterizes statistical dependencies among species.
Efficient training is achieved through second-order optimization techniques, in particular the natural gradient (NG) descent, which rescales parameter updates using the Fisher information matrix~\cite{Amari1998naturalgradient, pascanu2014natural, liu_efficient_2025}.
We also apply the time-dependent variational principle (TDVP) method~\cite{reh_time-dependent_2021}, which applies analogous ideas based on variational manifolds. These second-order methods have recently been developed to be scalable for high-dimensional neural networks through stochastic estimators and low-rank solvers for the metric tensor~\cite{chen_empowering_2024,rende_simple_2024,liu_efficient_2025}.
Furthermore, inspired by developments in the machine learning community~\cite{misery_looking_2025,karan2025reasoningsamplingbasemodel}, we develop enhanced-sampling strategies that augment vanilla neural-network sampling and enable accurate exploration of low-probability regions, resulting in efficient estimation of rare transitions between metastable states.

To demonstrate the capabilities of our approach, we first apply it to the genetic toggle switch~\cite{gardner_construction_2000} with multimodal distributions~\cite{terebus_discrete_2019}.
We benchmark computational performance and observe substantial acceleration, a 5- to 22-fold speedup over the previous implementation, while maintaining accuracy.
We next evaluate the method on more demanding systems, including the mitogen-activated protein kinase (MAPK) cascade~\cite{1996PNAS_MAPK,johnson_mitogen_2002,cao_accurate_2016}, a reaction network consisting of 16 species and 35 reactions with densely intertwined interactions. To the best of our knowledge, this system is the largest biochemical reaction network that has been studied by existing methods for solving the CME.
We further consider the spatially extended Schl\"{o}gl model in 1D (up to 8 sites)~\cite{vellela_stochastic_2009, nicholson_quantifying_2023} and 2D (up to a $2\times 4$ lattice). The enhanced-sampling strategy is particularly crucial in these spatial systems, as it enables the model to capture low-probability peaks and the associated rare transitions between metastable states.
Together, these advances make high-dimensional stochastic reaction networks and rare-event dynamics more tractable in practice.

\subsection{Chemical master equation}
We consider chemical reaction networks consisting of $L$ species and $K$ reactions. The state vector of species is $\mathbf{n}=(n_1,\dots,n_L)$, where $n_{i}\in[0,M-1]$ denotes the count of species $i$ ($i=1,\dots,L$).
Each reaction is specified by its stoichiometric change $s_{k}$ and propensity function $a_{k}(\mathbf{n})$, which gives the probability that reaction $k$ occurs in an infinitesimal time interval when the system is at the state $\mathbf{n}$. 
In our work, the propensity functions are calculated following the law of mass action~\cite{ge2012stochastic}, by multiplying the reaction rate constant by the copy numbers of the reactant species raised to their corresponding stoichiometric powers.

The probability distribution $P_{t}(\mathbf{n})$ then evolves according to the CME~\cite{Gillespie_stochastic_2007,van2007stochastic}:
\begin{align}
\label{CME}
\partial_{t}P_{t}(\mathbf{n})  = \sum_{k=1}^{K} \big[a_{k}(\mathbf{n}-s_{k})P_{t}(\mathbf{n}-s_{k}) - a_{k}(\mathbf{n})P_{t}(\mathbf{n})\big].
\end{align}

Given reaction rates, stoichiometry, and an initial distribution $P_{0}(\mathbf{n})$, the CME fully characterizes the stochastic dynamics of the reaction network. 
However, the state space grows exponentially with the number of species as $M^{L}$, making direct solutions computationally prohibitive. 
This motivates the neural-network approach introduced in the next section.

\section{Framework}
Fig.~\ref{Fig1} provides a schematic overview of the framework used throughout this work. For clarity, we refer to the present implementation as NNCME-2 to distinguish it from an earlier autoregressive formulation~\cite{tang_neural-network_2023}.
We now introduce the framework and its key components in detail. 

\subsection{Design of the VAN by NADE}

We leverage the VAN~\cite{wu_solving_2019, hibat-allah_recurrent_2020, PhysRevLett.124.020503, barrett2022autoregressive, PhysRevLett.128.090501, shin2021protein, tang_neural-network_2023} to represent the joint probability distribution of the state space. The VAN factorizes the joint distribution into a product of conditional probabilities,
\begin{align}
\label{VAN}
\hat{P}^{\theta}(\mathbf{n})=\prod^{L}_{i=1}\hat{P}^{\theta}\!\left(n_i \mid \mathbf{n}_{<i}\right),
\end{align}
where $n_i$ denotes the count of the $i$-th species, $\mathbf{n}_{<i}$ represent the inputs from previous species $\{n_1,\dots,n_{i-1}\}$, and $\theta$ denotes trainable parameters of the VAN. The VAN parameterizes an automatically normalized joint probability distribution, from which configurations can be sampled efficiently.

The VAN can be implemented with different architectures, including the masked autoencoder for distribution estimation (MADE)~\cite{germain_made_2015}, the neural autoregressive distribution estimator (NADE)~\cite{larochelle_neural_2011, uria_neural_2016}, recurrent neural network (RNN)~\cite{Elman1990, LSTM_1997, cho_learning_2014,sutskever_sequence_2014, hibat-allah_recurrent_2020}, and the transformer~\cite{vaswani_attention_2017}. These architectures differ in their expressive power and parameter complexity, which determine the trade-off between computational efficiency and the ability to capture correlations in the distribution~\cite{tang_neural-network_2023}.

For example, RNNs incorporate autoregressive dependencies through sequential state updates, making them well-suited for capturing temporal correlations~\cite{Elman1990, LSTM_1997, cho_learning_2014,sutskever_sequence_2014}. However, their inherently sequential updates limit computational parallelism during training and sampling~\cite{vaswani_attention_2017}. NADE and MADE implement feed-forward autoregressive formulations that avoid explicit recurrent iterations, thereby improving training efficiency while preserving the conditional dependency structure among species~\cite{larochelle_neural_2011, germain_made_2015}. The transformer implements conditional dependencies through self-attention mechanisms~\cite{vaswani_attention_2017}, offering the highest representational flexibility for large or complex reaction networks, albeit at the cost of increased computational demand and optimization sensitivity.

In this work, we primarily adopt the NADE architecture~\cite{larochelle_neural_2011}, which achieves a favorable balance between efficiency and expressivity for the CME. By enforcing autoregressive dependencies through masked feed-forward connections, NADE computes the conditional probabilities with shared parameters, making it lightweight and scalable. We have also tested the transformer as a backbone of the VAN, which typically requires longer computational time due to its attention-based architecture. Since the transformer has a strong representative power, we anticipate that it can have more applications in systems with complex distributions and have included it in our code package. Next, we introduce NADE in more detail.

\subsubsection{Neural Autoregressive Distribution Estimator (NADE)}

NADE models the conditional distributions in Eq.~\eqref{VAN} by a fully connected feedforward neural network with shared parameters. The $i$-th conditional distribution $\hat{P}^{\theta}(n_i| \mathbf{n}_{<i})$ is:
\begin{align}
\hat{P}^{\theta}(n_i| \mathbf{n}_{<i}) &= \mathrm{softmax}\left(V_{i} \mathbf{h}_i + \mathbf{b}_i\right), \\
\mathbf{h}_i &= \sigma\left(W_{<i} \mathbf{n}_{<i} + \mathbf{c}\right),
\end{align}
where $\sigma(\cdot)$ is the sigmoid activation function, and $V =\{V_1,\dots,V_L\}\in \mathbb{R}^{M \times H}$, $W \in \mathbb{R}^{H \times L}$, $\mathbf{b} \in \mathbb{R}^{L \times M}$,  $\mathbf{c} \in \mathbb{R}^{H}$ are shared parameters across all species. As defined above, $L$ is the number of species, $M$ is the upper bound each species can reach, and $H$ is the hidden-layer dimension of the NADE.

\subsection{Optimization with second-order methods}
\subsubsection{Training objectives: reverse or forward KL-divergence}
To track the distribution over time, we train the VAN~\cite{tang_neural-network_2023} by minimizing the reverse Kullback-Leibler (KL)-divergence between the updated distribution $\hat{P}^{\theta_{t+\delta t}}_{t+\delta t}$ parameterized by the VAN, and the propagated distribution $\mathbb{T}\hat{P}^{\theta_t}_{t}$, i.e., the one-step time evolution of the distribution at time $t$:
\begin{align}
\label{Loss}
\mathcal{L}= D_{KL}\!\left[\hat{P}^{\theta_{t+\delta t}}_{t+\delta t}\,\big|\big|\,\mathbb{T}\hat{P}^{\theta_{t}}_{t}\right],
\end{align}
where $\mathbb{T}=e^{\delta t \mathbb{W}}$ is the one-step transition kernel of the Eq.~\eqref{CME}, and $\mathbb{W}$ is its generator corresponding to the transition rate matrix. 
The loss can be estimated from samples as
\begin{align}
\label{loss2}
\mathcal{L}=
\mathbb{E}_{\mathbf{s}\sim\hat{P}^{\theta_{t+\delta t}}_{t+\delta t}}
\big[\ln \hat{P}^{\theta_{t+\delta t}}_{t+\delta t}(\mathbf{s})-
\ln(\mathbb{T}\hat{P}^{\theta_{t}}_{t})(\mathbf{s})\big],
\end{align}
where $\mathbf{s}$ denotes samples drawn from distribution $\hat{P}^{\theta_{t+\delta t}}_{t+\delta t}$. The parameters are updated by the gradient
\begin{align}
\nabla_{\theta_{t+\delta t}} \mathcal{L}
&= \mathbb{E}_{\mathbf{s}\sim\hat{P}^{\theta_{t+\delta t}}_{t+\delta t}}\!\Big[
\nabla_{\theta_{t+\delta t}}\ln \hat{P}^{\theta_{t+\delta t}}_{t+\delta t}(\mathbf{s}) \nonumber\\
&\quad
\cdot \big(\ln \hat{P}^{\theta_{t+\delta t}}_{t+\delta t}(\mathbf{s}) - \ln (\mathbb{T}\hat{P}^{\theta_t}_{t})(\mathbf{s})\big)
\Big].
\end{align}

This reverse KL-divergence was employed in our previous work~\cite{tang_neural-network_2023}, where training was performed by standard stochastic gradient descent (SGD). Because the KL-divergence is asymmetric, different choices of KL interact with how training samples are obtained, i.e., from $\hat{P}^{\theta_{t+\delta t}}_{t+\delta t}$ or not. This leads to different optimization behaviors during temporal propagation. 

In general, forward KL-divergence is known to be mode-covering, whereas reverse KL-divergence is mode-seeking~\cite{pmlr-diffusiveKL}. In VAN-based time evolution, the target distribution often transitions gradually from unimodal to multimodal, as in the toggle switch model, so the training behavior of different KL formulations is not obvious a priori.
To examine these effects, we consider two alternatives in Appendix~\ref{sec:kl_analysis}: the forward KL-divergence and a measure-transformed reverse KL-divergence. 
For robustness and simplicity, we adopt the standard reverse KL-divergence in the main text.

\subsubsection{Natural gradient descent}
We improve the optimization procedure used in the first version of our method by incorporating 
natural gradient (NG) descent~\cite{Amari1998naturalgradient, pascanu2014natural, liu_efficient_2025}. 
In contrast to standard SGD, which updates the parameters along the Euclidean gradient, the NG rescales the update direction using the inverse Fisher information matrix, yielding the steepest descent direction on the statistical manifold. 
This geometry-aware update leads to more stable training and faster convergence, as shown in the examples.

The Fisher information matrix (FIM) is given by
\begin{align}
\label{eq:Fisher}
S(\theta_t)
=
\mathbb{E}_{\mathbf{s}\sim \hat{P}^{\theta_t}_t}
\!\left[
\nabla_{\theta_t} \ln \hat{P}^{\theta_t}_t(\mathbf{s})\,
\nabla_{\theta_t} \ln \hat{P}^{\theta_t}_t(\mathbf{s})^{\top}
\right],
\end{align}
where $\mathbf{s}$ denotes samples drawn from the current variational distribution $\hat{P}^{\theta_t}_t$, and $\theta_t$ denotes the trainable parameters of the VAN at time.
FIM captures the local curvature of the parameter manifold and acts as a Riemannian metric tensor in the distribution space.

Let $\theta$ denote the parameters being optimized at time $t+\delta t$. Given the loss function $\mathcal{L}$ defined in Eq.~\eqref{loss2}, the NG update rule takes the form:
\begin{align}
\label{eq:natgrad}
\theta^{(k+1)}
= \theta^{(k)} - \eta\, S(\theta^{(k)})^{-1} \nabla_\theta \mathcal{L}.
\end{align}
where $\eta$ is the learning rate and $k$ denotes the iteration step of training epochs. The preconditioned gradient $S^{-1} \nabla \mathcal{L}$ ensures that parameter updates are appropriately scaled relative to the geometry of the model distribution.

While the NG update improves convergence, its computational cost is dominated by the inversion of the FIM, which scales as $\mathcal{O}(N_p^3)$ with the number of parameters $N_p$. To overcome this limitation, a broad range of efficient approximation strategies has been developed to accelerate NG updates. One line of work focuses on exploiting structure in the FIM: structured approximations such as Kronecker-factored methods~\cite{pauloski_convolutional_2020, kaul_projective_2023, zhang_kronecker-factored_2023} and sparse graphical models~\cite{grosse_scaling_2015} leverage factorization and conditional independence properties to significantly reduce computational cost. Beyond structural approximations, resource-reduction techniques~\cite{kolotouros_random_2024, liu_efficient_2025} further lower computational demands through random projections or stochastic probing of the FIM. Finally, for large-scale neural networks, regularization-based strategies such as SOFIM~\cite{sen_sofim_2024} and adaptive regularization schemes~\cite{wu_convergence_2024} address ill-conditioning in the FIM, thereby improving convergence stability and training efficiency.

In this work, we adopt the stochastic low-rank approximation proposed in~\cite{liu_efficient_2025}, which is specifically tailored to VAN and therefore naturally aligns with the present framework.
This method leverages matrix identities to avoid direct inversion of the FIM, reducing the overall complexity to $\mathcal{O}(N_b^3 + N_p N_b^2)$, where $N_b$ denotes the number of samples used for estimating the stochastic gradient. 
For completeness, detailed derivations and implementation details are provided in Appendix~\ref{app:natgrad}, in which we also investigate how to choose the learning rate $\eta$ in practice. We find that values in the range $\eta \in [0.1, 0.8]$ generally yield better performance, with 5-10 training epochs per time step providing a good balance between efficiency and stability for CME problems.

\begin{algorithm*}[t]
\label{algorithm1}
\begin{algorithmic}
\\\hrulefill
\State \textbf{Input:} System setup (stoichiometric matrix, propensities, initial distribution $P_{0}(\mathbf{n})$;
hyperparameters (time step $\delta t$, total steps, VAN size, learning rate, batch size, epochs);
optimizer (Natural gradient or TDVP);
sampling scheme (vanilla, mixture ES, diffusive ES, or $\alpha$ ES). 
\State \textbf{Output:} Learned joint probability distributions $\hat{P}^{\theta_t}_t$ over time.
\\\hrulefill

\State Initialize VAN parameters $\theta_0$ to match $P_{0}(\mathbf{n})$.
\FOR{each time step $t \to t+\delta t$}
    \State Learn the next-step VAN $\hat{P}^{\theta_{t+\delta t}}_{t+\delta t}$.
    \FOR{each epoch (for TDVP, typically one epoch suffices)}
    
        \State (1) Sample $\mathbf{s}\!\sim\!\hat{P}^{\theta_{t+\delta t}}_{t+\delta t}$.
        \State \hspace{1.5em}(Optional) Augment with enhanced samples:
        \State \hspace{2.5em}-- \textit{mixture ES}: mix a small fraction of uniformly drawn states $P_u(s)=|S|^{-1}$ to promote exploration of low-probability regions;
        \State \hspace{2.5em}-- \textit{diffusive ES}: apply a smoothing kernel $\mathcal{K}$ (such as a uniform kernel) to the distribution to generate nearby diffused samples;
        \State \hspace{2.5em}-- \textit{$\alpha$ ES}: sample from an overdispersed distribution $q_t^{(\alpha)}$ obtained by raising $\hat{P}^{\theta_{t+\delta t}}_{t+\delta t}$ to the power $\alpha\in(0,1)$.

        \State (2) Build local neighborhoods:
        \State \hspace{1.5em}For each sampled configuration $\mathbf{s}$, enumerate all states reachable by any chemical reaction or diffusion event.
        \State \hspace{1.5em}Evaluate $(\mathbb{T}\hat{P}^{\theta_t}_t)(\mathbf{s})$ via the transition operator $\mathbb{T}$ defined by the CME generator.
        
        \State (3) Loss evaluation and reweighting:
        \State \hspace{1.5em}Form the reverse KL-divergence objective
        $\mathcal{L} = \mathbb{E}_{\mathbf{s}\!\sim\!\hat{P}^{\theta_{t+\delta t}}_{t+\delta t}}[\ln \hat{P}^{\theta_{t+\delta t}}_{t+\delta t}(\mathbf{s}) - 
        \ln (\mathbb{T}\hat{P}^{\theta_t}_t)(\mathbf{s})]$.
        \State \hspace{1.5em}(Optional) If enhanced sampling is used, apply importance weights to correct sampling bias.
        
        \State (4) Estimate gradients of loss function: 
        \State \hspace{1.5em}Evaluate $\nabla_{\theta_{t+\delta t}} \ln \hat{P}^{\theta_{t+\delta t}}_{t+\delta t}(\mathbf{s})$ for all sampled states and compute the gradient of the reverse KL loss $\nabla_{\theta_{t+\delta t}} \mathcal{L}$ (Eq.~\eqref{loss2}).
        
        \State (5) Update parameters:
        \State \hspace{2.5em}-- 
        \textit{Natural gradient:}
           apply 
           $\theta_{t+\delta t} \leftarrow 
            \theta_{t+\delta t} 
            - \eta\, S(\theta_{t+\delta t})^{-1} \nabla_{\theta+\delta t}\mathcal{L}$,
           where $S(\theta)$ is the Fisher information matrix (Eq.~\eqref{eq:Fisher}).

        \State \hspace{2.5em}-- \textit{TDVP:}
        evolve by $\dot{\theta}_t = {S(\theta_t)}^{-1}
        \mathbb{E}_{\mathbf{s}\!\sim\!\hat{P}^{\theta_t}_t}
        [\nabla_{\theta_t}\ln\hat{P}^{\theta_t}_t(\mathbf{s})
        \,\ln(1+\partial_t\ln\hat{P}^{\theta_t}_t(\mathbf{s})\delta t)\,]$, then $\theta_{t+\delta t}=\theta_t+\eta\,\dot{\theta}_t$ $(\eta = 1)$.
        
    \ENDFOR
    \State Save joint probability distribution $\hat{P}^{\theta_{t+\delta t}}_{t+\delta t}$ and estimate statistics (means, variances, marginals, rare-event probabilities).
\ENDFOR
\\\hrulefill
\end{algorithmic}

\caption{Algorithmic framework for solving the CME with a VAN. 
NNCME-2 integrates natural gradient or TDVP optimization with enhanced sampling to accelerate convergence and accurately capture rare-event statistics.}
\end{algorithm*}

\subsubsection{Time-dependent variational principle (TDVP)}

In addition to discrete-time optimization via the natural gradient, the time-dependent variational principle (TDVP) offers a continuous-time formulation for evolving variational parameters on a statistical manifold. Originally developed and widely employed in quantum many-body dynamics, TDVP derives equations of motion by projecting the exact dynamics onto the tangent space of the statistical manifold. In this formulation, the evolution of the variational parameters is governed by the information geometry of the distribution space, ensuring motion along the steepest-descent direction defined by the Fisher--Rao metric~\cite{haegeman_time-dependent_2011,reh_time-dependent_2021}.

Instead of repeatedly optimizing the loss function at each time step, TDVP minimizes the reverse KL-divergence between the propagated distribution $\mathbb{T} \hat{P}^{\theta_t}_t$ and the variational distribution $\hat{P}^{\theta_{t+\delta t}}_{t+\delta t}$, leading to the evolution equation:
\begin{equation}
\dot{\theta}_t = S^{-1}(\theta_t)\,
\mathbb{E}_{\mathbf{s}\sim \hat{P}^{\theta_t}_t}
\!\left[
\nabla_{\theta_t}\ln\hat{P}^{\theta_t}_t(\mathbf{s})\,
\partial_t\ln\hat{P}^{\theta_t}_t(\mathbf{s})
\right],
\label{eq:TDVP_final}
\end{equation}
where $S(\theta_t)$ is the Fisher information matrix (Eq.~\eqref{eq:Fisher}). 
Introducing the projected probability flow
\begin{equation}
F=\mathbb{E}_{\mathbf{s}\sim \hat{P}^{\theta_t}_t}
\!\left[\nabla_{\theta_t}\ln\hat{P}^{\theta_t}_t(\mathbf{s})\,
\partial_t\ln\hat{P}^{\theta_t}_t(\mathbf{s})\right],
\end{equation}
then the TDVP equation becomes
\begin{align}
\dot{\theta}_t &= S^{-1}(\theta_t)\,F .
\end{align}
Unlike natural gradient descent, which also minimizes a static loss landscape, TDVP evolves the parameters dynamically such that each infinitesimal update already represents the locally optimal change in time. 
Practically, this means that at each time step, the model parameters need to be updated only once (i.e., one training epoch per step) while still following the information-geometrically optimal trajectory. 
TDVP therefore achieves comparable accuracy to natural gradient optimization, with reduced training cost. 
The full derivation and connections to the standard TDVP~\cite{haegeman_time-dependent_2011,reh_time-dependent_2021} formalism are provided in Appendix~\ref{app:tdvp}.

\subsection{Enhanced sampling for tracking rare events}

Beyond computational efficiency, achieving high accuracy is equally critical for modeling complex stochastic systems. These systems often exhibit multiple metastable states, and rare transitions between them determine the rates of transition among macroscopic phases.
Such events are central to phenomena such as phase transitions and conformational changes in biomolecules. Accurately capturing rare events is demanding and requires high precision in low probability regions~\cite{tang_learning_2024}.

Training a VAN for the CME requires sampling from the variational distribution at each time step to compute the loss. However, when rare configurations have yet to be explored, vanilla sampling tends to draw predominantly from high-probability regions. As a result, low-probability regions receive little learning signal, leading to biased estimates and poor representation of rare events.
To address this, we introduce enhanced-sampling strategies, which can be viewed heuristically as an ``explore and conquer" approach. The idea is to enrich the training set with additional exploratory samples, allowing the neural network to discover low-probability regions. Once these rare configurations are included, the neural network naturally learns to fit the corresponding regions of the probability distribution, yielding a globally consistent approximation.

This approach is general and does not depend on prior knowledge of specific systems. In practice, one needs to set the hyperparameters, such as the number of exploratory samples or the intensity of exploration, which requires care: overly aggressive exploration can reduce the role of other effective training samples and bias the training. We find from our examples that the proper choice of hyperparameters can hold across different lattice sizes for spatially extended systems. Overall, the guiding principle is that enhanced sampling should not compromise the training accuracy in the main high-probability regions. We next introduce the enhanced-sampling (ES) schemes used in this work.

\subsubsection{Mixture ES}

To increase the diversity of training samples, we first employ a uniformly random augmentation strategy based on a mixture sampling scheme. At each iteration, the training set is constructed from a mixture of model-generated (vanilla) samples $s \sim P_\theta$ and uniformly drawn configurations $s \sim P_u$ from the full state space $\mathcal{S}$.

The combined sampling process reads
\begin{equation}
    q_{\mathrm{mix}}(s) = (1-\alpha_r)\, \hat{P}^{\theta}(s) + \alpha_r\, P_u(s),
\end{equation}
where $\alpha_r$ controls the ratio of uniformly drawn samples. 
The uniform distribution $P_u(s)=|\mathcal{S}|^{-1}$ provides sparse yet broad coverage, occasionally populating low-probability regions that the model has not yet explored. 
Although these random samples are uninformative individually, they effectively prevent mode collapse and promote coverage of the global state space.

The corresponding training objective remains unchanged:
\begin{equation}
    \mathcal{L}_{\mathrm{mix}}
    = \mathbb{E}_{s\sim q_{\mathrm{mix}}}
      \!\left[\ln \hat P^{\theta_{t+\delta t}}_{t+\delta t}(s) - \ln (\mathbb{T} \hat{P}^{\theta_t}_t)(s)\right],
\end{equation}
and no importance reweighting is applied.
This is because $\alpha_r$ is chosen to be small (typically $5\%-15\%$), so that the uniform samples from $P_u$ have a negligible impact on the dominant training regions.
Empirically, we find that even this simple augmentation can improve the representation of rare probability regions in bistable systems such as the Schl\"{o}gl model, while incurring negligible computational overhead.

\begin{table*}[t]
\caption{The models solved by NNCME-2 and the corresponding computational cost under the chosen hyperparameters. The time-step length $\delta t$ is measured in units of the inverse reaction rates, and the physical time is $t=\delta t\,T_{\mathrm{step}}$. The upper count limit is $M$. Depth and width refer to the NADE architecture, and $\eta$ is the learning rate for optimization algorithms. NG denotes natural gradient and TDVP denotes the time-dependent variational principle. 
All computations were performed on a single Tesla V100 GPU.}
\squeezetable 
\begin{ruledtabular}
\begin{tabular}{lcccccccccccc}
Example & Optimize by & Species & Reactions & $M$ & $T_{step}$ & $\delta t$ & Depth & Width & Batch & Epochs per step  & $\eta$ & Time (hr) \\
\hline
Toggle switch & SGD       & 4  & 8  & 80 & 8001     & 0.005   & 1 & 16 & 2000  & 100 & 0.005 & 3.234 \\
   & NG   & 4  & 8  & 80 & 8001     & 0.005   & 1 & 16 & 2000  & 5   & 0.5   & 0.806 \\
   & TDVP      & 4  & 8  & 80 & 8001     & 0.005   & 1 & 16 & 2000  & 1   & 1.0   & 0.185 \\
MAPK cascade & NG   & 16 & 35 & 10 & $10^{6}$ & 0.01    & 1 & 8  & 2000  & 5   & 0.5   & 127.584  \\

1D Schl\"{o}gl (8 Sites) & NG  & 8  & 46 & 85 & $1.25\times10^{5}$ & $8\times10^{-6}$ & 1 & 16 & 5000  & 5   & 0.8   & 79.433 \\

2D Schl\"{o}gl (2x4 Sites) & NG  & 8  & 52 & 85 & $1.25\times10^{5}$ & $8\times10^{-6}$ & 1 & 16 & 5000  & 5   & 0.8   & 102.82 
\\
\end{tabular}
\end{ruledtabular}
\label{table1}
\end{table*}

\subsubsection{Diffusive ES}
Recent works have shown that convolving the proposal distribution with a suitable kernel can improve sampling efficiency. 
For example, several theoretical analyses demonstrate that convolution-based importance sampling can converge faster than the commonly used geometric-mean constructions~\cite{guo2025complexityanalysisnormalizingconstant,chehab2025provableconvergencelimitationsgeometric}, and reverse diffusive KL (DiKL)~\cite{pmlr-diffusiveKL} uses multiscale Gaussian convolutions to reduce the mode-seeking tendency of reverse KL.

Motivated by these insights, we adopt a diffusive sampling scheme in the CME setting, implemented through a local convolution kernel to broaden exploration and improve coverage of low-probability configurations.
Given the current variational distribution $\hat{P}_t^{\theta_t}(\mathbf{n})$, we define the diffusive proposal:
\begin{align}
\label{eq:diffusive}
q_t^{(\mathrm{diff})}(\mathbf{n})
&= \prod_{i=1}^{L}
q_t^{(\mathrm{diff})}(n_i \mid \mathbf{n}_{<i}), \\
q_t^{(\mathrm{diff})}(n_i \mid \mathbf{n}_{<i})
&\propto
\sum_{n_i'}
\mathcal{K}(n_i \mid n_i')\,
\hat{P}_t^{\theta_t}(n_i' \mid \mathbf{n}_{<i}),
\end{align}
where $\mathcal{K}$ is a local kernel that redistributes probability mass along coordinate $n_i$. 

The kernel $\mathcal{K}$ may take various forms, such as kernels constructed from discrete distributions (e.g., Poisson-type smoothers). 
In this work, we adopt a uniform diffusive kernel for the diffusive ES. Detailed implementation details, together with a study of the effect of different kernel sizes in the Schl\"ogl (2 sites) model, are provided in Appendix~\ref{subapp:diffusive}.

\subsubsection{\texorpdfstring{$\alpha$ ES}{alpha ES}}

Exponentiating a probability distribution to introduce tempering is a common strategy in variational Monte Carlo and annealing-based methods~\cite{zhang2025weightedVMC, misery_looking_2025}, where it effectively flattens sharp probability peaks and enhances the visibility of low-probability yet informative regions of the state space~\cite{misery_looking_2025}. Motivated by this idea, we adopt an exponent-based reweighting scheme to construct an exploratory proposal distribution. Importantly, although inspired by annealing concepts, our approach does not rely on Monte Carlo sampling.

Related exponent-based adjustments have also been used to enhance search and reasoning dynamics in large language models~\cite{karan2025reasoningsamplingbasemodel}.
In our setting, the same idea leads to a simple variant of the model:
\begin{align}
\label{eq:q_alpha}
q_t^{(\alpha)}(\mathbf{n})
&= 
\frac{[\hat{P}_t^{\theta_t}(\mathbf{n})]^{\alpha}}
     {\sum_{\mathbf{n}'}[\hat{P}_t^{\theta_t}(\mathbf{n}')]^{\alpha}}
= \prod_{i=1}^{L}
  \frac{
    [\hat{P}_t^{\theta_t}(n_i|\mathbf{n}_{<i})]^{\alpha}
  }{
    \sum_{n_i'}[\hat{P}_t^{\theta_t}(n_i'|\mathbf{n}_{<i})]^{\alpha}
  },
\end{align}
where $\alpha\in(0,1)$. Smaller values of $\alpha$ lead to broader, more over-dispersed proposal distributions, thereby improving coverage of low-probability configurations while preserving the autoregressive normalization at each conditional level.

By amplifying the contribution of rare configurations through $q_t^{(\alpha)}$, the $\alpha$ ES exploration enables the VAN to better capture the tails of the probability distribution while maintaining consistency in high-probability regions. The details for $\alpha$ ES and the range for setting $\alpha$ are provided in Appendix~\ref{subapp:alpha}.
As suggested in~\cite{misery_looking_2025}, this scheme can be further extended by adaptively tuning the overdispersion factor $\alpha$ based on the gradient of an importance-sampling objective, which dynamically adjusts the sampling breadth during training. 
The adaptive formulation provides a principled extension that may offer advantages for systems with sharper multimodal distribution or stronger localization. 

\begin{figure*}[ht!]
{\includegraphics[width=1\textwidth]{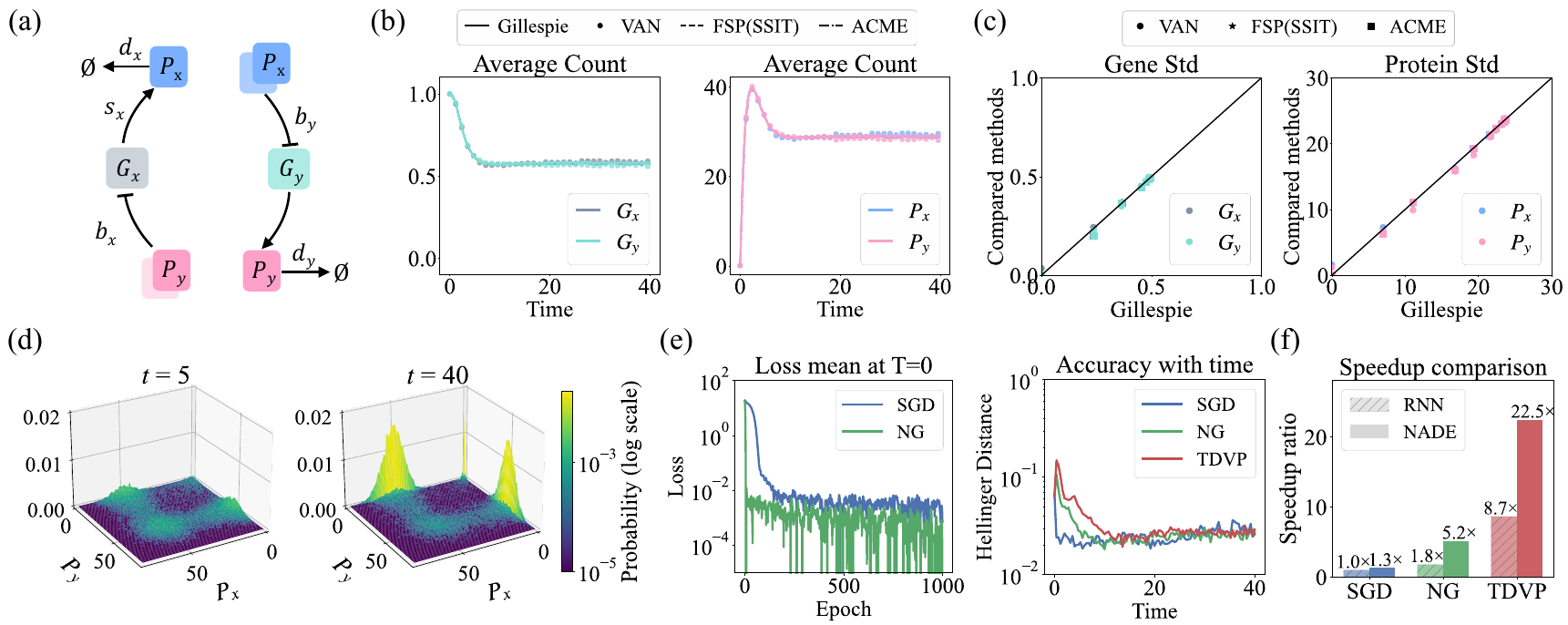}}
\centering
\caption{
NNCME-2 achieves at least a 5-fold acceleration while maintaining high accuracy, as demonstrated in the genetic toggle switch.
(a) A schematic of the chemical reactions. 
(b) The time evolution of the average counts for the genes and proteins, comparing the VAN (dots), Gillespie simulation (solid line), FSP results from SSIT (dashed), and ACME results (dash-dotted).
(c) Comparison of the standard deviations obtained from VAN, FSP (SSIT), and ACME against Gillespie at time points $t = 0, 2, ..., 38, 40$. The Gillespie simulation has $1\times10^{4}$ trajectories.
(d) Joint distribution of the two proteins from VAN at time points $t=5, 40$, showing four stable states in the long-time limit.
The initial distribution assigns equal probability to the bound and unbound promoter states for each gene, with zero initial protein counts. The results here are from the NADE architecture with natural gradient optimization. Parameters are $s_x=s_y=50$, $d_x=d_y=1$, $b_x=b_y=10^{-4}$, and $u_x=u_y=0.1$, with hyperparameters in Table~\uppercase\expandafter{\romannumeral1}. 
(e) Training performance of different optimizers on the NADE architecture. 
(f) Comparison of computational cost. NADE reduces overall training time compared to RNN, and both NG and TDVP significantly accelerate training.
} 
\label{toggle}
\end{figure*}

\section{Examples}
We demonstrate the performance of NNCME-2 on three chemical reaction networks of increasing complexity.  
These examples are chosen to test multiple aspects of the method, including computational efficiency, scalability in high-dimensional state spaces, and the ability to recover rare events. 
Starting from a multistable toggle switch~\cite{gardner_construction_2000, terebus_discrete_2019}, we then consider the MAPK cascade \cite{1996PNAS_MAPK, cao_accurate_2016} with complex reaction connectivity, followed by 1D and 2D spatially extended Schl\"ogl models~\cite{vellela_stochastic_2009, nicholson_quantifying_2023} that showcase the method in high-dimensional lattices.

To benchmark our method, we use Gillespie simulations generated with the Python package biocircuits as a common and reproducible reference across all examples. We further include two independent numerical tools for additional comparison. The stochastic system identification toolkit (SSIT)~\cite{popinga2026stochastic} is a MATLAB-based package for stochastic reaction networks; its FSP solver is used for the genetic toggle-switch model, while its efficient Gillespie implementation provides a modern SSA reference for the spatial Schl\"ogl models. We also perform ACME calculations using BIOACME~\cite{bioacme}. For the toggle-switch model, we use the example implementation distributed with the package. For the MAPK cascade, since no model-specific file or buffer configuration is provided, we reconstruct the reaction network and its two-buffer configuration from the detailed specification in the original ACME study~\cite{cao_accurate_2016}.
For the Schl\"ogl model, the tensor-network approach~\cite{nicholson_quantifying_2023} does not provide publicly available implementation code, and its evaluation was also performed against Gillespie simulations; we therefore adopt the same reference here. Overall, the distributions obtained by NNCME-2 are consistent with the results reported in these earlier studies.

The computational time and the hyperparameters for all the examples are summarized in Table~\ref{table1}. 
Typical widths for VAN range from 8 to 64: larger widths can represent more complex distributions but increase training cost (Appendix~\ref{sec:appendix_NADE_width}). For the MAPK cascade, we use a width $8$ to limit the number of parameters for this large system. Larger batch sizes generally improve accuracy, with the cost of more computation; in the Schl\"ogl example, larger batches are used to capture rare-event statistics more reliably. Enhanced sampling is applied in the Schl\"ogl model, where mixture sampling is used, and random exploratory samples account for 15\% of each batch.
For SGD, smaller learning rates $\eta$ and more epochs per time step are used, while for CME problems, NG is empirically stable with $\eta$ in the range $0.1$--$0.8$ and $5$--$10$ epochs per step. TDVP updates the parameters by solving a projected evolution equation, so in practice we set $\eta=1$ and use a single epoch per step. 

In practice, the architecture and optimization settings are largely transferable across the systems considered here. The main problem-dependent choices are the time step $\Delta t$ and the count limit $M$, which should resolve the relevant temporal scale and contain the significant probability mass, respectively. An excessively large $\Delta t$ may introduce temporal-discretization errors, whereas insufficient $M$, network width, batch size, or sampling coverage can lead to truncation bias, underfitting or noisy estimates. More detailed guidance on network width, upper-count selection and training convergence can also be found in the Methods section of NNCME-1~\cite{tang_neural-network_2023}.

\subsection{Genetic toggle switch}
We first examined the genetic toggle switch system~\cite{gardner_construction_2000} which exhibits a multimodal probability distribution and serves as a testbed for evaluating computational efficiency and accuracy. The model consists of six molecular species, as illustrated in Fig.~\ref{toggle}(a). Details of the reaction network are in Appendix~\ref{subsec:Toggle}. In our previous implementation (NNCME-1), we employed an RNN-based VAN with GRU units and optimized the model using SGD to accurately capture the joint distribution. In the present work, we adopt the NADE framework and update the VAN parameters using NG. As shown in Fig.~\ref{toggle}(b) and Fig.~\ref{toggle}(c), the mean counts and standard deviations obtained from NNCME-2 agree closely with the Gillespie reference and are also consistent with the FSP(SSIT) and ACME results. Moreover, the four stable genetic states arising from the mutual inhibition between the two genes are faithfully captured, as reflected in the joint probability distributions shown in Fig.~\ref{toggle}(d). Detailed distribution comparisons are provided in Appendix~\ref{app:other_methods} and Fig.~\ref{fig:appendix_toggle}.

We further benchmarked different optimization algorithms, including SGD, NG and TDVP, as summarized in Fig.~\ref{toggle}(e). 
Because the TDVP formulation relies on time-evolution operators that are not well defined at the initial time ($t=0$), we initialize the dynamics using NG optimization for the first time step, and subsequently switch to TDVP for temporal propagation. The loss plot shows that NG achieves faster convergence than SGD, owing to its consideration of the curvature of the parameter manifold. In practice, the NG optimization converges within about 50 epochs at $t=0$, whereas SGD typically requires on the order of $10^3$ epochs to reach a comparable loss.

\begin{figure*}[ht]
{\includegraphics[width=1\textwidth]{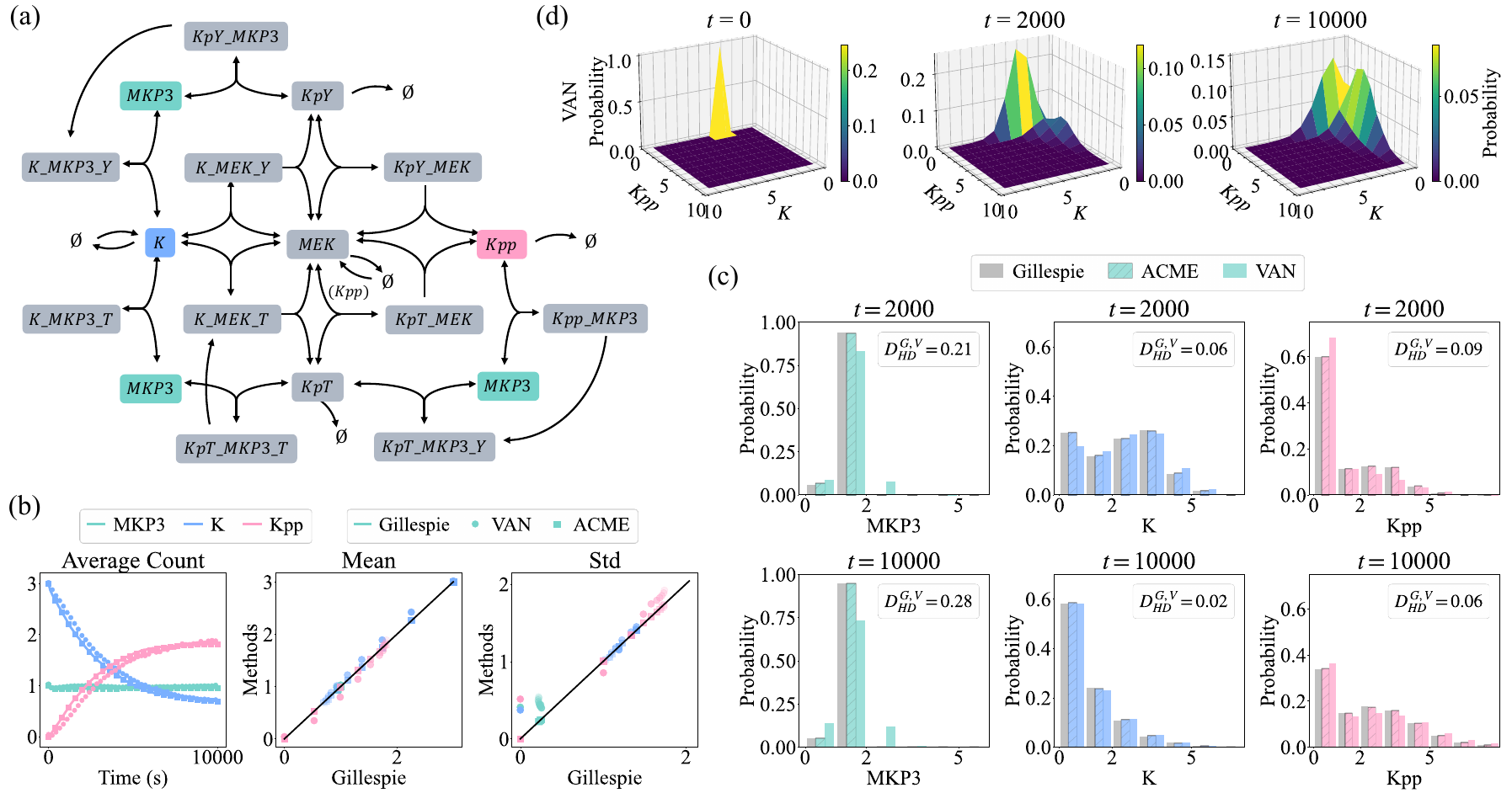}}
\centering
\caption{NNCME-2 handles large reaction networks efficiently.
(a) A schematic of a representative large reaction network (MAPK cascade, 16 species). Bidirectional arrows indicate reversible reactions, and the detailed reactions and rates are in Appendix~\ref{app:mapk}. 
(b) The left panel displays the time evolution of the mean counts of the principal species (MKP3, K, and Kpp), comparing Gillespie simulations (lines), VAN (circles), and ACME (squares). The two panels on the right compare the means and standard deviations obtained from VAN and ACME with Gillespie at time points $t=0, 1000, \ldots, 10000$.
(c) Marginal distributions of MKP3, K, and Kpp at $t=2000$ and $t=10000$, obtained from the three methods. The inset reports the Hellinger distance $D_{\mathrm{HD}}^{G,V}$ between the Gillespie and VAN distributions.
(d) Joint distributions of K and Kpp from the VAN at $t = 0, 2000,$ and $10000$. The system evolves from the initial distribution $P(\text{K} = 3, \text{MKP3} = 1, \text{others} = 0) = 1$ to the bimodal state. The Gillespie results are based on $10^{4}$ trajectories. 
The hyperparameters of the VAN are in Table~\uppercase\expandafter{\romannumeral1}.}
\label{MAPK}
\end{figure*}

To evaluate model accuracy, we computed the mean Hellinger distance between the marginal distributions obtained from VAN and those from the Gillespie simulations, as shown in Fig.~\ref{toggle}(e). Compared with SGD, both NG and TDVP exhibit a modest accuracy loss at early times, due to the rapid expansion of the initial delta distribution. This can be reduced by using larger batch sizes, which improve gradient estimates for NG and TDVP. All three optimization methods achieve comparable accuracy in most stages of the system's evolution, with the Hellinger distance remaining around 0.02. 
Fig.~\ref{toggle}(f) further reports the computational speedup under three optimization schemes using both RNN and NADE architectures. The results show that NADE achieves significant speedups over RNN, and this improvement is particularly pronounced for NG and TDVP. Considering the balance between computational efficiency, stability, and minimal accuracy loss, we adopt the NADE architecture with NG as the default setting in all main examples. 
Detailed comparisons of the learning curves and distributions for different optimizers are provided in Appendix~\ref{app:optim}.

We also examined different formulations of the training objective based on the KL divergence, including the forward KL divergence ($\mathcal{L}_{F}$), the reverse KL divergence ($\mathcal{L}_{R}$), and the measure-transformed reverse KL divergence ($\mathcal{L}_{R2}$), as detailed in Appendix~\ref{sec:kl_analysis}. We further presented the results of NADE-based architectures under different optimization schemes (Appendix~\ref{app:optim}), as well as the effect of varying network widths (Appendix~\ref{sec:appendix_NADE_width}). Finally, we compared the PyTorch and JAX implementations of NNCME-2. Both deep-learning frameworks yield consistent results, with PyTorch exhibiting slightly higher GPU efficiency, while JAX offers improved scalability and advantages in automatic differentiation (Appendix~\ref{sec_jax_vs_torch}).

\subsection{MAPK cascade: a large reaction network}

We investigate the mitogen-activated protein kinase (MAPK) signaling cascade~\cite{johnson_mitogen_2002, cao_accurate_2016}, a central pathway in cellular regulation involving two tiers of kinases, MEK and ERK, together with the phosphatase MKP3. ERK undergoes successive phosphorylation, forming singly and doubly phosphorylated intermediates (KpY, KpT, Kpp), while MKP3 mediates dephosphorylation, creating feedback that can generate bistability. The model consists of 16 molecular species and 35 reactions. The structure of the MAPK reaction network is illustrated in Fig.~\ref{MAPK}(a), and the detailed reaction equations and parameter settings are provided in Appendix~\ref{app:mapk}.

\begin{figure*}[ht!]
{\includegraphics[width=1\textwidth]{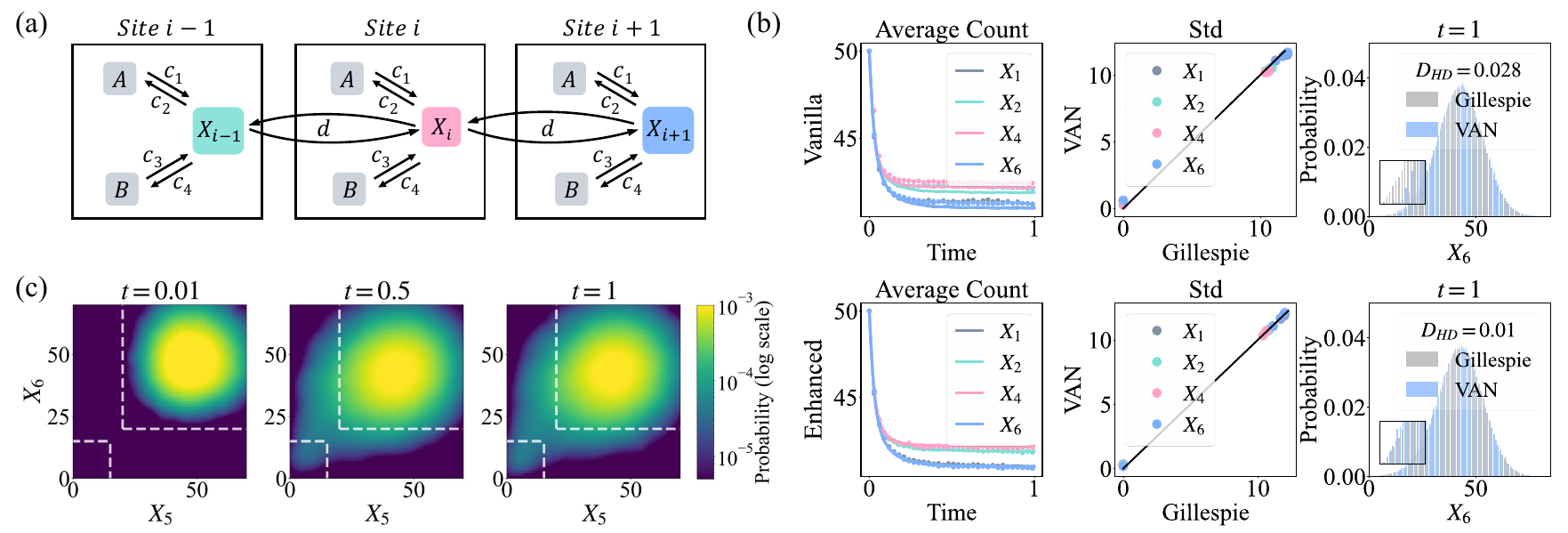}}
\centering
\caption{NNCME-2 captures rare events in the one-dimensional spatially extended Schl\"{o}gl model (6 sites). (a) A schematic of the reaction-diffusion system, where each site hosts a well-mixed Schl\"{o}gl model with fixed species $A$ and $B$, and stochastic species $X_i$ diffusing between neighboring sites with rate $d$.
(b) Comparison between the vanilla VAN (top) and the one with enhanced sampling (bottom). 
The results obtained from the VAN (dots) closely match those from Gillespie simulations (lines) for the time evolution of the mean counts at sites $X_1$, $X_2$, $X_4$, and $X_6$ (left), standard deviations at time points $t = 0, 0.1, ..., 1.0$ (middle), and the marginal distribution of $X_6$ at $t = 1$ with the Hellinger distance (right). The inset highlights the low-probability regime ($X \in [0, 20]$, probability $< 5 \times 10^{-4}$).
(c) Joint distributions of $X_5$ and $X_6$ from the VAN with enhanced sampling, with probabilities shown by color in a logarithmic scale. White boxes indicate the transition of the probability peak. 
The initial distribution is a delta peak at $X=50$. The Gillespie results are based on $5 \times 10^5$ trajectories. The hyperparameters of the VAN are in Table~\uppercase\expandafter{\romannumeral1}. Mixture ES is used in this example, with exploratory samples comprising 15\% of each batch.
}
    \label{Schlogl}
\end{figure*}

We applied the present approach to estimate the evolution of the joint probability distribution over time. As a comparison, the average counts of the three species MKP3, K, and Kpp estimated from the VAN closely match with those obtained from Gillespie simulations and ACME method~\cite{cao_accurate_2016} [Fig.~\ref{MAPK}(b)]. The corresponding standard deviations [Fig.~\ref{MAPK}(b)] and the marginal distributions [Fig.~\ref{MAPK}(c)] also show agreement between the three methods. A slightly larger standard deviation of the VAN near $t=0$ arises because an excessively sharp delta initialization (with nearly zero variance) would make the subsequent temporal evolution difficult to initiate and numerically unstable.
We further showed the joint distribution of K and Kpp obtained from the VAN at representative time points [Fig.~\ref{MAPK}(d)]. The distribution evolves from the delta-initialized state ($\text{K}=3$, $\text{MKP3}=1$) to a steady-state distribution with two peaks at $(\text{K}=1,\text{Kpp}=0)$ and $(\text{K}=0,\text{Kpp}=2)$, with probabilities 0.1229 and 0.1169, respectively, in close agreement with ACME~\cite{cao_accurate_2016}. These results demonstrate that the VAN can accurately capture bimodal long-time states in this high-dimensional network. Further joint-distribution comparisons are provided in Appendix~\ref{app:other_methods} and Fig.~\ref{fig:appendix_mapk}.

In ACME method~\cite{cao_accurate_2016}, the CME is propagated using matrix-based exponential integration with built-in error control, which often permits relatively large time steps (e.g., $\Delta t=10$ in the MAPK example). Its scalability relies on organizing the reachable state space into dynamically constructed buffers, together with prescribed upper bounds for species populations. For high-dimensional or densely coupled systems, determining these bounds and managing the resulting buffer structure may become increasingly involved. In contrast, NNCME-2 advances the CME by training a VAN to match the one-step propagated distribution at each time point. Because this propagation is based on the update $(I+\delta t\,\mathbb{W})P$, numerical stability requires a relatively small time step to ensure that probability increments remain nonnegative. Although this necessitates retraining at every step, the approach avoids assumptions about reaction-network topology and only requires constraints on species upper counts.

Also, for tensor-network formulation~\cite{nicholson_quantifying_2023}, it typically relies on explicit factorizations that align with local interaction patterns, which can become difficult to construct for dense or highly nonlocal topologies such as the MAPK cascade. Instead, the VAN can be trained directly on these systems without requiring structural simplifications, well-suited to reaction systems with complex or irregular connectivity.

\subsection{Spatially extended Schl\"{o}gl model: rare probability}
We next studied the spatially extended reaction-diffusion systems, using the Schl\"{o}gl model~\cite{schlogl_chemical_1972, vellela_stochastic_2009, nicholson_quantifying_2023} with bistability as an example. The well-mixed case of its reaction network involves three chemical species, where the concentrations of species $A$ and $B$ are held constant at $[A]=[B]=1$, while the count of species $X$ evolves stochastically through the following reversible reactions:
\begin{equation}
    2X + A \xrightleftharpoons[\tilde{c}_2]{\tilde{c}_1} 3X, \quad
    B \xrightleftharpoons[\tilde{c}_4]{\tilde{c}_3} X.
\end{equation}

We first consider the system on a one-dimensional lattice consisting of $L_{\text{lattice}}$ voxels. Each voxel hosts a well-mixed copy of the Schl\"{o}gl reaction network and is coupled to its nearest neighbors via diffusion of species $X$. The dynamics of the system are thus governed by both local chemical reactions and inter-voxel diffusion events:
\begin{equation}
    X_i \xrightleftharpoons[d]{} X_{i\pm1},
\end{equation}
where $X_i$ denotes the number of $X$ molecules in voxel $i$, and $d$ is the diffusion rate. The corresponding schematic of the reactions is in Fig.~\ref{Schlogl}(a). The parameters are $c_{1}=2.676$,  $c_{2}=0.040$, $c_{3}=108.102$, $c_{4}=37.881$ and $d=8.2207$~\cite{nicholson_quantifying_2023}.

\begin{figure}[ht!]
{\includegraphics[width=1\linewidth]{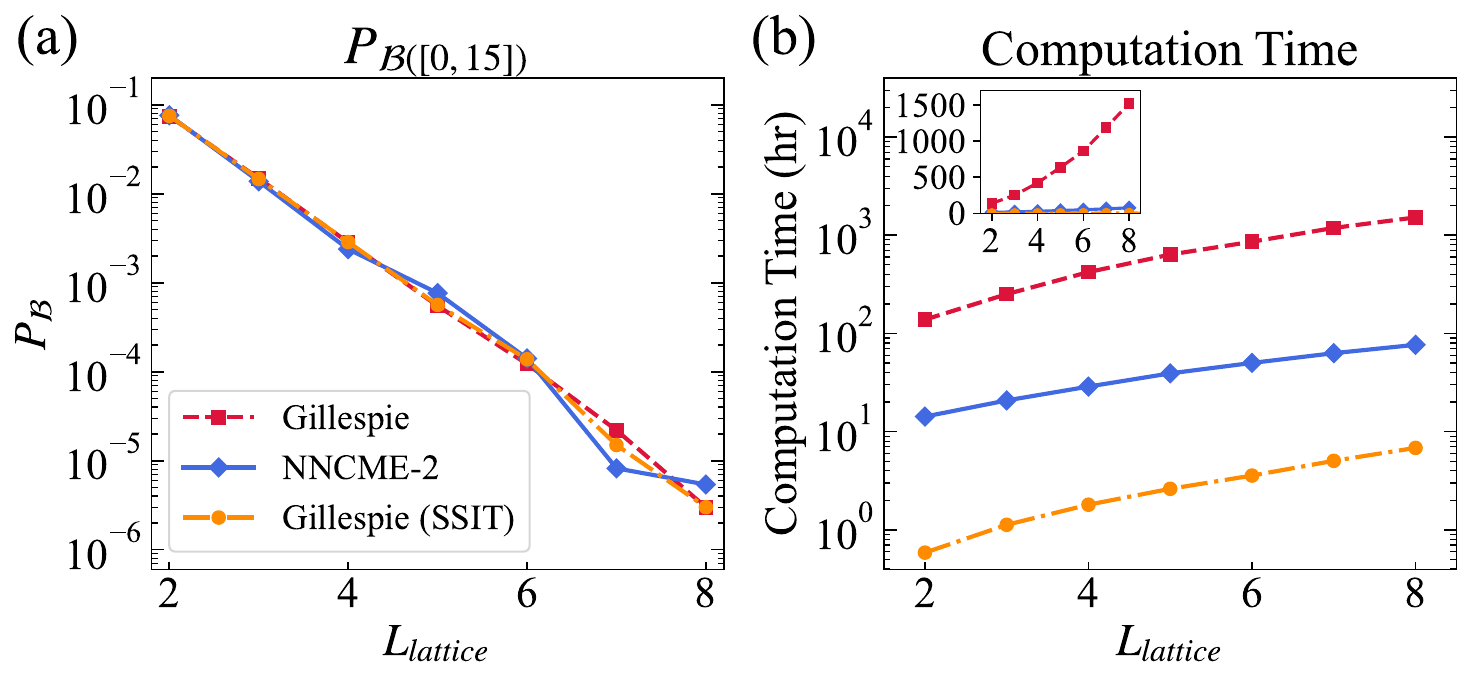}}
\centering
\caption{The scaling of rare probabilities and computational time for the one-dimensional Schl\"{o}gl model. 
(a) Probability mass $P_\mathcal{B}$ in $\mathcal{B}=\{\mathbf{X}\mid 0 \le X_i \le 15,\ \forall i\}$ across lattice size $L_{\text{lattice}}$, from vanilla Gillespie simulations (red), NNCME-2 (blue), and Gillespie (SSIT~\cite{popinga2026stochastic}) with an optimized SSA implementation (orange).
(b) Total computation time (log scale), with a linear-scale view shown in the inset. 
Both Gillespie implementations used $1\times10^6$ trajectories. The original Gillespie simulations were run on an Intel Xeon Gold 6248 CPU node, SSIT calculations were performed in MATLAB on a single-processor MacBook Pro, and NNCME-2 computations were performed on an NVIDIA Tesla V100 GPU node. Enhanced sampling is applied to all lattice sizes using the mixture ES scheme, with 15\% exploratory samples per training batch.
}
    \label{Schlogl_k}
\end{figure}

Because the probability distribution in these bistable systems is dominated by high-occupancy states, direct sampling is inefficient in probing the transition region. We therefore used the Schl\"{o}gl system to examine how enhanced sampling improves the exploration of rare configurations.
Fig.~\ref{Schlogl} shows the results for a six-site lattice using mixture enhanced sampling. The enhanced sampling improves the performance of the model while preserving its accuracy in well-sampled regions. The mean and variance of species counts across sites are consistent with those obtained from Gillespie simulations, indicating that the enhanced sampling does not distort the statistics of the dominant metastable state. The model can recover the small secondary peak in the marginal distribution, corresponding to the rare transition between the two stable states [Fig.~\ref{Schlogl}(b)]. 

\begin{figure}[ht]
\centering
\includegraphics[width=0.65\linewidth]{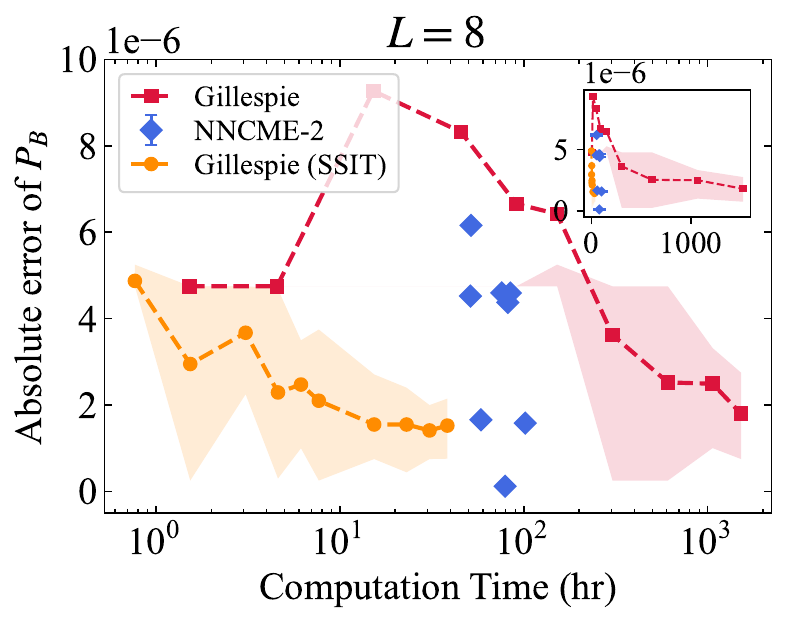}
\caption{
Absolute error of the rare-event probability $P_B$ for the one-dimensional Schl\"ogl model with $L=8$ as a function of computation time. Results from the vanilla Gillespie implementation, Gillespie (SSIT), and NNCME-2 are compared with the extrapolated reference $P_B^{\mathrm{ref}}=4.75\times10^{-6}$. The Gillespie and SSIT curves with shaded regions show the variation over 20 resampled estimates. The NNCME-2 points correspond to different training settings. The inset shows the same results on linear axes. The computational platforms follow those described for Fig.~\ref{Schlogl_k}.
}
\label{fig:schlogl_error}
\end{figure}

\begin{figure*}[ht!]
    \centering
    \includegraphics[width=\textwidth]{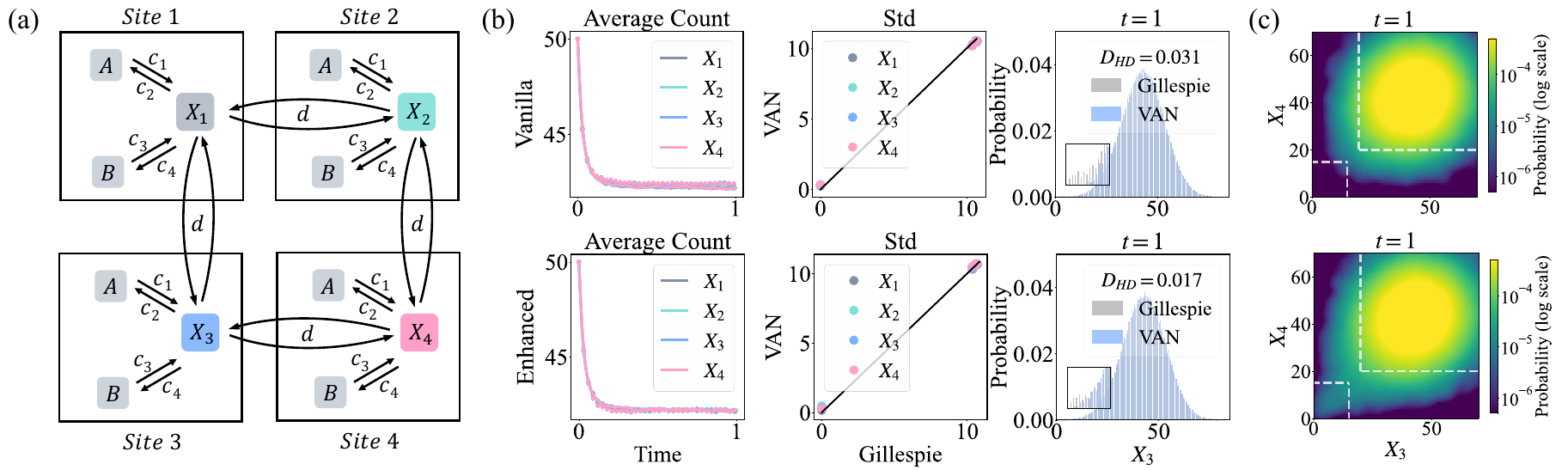}
\caption{NNCME-2 captures rare events in the two-dimensional spatially extended Schl\"{o}gl model ($2\times 2$ grid).
(a) A schematic of the 2d well-mixed Schl\"{o}gl model. The parameters are the same as Fig.~\ref{Schlogl}.
(b) Comparison between the vanilla VAN (top) and the enhanced VAN (bottom), using mixture enhanced sampling with 15\% exploratory samples per training batch.
The left and middle panels show the time evolution of the mean counts and the corresponding standard deviations of sites $X_1$-$X_4$, respectively. 
The right panels display the marginal distribution of $X_3$ at $t=1$, with the Hellinger distance $D_{\mathrm{HD}}$ between VAN and Gillespie results indicated. 
(c) Joint distributions of $X_3$ and $X_4$ from the enhanced VAN at $t=1$, shown on a logarithmic probability scale. 
}
    \label{Schlogl_2d}
\end{figure*}

The joint distributions in Fig.~\ref{Schlogl}(c) demonstrate that the enhanced VAN accurately captures the evolution of the probability distribution. 
Starting from a narrow distribution around the initial state, the probability gradually broadens and shifts toward the alternative state, where finite density appears in the low-$X$ region, indicating rare transition between metastable states. 
The enhanced sampling allows the VAN to recover both dominant and low-probability configurations while reducing computational cost compared with Gillespie simulations. We also examined alternative enhanced-sampling schemes, such as the diffusive ES and the $\alpha$ ES, as exemplified in the Schl\"{o}gl (2 sites) model. Detailed implementations and the corresponding hyperparameters are in Appendix~\ref{app:enhanced_sampling}.

\subsubsection{Rare probability and transition rate}
The bistable Schl\"{o}gl system also provides a useful testbed for evaluating how well the neural-network approach can capture rare events and estimate transition rates, particularly as the lattice size increases. 
Similar to the rate calculations for the Schl\"ogl model in~\cite{nicholson_quantifying_2023}, we define two macrostates corresponding to the two stable states. For a lattice configuration $\mathbf{X}=(X_1,\ldots,X_{L_{\mathrm{lattice}}})$, the high-occupancy macrostate is defined as $\mathcal{A}=\{\mathbf{X}\mid 25 \le X_i \le 85,\ \forall i\}$, and the low-occupancy macrostate as $\mathcal{B}=\{\mathbf{X}\mid 0 \le X_i \le 15,\ \forall i\}$. Starting from configurations in $\mathcal{A}$, the probability of finding the system in $\mathcal{B}$ at time $t$ is denoted by $P_{\mathcal{B}|\mathcal{A}}(t)$.
The unidirectional transition rate from $\mathcal{A}$ to $\mathcal{B}$ is defined from the growth of this probability after the microscopic transient,
\begin{equation}
  k_{\mathcal{B}\mathcal{A}} \;=\; \left.\frac{d}{dt}\, P_{\mathcal{B}|\mathcal{A}}(t)\right|_{t>\tau_{\rm mol}},
  \label{eq:kBA_def}
\end{equation}
so that in the intermediate-time window with clear time-scale separation,
\begin{equation}
  \tau_{\rm mol} \ll t \ll k_{\mathcal{B}\mathcal{A}}^{-1}
  \qquad\Rightarrow\qquad
  P_{\mathcal{B}|\mathcal{A}0}(t)\;\approx\; k_{\mathcal{B}\mathcal{A}}\, t .
  \label{eq:linear_window}
\end{equation}
The slope of this linear increase provides an estimate of $k_{\mathcal{B}\mathcal{A}}$, while the steady-state probability mass in $\mathcal{B}$ reflects the relative weight of the low-$X$ state.

We computed the total probability in the region $\mathcal{B}$ for lattice sizes ranging from $L_{\text{lattice}}=2$ to $8$. 
The results are shown in Fig.~\ref{Schlogl_k}(a) together with reference Gillespie simulations based on $1\times10^6$ trajectories. We additionally include the optimized Gillespie simulations using SSIT~\cite{popinga2026stochastic} as a modern implementation.
As $L_{\text{lattice}}$ increases, $P_\mathcal{B}$ decreases rapidly, consistent with the exponential suppression of rare configurations in spatially coupled systems. 
The scaling of $P_\mathcal{B}$ with $L_{\text{lattice}}$ also closely resembles the scaling of the transition rate estimated using the tensor-network approach~\cite{nicholson_quantifying_2023}. At smaller lattice sizes, all three methods give similar estimates of $P_\mathcal{B}$. At larger lattice sizes, the increasing rarity of $\mathcal{B}$ leads to larger fluctuations and errors: NNCME-2 shows some loss of accuracy, while the estimate from $10^6$ Gillespie trajectories also becomes underestimated at $L_{\text{lattice}}=8$.
Fig.~\ref{Schlogl_k}(b) reports the corresponding computation time.
The computational cost of the vanilla Gillespie implementation increases rapidly, while the optimized SSIT implementation substantially reduces the absolute simulation time. Notably, the runtime of NNCME-2 increases more gradually with lattice size. This different growth trend is encouraging for extending NNCME-2 to larger lattices.

To compare accuracy and computation time more directly, we examine the error in estimating the rare-event probability $P_B$ for the $L=8$ system. Since direct Gillespie sampling becomes poorly resolved in this regime, we use $P_B^{\mathrm{ref}}=4.75\times10^{-6}$ as the reference value, obtained by linearly fitting $\log_{10} P_B$ versus $L$ using the Gillespie results at $L=2$ and $L=3$ and extrapolating to $L=8$.
Fig.~\ref{fig:schlogl_error} plots the corresponding absolute error against computation time. For the vanilla Gillespie implementation, we increase the number of trajectories up to $10^6$, while for Gillespie (SSIT) we use up to $5\times10^6$ trajectories. The shaded regions show the variation over 20 resampled estimates. 
For NNCME-2, the points correspond to different training settings with batch sizes ranging from 2000 to 5000. For each setting, we use the networks saved at every training epoch of the final time step and estimate $P_B$ by sequential Monte Carlo (SMC). SMC estimates the rare-region probability through sequential sampling and weighting, avoiding the need for a very large number of direct network samples to fall into $\mathcal{B}$. The plotted value is obtained from the errors across these saved networks, while the error bars indicate their variation and are mostly too small to be visible beyond the markers.

As illustrated, for both Gillespie implementations, increasing the number of trajectories generally reduces the error at the cost of longer computation time. The optimized SSIT implementation reaches good accuracy much more rapidly than the vanilla Gillespie code, although increasing the trajectory number further from $10^6$ to $5\times10^6$ gives only a modest additional reduction in error for this very rare event.
The NNCME-2 results achieve comparable accuracy and computation time to the optimized SSIT results, with some settings reaching small errors at similar computation times. At the same time, the spread among the NNCME-2 points reveals that its accuracy is sensitive to the training settings.
For NNCME-2, higher accuracy generally requires larger networks or batches, stronger sampling, or smaller propagation time steps, all of which increase computational cost.
Future work may reduce this cost by exploiting generalization across lattice sizes. Transformer-based models may also learn the full time-dependent distribution directly and support parameter inference through the differentiability of neural networks. In addition, automated hyperparameter selection, for example through feedback-driven optimization, may help improve training robustness when no accurate ground truth is available.

All original Gillespie simulations were executed on a single Intel Xeon Gold CPU node, while NNCME-2 computations were performed on a single NVIDIA Tesla V100 GPU. 
The Gillespie simulations implemented using SSIT were performed on a MacBook Pro 16GB using a single processor.
Because these calculations were performed using different software implementations and hardware platforms, their absolute runtimes should not be interpreted as a direct comparison of efficiency. The same consideration applies to the tensor-network approach~\cite{nicholson_quantifying_2023}, which was performed on CPU hardware.

\subsubsection{Two-dimensional lattice}
Since our framework extends naturally to higher-dimensional spatial systems, we next investigate the two-dimensional Schl\"ogl model. 
The system is placed on a 2D grid, where each voxel hosts a well-mixed Schl\"ogl reaction network and species $X$ diffuses between neighboring voxels in both horizontal and vertical directions. 
The results for the 2D Schl\"ogl model on a $2\times 2$ grid  are illustrated in Fig.~\ref{Schlogl_2d}(a).
Fig.~\ref{Schlogl_2d}(b) compares the vanilla and enhanced VAN: the enhanced version yields more accurate means and standard deviations, and reduces the Hellinger distance of the marginal distribution. The joint distribution of $(X_3, X_4)$ in Fig.~\ref{Schlogl_2d}(c) further demonstrates that NNCME-2 can capture the tail structure of the distribution on the two-dimensional lattice.

\begin{figure}[ht!]
{\includegraphics[width=1\linewidth]{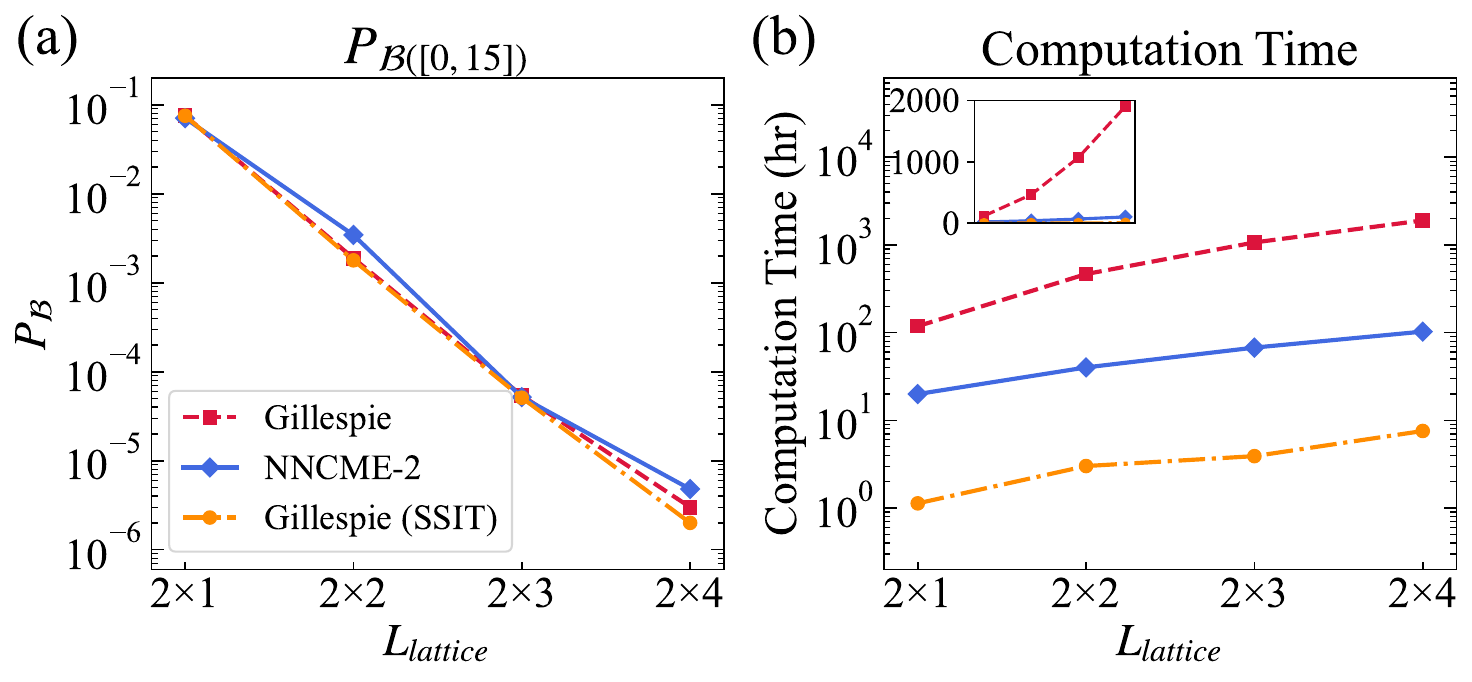}}
\centering
\caption{The scaling of rare probabilities and computational time for the two-dimensional spatially extended Schl\"{o}gl model.
(a) Probability mass $P_\mathcal{B}$ obtained using the vanilla Gillespie implementation, NNCME-2, and Gillespie (SSIT). Both Gillespie implementations use $1\times10^6$ trajectories.
(b) The total computation times of the three methods on a logarithmic scale, with the inset showing the same data on a linear scale. The computational platforms follow those described for Fig.~\ref{Schlogl_k}.
All NNCME-2 results use mixture enhanced sampling with 15\% exploratory samples per training batch.
}
\label{Schlogl2d_k}
\end{figure}

We further investigate the rare-event probability and computation time for the two-dimensional Schl\"ogl model in Fig.~\ref{Schlogl2d_k}, as the lattice size increases from $2\times1$ to $2\times4$.
The results show trends similar to the one-dimensional case: NNCME-2 reproduces the rare-region probability $P_\mathcal{B}$ over the lattice sizes considered, with the corresponding computation times shown in Fig.~\ref{Schlogl2d_k}(b).
As discussed previously, the tensor-network method~\cite{nicholson_quantifying_2023} produced highly accurate transition-rate estimates for one-dimensional Schl\"ogl systems, owing to the efficiency of MPS representations for 1D chains.
However, extending these approaches beyond one dimension can be challenging: lattice loops and higher connectivity in 2D drive rapid growth in the required bond dimension and make tensor contraction computationally expensive~\cite{nicholson_quantifying_2023, PhysRevLett.98.140506, PhysRevB.109.235102}.
Since NNCME-2 does not depend sensitively on the network structure to construct the CME operator for neural-network training, it extends more naturally to higher-dimensional and densely coupled systems.
These results thus highlight the structural flexibility of the neural-network representation for higher-dimensional stochastic dynamics.

\section{Discussion}

NNCME-2 provides a general and efficient framework for learning the joint probability distribution of chemical reaction networks. The method integrates faster neural architectures with more effective optimization and sampling strategies, addressing the two challenges of high dimensionality and rare events within a unified framework. In examples, it is flexible for large and complex systems such as the MAPK cascade, which contains multiple couplings and feedback, without the need for specific operator designs for the implementation. The probability distribution in CME systems can vary rapidly over time, particularly in models such as MAPK, which constrains the allowable timestep $\delta t$ through the reaction propensities. While our experiments employed a fixed timestep, adaptive strategies are available to reduce the total number of training steps~\cite{tang_neural-network_2023}, for example, the timestep can be enlarged when the distribution evolves gradually. 

It is worth noting that neural-network models do not automatically overcome the challenge of rare-event exploration. The vanilla VAN learns accurately only in regions where training samples are available. The enhanced sampling is therefore essential: once rare configurations are included in the training set, the model can fit to those regions and provide a consistent global approximation of the full probability state space. This is demonstrated extensively in the example of Schl\"ogl system with rare events. Future developments of enhanced-sampling strategies can also be incorporated to improve the efficiency of capturing the rare events. Especially, the tensor networks can be helpful for drawing sufficient samples~\cite{chen2025tensor} to better train the neural network.

For reaction-diffusion systems, NNCME-2 can handle the Schl\"ogl model on both one- and two-dimensional lattices, where the accuracy in very low-probability regions may be further improved. On the other hand, while tensor-network approaches are highly accurate in one-dimensional systems, extending them to higher-dimensional lattices and general biochemical reaction networks is challenging, as the tensor structure needs to reflect the reaction topology, and the bond dimension can increase rapidly. Thus, combining neural and tensor-network methods by integrating the scalability of neural networks with the accuracy of tensor-network representations is a promising direction for studying more complex stochastic reaction-diffusion dynamics.

Also, several trajectory-driven learning approaches have been developed for stochastic reaction networks.
DeepCME~\cite{gupta_deepcme_2021} reformulates the CME via backward martingale representations along stochastic trajectories, enabling estimation of selected observables without modeling the full distribution.
From an operator-theoretic viewpoint, spectral Koopman methods~\cite{gupta_spectral_2025}, and their neural extension DeepSKA~\cite{badolle_interpretable_2025} extract dominant dynamical modes from ensembles of stochastic trajectories, thereby avoiding state-space truncation and enabling scalable treatment of large systems.
Here, NNCME-2 directly learns the full time-evolving joint distribution, capturing correlations, multimodality, and rare-event mass beyond trajectory-averaged information.
Future work may incorporate spectral structures or variance reduction techniques from Koopman-based methods into distribution-learning frameworks, achieving both accurate global distributions and efficient statistical estimation.

Another promising direction is to improve the probability decomposition for systems with strong cyclic probability flows. In the present NADE implementation, a fixed autoregressive ordering provides a valid factorization of the joint probability distribution, and works well for the examples studied here. However, for oscillatory reaction networks or systems with circulating probability currents, the physical dependencies among species may form loops rather than a single directed sequence. In such cases, the accuracy may be further improved by using order-averaged or adaptive-ordering autoregressive models, where the ordering is chosen according to reaction connectivity, spatial adjacency, or learned species correlations. More flexible decompositions, for example transformer-based architectures with adaptive dependency structures~\cite{LO_ARM}, may also provide an effective way to represent such cyclic probability flows.

Beyond chemical reaction networks, the present framework may be extended to a broader class of nonequilibrium stochastic dynamics. Representative examples include reaction--diffusion and pattern-formation processes in soft and condensed-matter systems~\cite{cross_pattern_1993}, interfacial and surface-reaction dynamics~\cite{2010stochastic_reaction-diffusion}, and nonequilibrium phenomena involving collective excitations~\cite{kamenev_field_2011}, such as skyrmion formation~\cite{nagaosa_topological_2013, fert_magnetic_2017} and dynamical phase transitions~\cite{zhu2025universaltwostagedynamicsphase,PhysRevB.111.184415}. Another promising direction is to analyze the learning dynamics in the neural-network parameter space itself, which may shed light on optimization efficiency and representational capacity from a dynamical systems perspective~\cite{li2025weightflow}. Together, these directions highlight several avenues through which neural-network-based approaches can be further developed to tackle general stochastic dynamics.

\section*{Acknowledgments}
We acknowledge Jie Liang, Ali Farhat and Online Club Nanothermodynamica for helpful discussions. 
We thank the anonymous reviewer for mentioning the SSIT~\cite{popinga2026stochastic} package and its efficient SSA implementation.
This work was supported by Projects No.12322501, No.12575035, No.12405047, No.12325501, No.12247104, No.12447101 and No.T2293771 of National Natural Science Foundation of China, and No.2026NSFSCZY0124 of the Natural Science Foundation of Sichuan Province. 
P.Z. is supported by the Strategic Priority Research Program of Chinese Academy of Sciences, Grant No.XDB1680000.
The HPC was supported by Dawning Information Industry Corporation Ltd, and the Center for HPC at University of Electronic Science and Technology of China.

\section*{Data availability}
The data supporting this study are available within the paper and in the NNCME repository on GitHub~\cite{NNCME2_2025}. The PyTorch implementation of NNCME-2 is available in the same repository.

\appendix 
\section*{Appendix}
The appendix provides detailed elaborations on the key methodological components of the main study. 
To facilitate navigation, its structure is organized as follows. 
We begin by summarizing all mathematical symbols used throughout the manuscript in Table~\ref{tab:symbols} for ease of reference. Appendix~\ref{sec:appendix_NADE_width} examines how the width hyperparameter of NADE affects model performance. 
Next, in Appendix~\ref{sec:kl_analysis}, we analyze the implications of employing different variants of the KL divergence as evaluation metrics. 
Subsequently, Appendix~\ref{app:optim} investigates the influence of optimization algorithms on computational precision and efficiency. 
Furthermore, Appendix~\ref{app:enhanced_sampling} evaluates four sampling methods and assesses how their hyperparameters influence the resulting outcomes. 
In addition, Appendix~\ref{sec_jax_vs_torch} compares the computational efficiency and numerical accuracy of implementations developed using the PyTorch and JAX frameworks. 
Finally, Appendix~\ref{sec:example} provides detailed descriptions of the chemical reaction systems discussed in the main text.

\section{Effect of NADE's width}
\label{sec:appendix_NADE_width}
In this section, we discuss how the width of the NADE network-defined as the number of hidden units per layer-affects the learning process in the toggle switch system. The network width fundamentally determines the model's performance, as a wider network, in principle, can capture more complex joint probability distributions. However, excessive width increases the number of trainable parameters, potentially leading to overfitting and more computational costs. To assess the trade-off, we compare network widths of $8$, $64$ and $128$, with depth fixed as $1$.

\begin{table}[t]
\centering
\renewcommand{\arraystretch}{1.15}
\begin{tabular}{ll}
\toprule
\textbf{Symbol} & \textbf{Description} \\
\midrule
$\eta$ & Learning rate used in optimization. \\
$\delta t$ & Discrete time step of CME propagation. \\
$T_{\mathrm{step}}$ & Number of discrete time steps in the evolution. \\
$t$ & Physical time, $t=\delta t\,T_{\mathrm{step}}$. \\
$N_b$ & Number of samples drawn per training step. \\
$N_p$ & Total number of trainable parameters. \\
$\mathbb{T}$ & One-step evolution operator, $\mathbb{T}=e^{\delta t\,\mathbb{W}}$. \\
$\mathbb{W}$ & Generator (transition-rate matrix) of the CME. \\
$a_k$ & Propensity of reaction $k$. \\
$s_k$ & Stoichiometric change vector of reaction $k$. \\
$\mathbf{n}$ & State vector of species copy numbers. \\
$L$ & Number of species in the reaction network. \\
$M$ & Upper count of species. \\
$H$ & Hidden-layer width of the NADE. \\
$\hat{P}^{\theta_t}_t$ & Variational distribution at time $t$. \\
$S$ & Fisher information matrix. \\
$L_{\mathrm{lattice}}$ & Size of the spatial lattice. \\
\bottomrule
\end{tabular}
\caption{Mathematical symbols used throughout the manuscript.}
\label{tab:symbols}
\end{table}

\begin{figure}[ht]
{\includegraphics[width=1\linewidth]{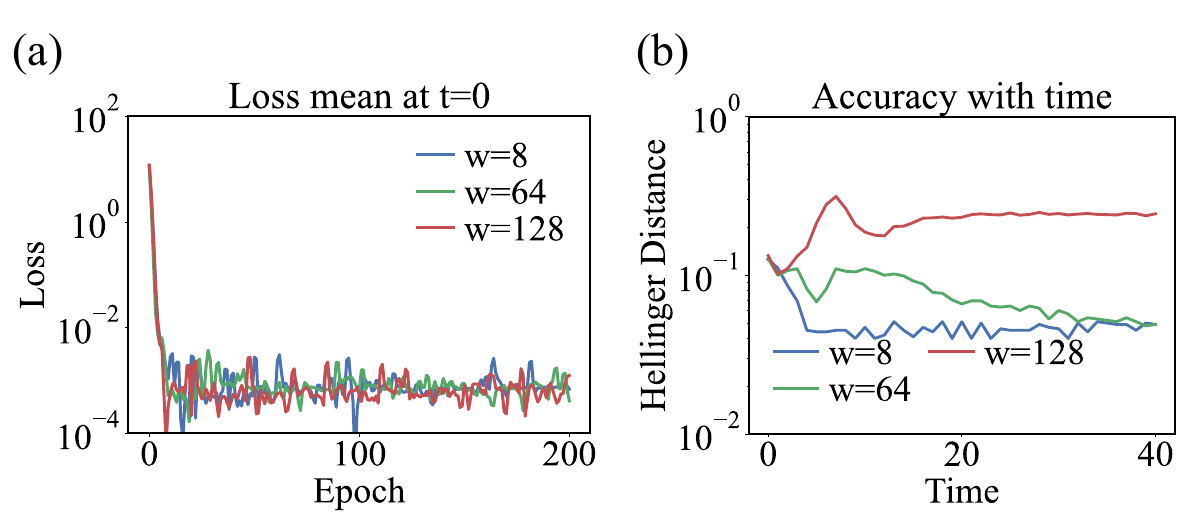}}
\caption{Effect of network width on the learning dynamics and accuracy of the toggle-switch system.
(a) Evolution of the average loss during training of the initial distribution ($t=0$) for different network widths. 
(b) Evolution of the Hellinger distance over time between the VAN-generated distribution and the Gillespie reference for different network widths.
In these experiments, the VAN model employs NADE architecture ($\text{depth}=1$ and $N_b=2000$). Training utilizes NG and reverse KL, with $\text{epoch}=50$, $\eta$ = 0.5 at $t=0$ and $\text{epoch}=5$, $\eta$ = 0.5 for $t>0$.
}
\label{fig:F_width}
\end{figure}

Fig.~\ref{fig:F_width} reveals that networks with different widths exhibit similar requirements for training epochs to achieve convergence. Notably, increasing the network width beyond a certain threshold does not necessarily result in improved performance. For example, the network with $\text{width}=128$ shows no great improvement over a $\text{width}=64$ but incurs higher computational cost. This finding underscores that while sufficient network capacity is essential for accurately modeling joint probability distributions, an overly large parameter space may hinder efficient training in the time evolution.

\section{KL-Divergence variants as loss function for VAN}
\label{sec:kl_analysis}
\begin{figure*}[t]
    \centering
    \includegraphics[width=\textwidth]{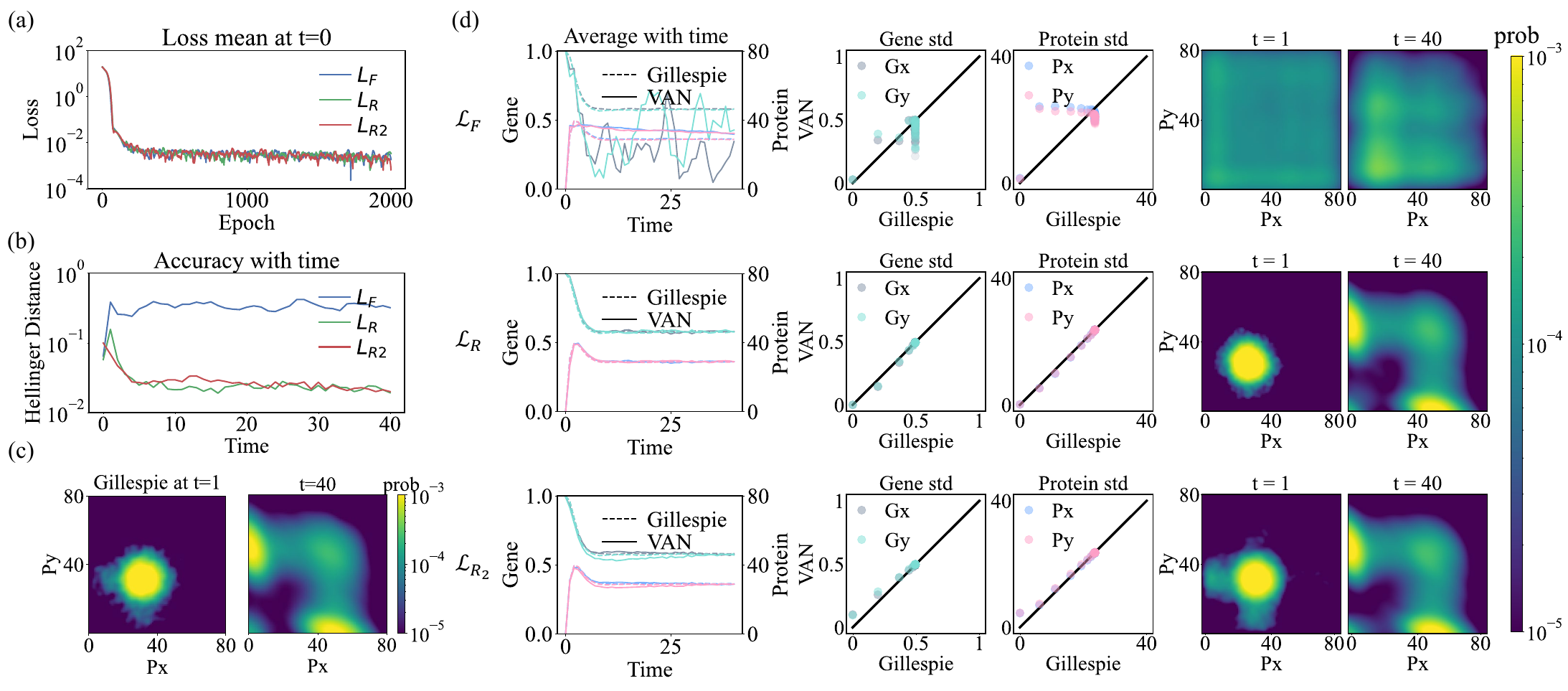}
    \caption{Evaluation of KL-divergence loss functions in learning the dynamics of the toggle switch system. 
    (a) Mean of training loss at the initial time step $t=0$. 
    (b) The accuracy of the VAN-generated distribution over time, quantified by the Hellinger distance from the benchmark $10^4$ Gillespie trajectories. 
(c) Joint probability heatmap of $P_x$ and $P_y$ for the benchmark Gillespie simulation at $t=1$ (left) and $t=40$ (right). 
(d)  The time evolution of the average count for the genes and proteins, from the VAN (lines) and the Gillespie simulation (dotted line) (left).
Standard deviations comparisons (middle). 
The heatmaps depict the joint distributions of $P_x$ and $P_y$ at time points $t=1$ and $t=40$ (right).
In these experiments, the VAN model employs NADE architecture ($\text{width}=16$, $\text{depth}=1$ and $N_b=2000$). Training utilizes SGD, with $\text{epoch}=2000$, $\eta$ = 0.005 at $t=0$ and $\text{epoch}=100$, $\eta$ = 0.005 for $t>0$.
}
    \label{fig:kl-losses}
\end{figure*}

Kullback-Leibler (KL) divergence~\cite{kullback_information_1951} is a suitable distance metric and widely applicable, defined as:
$$
D_{\mathrm{KL}}(P \,\|\, Q)
= \sum_{x} P(x)\,\log\!\left(\frac{P(x)}{Q(x)}\right),
$$
where $P$ and $Q$ denote two probability distributions, and the divergence quantifies the information loss incurred when $Q$ is used to approximate $P$. Notably, it is important to note that the forward KL-divergence $D_{KL}[\mathbb{T}\hat{P}^{\theta_t}_t\,\big|\big|\,\hat{P}^{\theta_{t+\delta t}}_{t+\delta t}] $ is not equivalent to the reverse KL-divergence $D_{KL}[ \hat{P}^{\theta_{t+\delta t}}_{t+\delta t}\,\big|\big|\,   \mathbb{T}\hat{P}^{\theta_t}_t ] $.  This subsection compares three KL-divergence variants in training VAN and analyzes their formulations and empirical performance in the toggle switch system.

\subsection{Forward KL-divergence}
\label{subsec:fkl}
Variational inference commonly employs reverse KL-divergence as the loss function because sampling from the VAN-generated distribution is more straightforward than sampling from posterior distributions. Nevertheless, serving reverse KL-divergence as loss function may lead to an underestimation of the posterior uncertainty~\cite{minka_divergence_2005, naesseth_markovian_2020}. Consequently, some recent studies started to employ forward KL-divergence as the loss function for variational inference~\cite{gu_neural_2015}. To address the sampling issue of the variational inference problem, these studies usually utilize Markov Chain Monte Carlo (MCMC) sampling from the posterior distribution~\cite{naesseth_markovian_2020}. Similarly, from the well-trained VAN distribution $\hat{P}^{\theta_t}_t$, we can calculate samples from $\mathbb{T}\hat{P}^{\theta_t}_t$ through the transition operator $\mathbb{T}=e^{\delta t \mathbb{W}}\approx (I+\delta t\mathbb{W})$, where the generator $\mathbb{W}$ is given by Eq.~\eqref{CME}. 

The forward KL-divergence in our problem is defined as:
\begin{align}
\label{eq:F_Loss}
\mathcal{L}_F &= D_{KL}\left[\mathbb{T}\hat{P}^{\theta_t}_t\,\big|\big|\,\hat{P}^{\theta_{t+\delta t}}_{t+\delta t}\right] \\
&= \mathbb{E}_{\mathbf{s}\sim\mathbb{T}\hat{P}^{\theta_t}_t}\left[\ln(\mathbb{T}\hat{P}^{\theta_t}_t)(\mathbf{s}) - \ln\hat{P}^{\theta_{t+\delta t}}_{t+\delta t}(\mathbf{s})\right].
\end{align}

Here, $\mathbb{T} = e^{\delta t \mathbb{W}} \approx I + \delta t \mathbb{W}$ represents the transition operator, preserving the total probability. The expectation is estimated through the VAN-generated distribution $\hat{P}^{\theta_t}_t$ of the previous time step.

The parameter of VAN updating rule derives from the gradient:
\begin{align}
\label{eq:F_Training}
\nabla_{\theta_{t+\delta t}}\mathcal{L}_F &= -\nabla_{\theta_{t+\delta t}} \mathbb{E}_{\mathbf{s}\sim\mathbb{T}\hat{P}^{\theta_t}_t}\left[\ln\hat{P}^{\theta_{t+\delta t}}_{t+\delta t}(\mathbf{s})\right].
\end{align}
Let $P$ denote the target distribution and $Q$ denote the variational distribution. 
The forward KL divergence $$
D_{\mathrm{KL}}(P \,\|\, Q) = \sum_{x} P(x)\,\log\!\frac{P(x)}{Q(x)}$$ exhibits mode-covering behavior because:
\begin{enumerate}
    \item The expectation is taken with respect to $P(x)$, so the divergence is dominated by regions where $P$ assigns significant probability mass.
    \item When $P(x) > 0$ but $Q(x) \approx 0$, the term $\log\!\bigl(P(x)/Q(x)\bigr)$ becomes very large, causing the divergence to blow up.
    \item This creates extremely high penalties when $Q$ fails to place probability mass in regions where $P$ has support.
    \item As a result, $Q$ is driven to ``cover'' all regions in which $P$ has substantial probability mass.
\end{enumerate}
Consequently, $Q$ tends to spread its probability mass across all modes of $P$, even if this implies assigning non-negligible probability to low-density regions between modes. When $Q$ has limited capacity, the resulting approximation can appear ``blurry,'' averaging over multiple modes rather than sharply resolving any single one.

At each time step $\delta t$, we expect to iteratively update the parameters of VAN to approximate the target distribution $\mathbb{T}\hat{P}^{\theta_t}_t$. When employing forward KL-divergence as loss, resampling is not required during the iterative training process, while reverse KL-divergence necessitates sampling from the updated parameter distribution at each epoch iteration.

\subsection{Reverse KL-divergence}
\label{subsec:rkl}

The reverse KL-divergence inverts the variational distribution and target distribution of Eq.~\eqref{eq:F_Loss}:
\begin{align}
\label{eq:R_Loss}
\mathcal{L}_R &= D_{KL}\left[\hat{P}^{\theta_{t+\delta t}}_{t+\delta t}\,\big|\big|\,\mathbb{T}\hat{P}^{\theta_t}_t\right] \\
&= \mathbb{E}_{\mathbf{s}\sim\hat{P}^{\theta_{t+\delta t}}_{t+\delta t}}\left[\ln\hat{P}^{\theta_{t+\delta t}}_{t+\delta t}(\mathbf{s}) - \ln\mathbb{T}\hat{P}^{\theta_t}_t(\mathbf{s})\right].
\end{align}

The gradient update rule becomes:
\begin{align}
\label{eq:R_Training}
\nabla_{\theta_{t+\delta t}}\mathcal{L}_R &= \mathbb{E}_{\mathbf{s}\sim\hat{P}^{\theta_{t+\delta t}}_{t+\delta t}}\left[\nabla_{\theta_{t+\delta t}}\ln\hat{P}^{\theta_{t+\delta t}}_{t+\delta t}(\mathbf{s}) \cdot \Delta(\mathbf{s})\right],
\end{align}
where $\Delta(\mathbf{s}) = \ln\hat{P}^{\theta_{t+\delta t}}_{t+\delta t}(\mathbf{s}) - \ln\mathbb{T}\hat{P}^{\theta_t}_t(\mathbf{s})$.

Let $P$ denote the target distribution and $Q$ denote the variational distribution. 
The reverse KL divergence
$$
D_{\mathrm{KL}}(Q \,\|\, P) = \sum_{x} Q(x)\,\log\!\frac{Q(x)}{P(x)}
$$
exhibits mode-seeking behavior because:
\begin{enumerate}
    \item The expectation is taken with respect to $Q(x)$, so the divergence is dominated by regions where $Q$ assigns significant probability mass.
    \item When $Q(x) > 0$ but $P(x) \approx 0$, the term $\log\!\bigl(Q(x)/P(x)\bigr)$ becomes very large.
    \item This produces strong penalties when $Q$ assigns probability mass to regions where $P$ has little or no support.
    \item By contrast, there is no direct penalty when $Q$ assigns negligible probability to regions where $P$ has mass (i.e., when $Q(x) \approx 0$).
\end{enumerate}
Consequently, minimizing the reverse KL divergence typically drives $Q$ to concentrate on a subset of the modes of $P$, often focusing on a single dominant mode rather than covering all modes. In multimodal settings, this mode-seeking behavior enables $Q$ to represent one mode of $P$ accurately while ignoring others and avoiding low-density regions between modes.

\subsection{Measure-transformed reverse KL-divergence}
\label{subsec:rkl2}

Forward KL offers a sampling advantage in our temporal setting, because samples can be obtained from $\mathbb{T}\hat{P}^{\theta_t}_t$ without resampling from the evolving model at each inner training step. However, its mode-covering tendency may lead to overly smoothed approximations in systems with pronounced multimodality. To combine the sampling convenience of the forward direction with the mode-seeking behavior of reverse KL, we consider a measure-transformed reverse KL based on importance sampling~\cite{schulman_proximal_2017}.

The transformed loss is
\begin{align}
\label{eq:R2_Loss}
\mathcal{L}_{R_2} 
&= \mathbb{E}_{\mathbf{s}\sim\mathbb{T}\hat{P}^{\theta_t}_t}\left[\frac{\hat{P}^{\theta_{t+\delta t}}_{t+\delta t}(\mathbf{s})}{\mathbb{T}\hat{P}^{\theta_t}_t(\mathbf{s})}\left(\ln\hat{P}^{\theta_{t+\delta t}}_{t+\delta t}(\mathbf{s}) - \ln\mathbb{T}\hat{P}^{\theta_t}_t(\mathbf{s})\right)\right],
\end{align}
where samples are drawn from $\mathbb{T}\hat{P}_t^{\theta_t}$, rather than the variational distribution $\hat{P}^{\theta_{t+\delta t}}_{t+\delta t}$.

The gradient becomes
\begin{align}
\label{eq:R2_Training}
\nabla_{\theta_{t+\delta t}}\mathcal{L}_{R_2} 
&= \mathbb{E}_{\mathbf{s}\sim\mathbb{T}\hat{P}^{\theta_t}_t}
\left[\frac{\hat{P}^{\theta_{t+\delta t}}_{t+\delta t}(\mathbf{s})}
{\mathbb{T}\hat{P}^{\theta_t}_t(\mathbf{s})}
\nabla_{\theta_{t+\delta t}}\ln\hat{P}^{\theta_{t+\delta t}}_{t+\delta t}(\mathbf{s})\,
\Delta(\mathbf{s})\right].
\end{align}
This construction keeps the reverse-KL structure in the integrand, while allowing sampling from a distribution aligned with the temporal evolution operator.

\subsection{Evaluate the performance of loss-variants applied to the toggle switch system}
\label{subsec:eval}

To evaluate these loss functions in practice, we apply the three KL variants to the genetic toggle switch system; more details of the setup are given in Sec.~\ref{subsec:Toggle}. The corresponding loss definitions are summarized in Table~\ref{tab:al-sum}.

\begin{table}[!ht]
\centering
\scriptsize 
\setlength{\tabcolsep}{4pt} 
\begin{tabular}{l@{\quad}c}
\toprule
\text{Loss Type} & \text{Loss} \\
\midrule
\text{ForwardKL}&
$\mathcal{L}_F = \mathbb{E}_{s \sim \mathbb{T}\hat{P}^{\theta_t}_t}
\big[ \ln \mathbb{T}\hat{P}^{\theta_t}_t(s)
- \ln \hat{P}^{\theta_{t+\delta t}}_{t+\delta t}(s) \big]$ \\

\text{ReverseKL}&
$\mathcal{L}_R = \mathbb{E}_{s \sim \hat{P}^{\theta_{t+\delta t}}_{t+\delta t}}
\big[ \ln \hat{P}^{\theta_{t+\delta t}}_{t+\delta t}(s)
- \ln \mathbb{T}\hat{P}^{\theta_t}_t(s) \big]$ \\

\text{ReverseKL2}&
$\mathcal{L}_{R_2} = \mathbb{E}_{s \sim \mathbb{T}\hat{P}^{\theta_t}_t}
\!\left[ \frac{\hat{P}^{\theta_{t+\delta t}}_{t+\delta t}(s)}{\mathbb{T}\hat{P}^{\theta_t}_t(s)}
\big( \ln \hat{P}^{\theta_{t+\delta t}}_{t+\delta t}(s)
- \ln \mathbb{T}\hat{P}^{\theta_t}_t(s) \big) \right]$ \\
\bottomrule
\end{tabular}
\caption{Summary of loss functions.}
\label{tab:al-sum}
\end{table}

Fig.~\ref{fig:kl-losses} compares the three KL-based objectives on the toggle-switch system.
Fig.~\ref{fig:kl-losses}(a) shows the training loss at the initial step $t=0$, where all three losses converge.
As shown in Fig.~\ref{fig:kl-losses}(b), reverse KL and reverse KL2 consistently achieve smaller Hellinger distances to the Gillespie benchmark than forward KL, indicating higher accuracy of the learned distributions over time. Fig.~\ref{fig:kl-losses}(d) further shows that reverse KL-based objectives more closely reproduce the statistics of the underlying dynamics, with improved agreement in both marginal moments and joint distributions. In particular, reverse KL and reverse KL2 successfully capture the bimodal equilibrium distribution of the toggle-switch system, whereas forward KL yields an over-smoothed approximation, consistent with its mode-covering behavior.

When used as the VAN loss, reverse KL-divergence can effectively learn multi-modal distributions due to its mode-seeking property, allowing the multi-modality of the toggle-switch system to emerge gradually during time evolution. As a result, the VAN-generated distribution progressively expands toward regions corresponding to multiple modes. 
Reverse KL2 offers additional computational advantages by avoiding re-sampling while retaining adequate accuracy in modeling the evolving distributions. Although this formulation inherently limits global distributional coverage, Fig.~\ref{fig:kl-losses}(c)-(d) shows that reverse KL2 provides a closer approximation to the joint distribution than reverse KL at early times ($t=1$).

\subsection{Comparison of evaluation metrics for distribution matching}

To assess the suitability of the Hellinger distance as an evaluation metric~\cite{hellinger_orthogonalinvarianten_1907}, we compare four probability distances: TV, KL divergence, JS divergence, and HD. Here, $P$ and $Q$ denote the two distributions under comparison.

\begin{figure}[ht]
    \centering
    \includegraphics[width=0.8\linewidth]{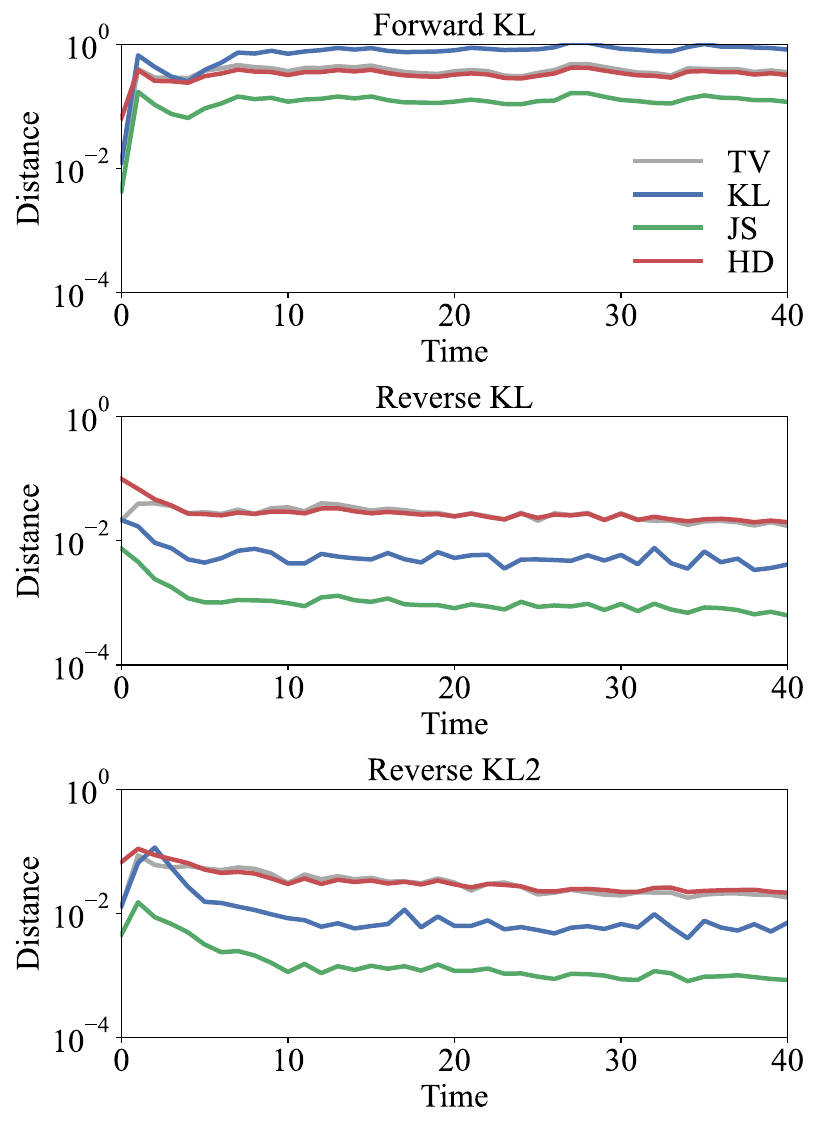}
    \caption{Validation of the Hellinger distance as a representative evaluation metric for the toggle-switch system. The VAN is trained using different loss functions based on variants of the KL divergence, and model performance is evaluated using multiple distance measures, including Total Variation (TV), Kullback--Leibler (KL), Jensen--Shannon (JS), and Hellinger distance (HD).
    The reference distribution is based on $10^4$ trajectories generated via the Gillespie algorithm.
In these experiments, the VAN model employs NADE architecture ($\text{width}=16, \text{depth}=1$ and $N_b=2000$). Training utilizes SGD, with $\text{epoch}=2000$, $\eta$ = 0.005 at $t=0$ and $\text{epoch}=100$, $\eta$ = 0.005 for $t>0$.}
    \label{fig:divergence_comparison}
\end{figure}
The four metrics are defined as follows:

\begin{itemize}
    \item TV distance:
    $$
    D_{\mathrm{TV}}(P, Q) = \frac{1}{2} \sum_{x} |P(x) - Q(x)|.
    $$
    TV measures the maximum discrepancy between two distributions, and it is bounded between 0 and 1.

    \item  KL divergence:
    $$D_{\mathrm{KL}}(P \| Q) = \sum_{x} P(x) \log \frac{P(x)}{Q(x)}.$$
    KL-divergence is asymmetric and penalizes underestimation by $Q$ relative to $P$.

    \item JS divergence:
    \begin{align}
    D_{\mathrm{JS}}(P \| Q) = \frac{1}{2} D_{\mathrm{KL}}(P \| M) + \frac{1}{2} D_{\mathrm{KL}}(Q \| M), \\ \quad \text{where } M = \frac{1}{2}(P + Q).
    \end{align}
    JS divergence symmetrizes KL-divergence and ensures boundedness between 0 and $\log 2$, which improves stability. It is especially useful when $P$ and $Q$ have disjoint supports.

    \item HD distance:
    $$
    D_{\mathrm{H}}(P, Q) = \frac{1}{\sqrt{2}} \left( \sum_{x} \left( \sqrt{P(x)} - \sqrt{Q(x)} \right)^2 \right)^{1/2}.
    $$
    Hellinger Distance is symmetric and bounded between 0 and 1. It maintains desirable geometric properties and is less sensitive to low-probability regions than KL, making it suitable for training generative models with sample noise.
\end{itemize}

Fig.~\ref{fig:divergence_comparison} demonstrates that, across the three independent training runs, the four distance metrics above exhibit a consistent tendency, thereby substantiating the adoption of the Hellinger distance as a reliable and representative measure.

\section{Optimization algorithms for VAN}
\label{app:optim}
This subsection provides details of NG and 
TDVP, and demonstrates their training effects in toggle switch system.

\subsection{Details of natural gradient}\label{app:natgrad}

To clearly illustrate the implementation of the efficient NG update within our framework, we follow the general principles of Ref.~\cite{liu_efficient_2025} while reformulating the derivation in a way suitable for VAN-based CME propagation.
The following derivation is conducted, taking the reverse KL-divergence as an example.
We denote the number of samples as $N_b$, and the number of VAN's parameters as $N_p$. Gradients can be estimated by samples $\mathbf{s}$ drawn from the variational distribution $\hat{P}^{\theta_{t+\delta t}}_{t+\delta t}$:
\begin{align}
\label{eq:NG_gradient}
\nabla_{\theta_{t+\delta t}}\mathcal{L} &= \mathbb{E}_{\mathbf{s}\sim\hat{P}^{\theta_{t+\delta t}}_{t+\delta t}}\left[\nabla_{\theta_{t+\delta t}}\ln\hat{P}^{\theta_{t+\delta t}}_{t+\delta t}(\mathbf{s}) \cdot \Delta(\mathbf{s})\right]\\
\label{eq:NG_gradient2}
&= \frac{1}{N_b}\sum_{i=1}^{N_b}\nabla_{\theta_{t+\delta t}}\ln\hat{P}^{\theta_{t+\delta t}}_{t+\delta t}(\mathbf{s}^i) \cdot  \Delta(\mathbf{s}^i).
\end{align}
We define $O_{ik} = ({1}/{\sqrt{N_b}}\nabla\ln \hat{P}^{\theta_{t+\delta t}}_{t+\delta t}(s^i))$ and $R_i=(1/{\sqrt{N_b}}\Delta(s^i))$ , where $O$ is a $N_b\times N_p$ matrix and R is a $N_b$ dimensional vector. Eq.~\eqref{eq:NG_gradient2} can be rewritten as:
\begin{equation}
\nabla_{\theta_{t+\delta t}}\mathcal{L} = O^T R.
\end{equation}
We consider the case where $N_p \gg N_b$, which is typical in the majority of neural-network models. The FIM from Eq.~\eqref{eq:Fisher} can be written as:
\begin{equation}
\label{eq:Fisher_simple}
S = O^TO,
\end{equation}
allowing the updated parameter $\delta \theta _t$ from Eq.~\eqref{eq:natgrad} to be represented as:
\begin{equation}
\label{eq:FIM_1}
\delta \theta_t = -\eta(O^TO)^{-1}O^T R.
\end{equation}
The computational complexity of Eq.~\eqref{eq:FIM_1} is $\mathcal{O}(N_p^3)$, so we leverage a linear algebra identity $(O^TO)^{-1}O^T=O^T(OO^T)^{-1}$~\cite{liu_efficient_2025}, then we have:
\begin{equation}
\label{eq:FIM2}
\delta \theta_t = -\eta O^T(OO^T)^{-1} R.
\end{equation}
The computational complexity of Eq.~\eqref{eq:FIM2} is $\mathcal{O}(N_b^3+N_pN_b^2)$, this method is more efficient than Eq.~\eqref{eq:FIM_1} when $N_p \gg N_b$. In the implementation, to ensure the stability of the gradient, a regularisation scheme needs to be introduced. Usually a non-negative damping factor $\xi$ is introduced $(OO^T+\xi I)^{-1}$ for numerical stability~\cite{liu_efficient_2025}. In our implementation, $(OO^T+\xi I)^{-1}$ is computed via Cholesky decomposition.

\begin{figure}[ht]
{\includegraphics[width=1\linewidth]{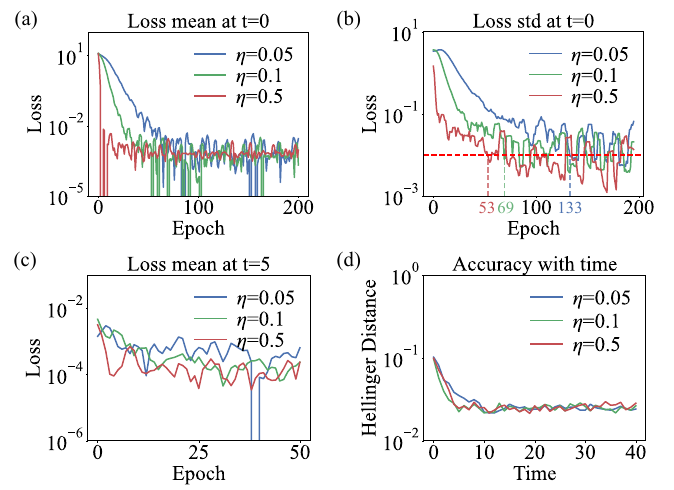}}
\caption{Effect of the learning rate $\eta$ on NG.
(a) Mean of loss at $t=0$ for different values of $\eta$.
(b) Standard deviation of the loss at $t=0$ for different values of $\eta$. The red dashed line marks a threshold of $1\times10^{-2}$, and the colored vertical lines indicate the first epoch at which the threshold is reached for the corresponding $\eta$.
(c) Mean of loss at $t=5$.  (d) The accuracy of the VAN-generated distribution for various $\eta$ over time, quantified by the Hellinger distance from the $10^4$ Gillespie trajectories. In these experiments, the VAN model employs NADE architecture ($\text{width}=8, \text{depth}=1$) and is optimized using the reverse KL-divergence. Training utilizes NG, with $\text{epoch}=50$ for $\eta \in \{0.05, 0.1\}$ and $\text{epoch}=5$ for $\eta = 0.5$, and $N_b=2000$.}
\label{fig:F_lrs}
\end{figure}

Fig.~\ref{fig:F_lrs} shows how $\eta$ impacts NG. A higher learning rate ($\eta$=0.5) leads to faster convergence, as evidenced by the rapid decline in both the mean and variance of the loss. However, this comes at the cost of reduced training epochs and a potential risk of overshooting optimal parameter regions. In contrast, smaller learning rates ($\eta =0.05,0.1$) require more training epochs to reach comparable performance, but provide a more stable and gradual convergence process. These results suggest that, provided an appropriate combination of learning rate and epoch, VAN can robustly learn dynamics across a range of learning rate settings.

\begin{figure*}[ht]
\centering
\includegraphics[width=\textwidth]{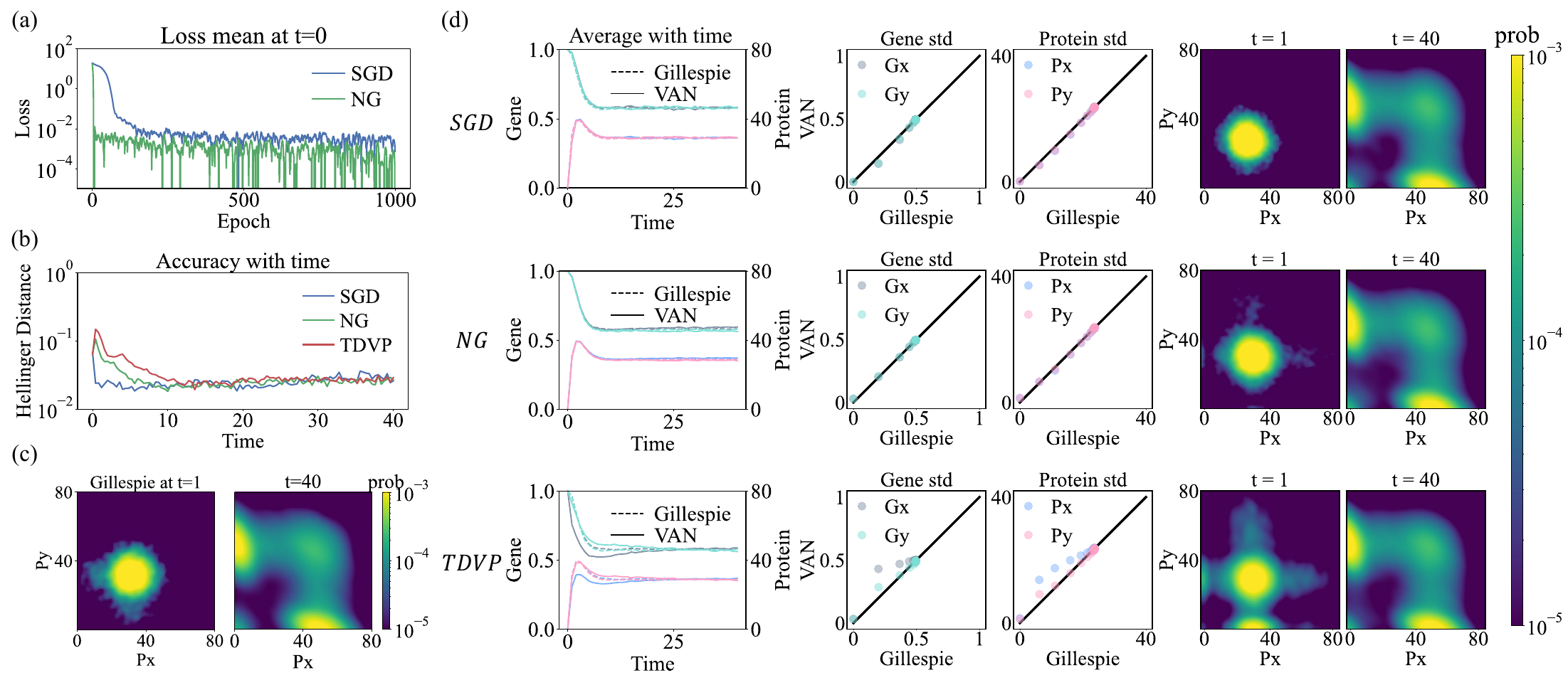}
\caption{Comparison of the training efficiency and convergence behavior of different optimization algorithms for the VAN in the toggle-switch system. 
(a) Means of training loss at the initial time step $t=0$ for SGD (green) and NG (blue) with 1000 epochs. 
(b) The accuracy of the VAN-generated distribution using three optimization methods over time, quantified by the Hellinger distance from the benchmark $10^4$ Gillespie trajectories. 
(c) Joint probability heatmap of $P_x$ and $P_y$ for the benchmark Gillespie simulation at $t=1$ (left) and $t=40$ (right). 
(d) The left panel shows the time evolution of the mean counts of the genes and proteins obtained from the VAN (lines) and from Gillespie simulations (dotted lines). The middle panel presents a comparison of the corresponding standard deviations. The right panel displays heatmaps of the joint distributions of $P_x$ and $P_y$ at time points $t=1$ and $t=40$.
All experiments were trained with reverse KL-divergence, the optimizer parameters: learning rate $\eta$ = 0.005, epoch = 100 at $t>0$,  epoch = 2000 at $t=0$ for SGD; $\eta$ = 0.5,  epoch = 5 for $t>0$,  epoch = 50 for $t=0$ for NG, in NADE architecture (width $16$, $N_b=2000$).}
\label{fig:optim-methods}
\end{figure*}

\subsection{Derivation of the TDVP update}\label{app:tdvp}

Here we outline the detailed derivation of the TDVP condition used in Eq.~\eqref{eq:TDVP_final}. 
The derivation follows Refs.~\cite{haegeman_time-dependent_2011,reh_time-dependent_2021}. 

At time t, the propagated distribution can be expanded as:
\begin{equation}
    \mathbb{T}_t\hat{P}^{\theta_t}_t = \hat{P}^{\theta_t}_t + \partial_t \hat{P}^{\theta_t}_t \delta t + \mathcal{O}(\delta t^2),
    \label{eq:Doob_Meyer}
\end{equation}
which is the first-order Taylor expansion of the transition operator. Concurrently, the variational distribution expands as:
\begin{equation}
    \hat{P}^{\theta_{t+\delta t}}_{t+\delta t} = \hat{P}^{\theta_t}_t + \dot{\theta}_t \cdot \nabla_{\theta_t} \hat{P}^{\theta_t}_t \delta t + \mathcal{O}(\delta t^2),
    \label{eq:param_expansion}
\end{equation}
where $\dot{\theta}_t$ is the updates of the VAN's parameter at t. Eq.~\eqref{eq:param_expansion} is the first-order change due to parameter evolution. 
\begin{widetext}
Substituting Eq.~\eqref{eq:Doob_Meyer} and Eq~\eqref{eq:param_expansion} into Eq~\eqref{Loss}, we derive the KL-divergence minimization:
\begin{align}
& D_{KL}\!\left[\hat{P}^{\theta_{t+\delta t}}_{t+\delta t} \middle\| \mathbb{T} \hat{P}^{\theta_t}_t\right] \nonumber \\
=~& \mathbb{E}_{\mathbf{s}\sim\hat{P}^{\theta_{t+\delta t}}_{t+\delta t}}
   \big[\ln \hat{P}^{\theta_{t+\delta t}}_{t+\delta t}(\mathbf{s})
   - \ln(\mathbb{T}\hat{P}^{\theta_{t}}_{t})(\mathbf{s})\big] \\
=~& \mathbb{E}_{\mathbf{s}\sim\hat{P}^{\theta_{t}}_{t}}
   \frac{\hat{P}^{\theta_{t+\delta t}}_{t+\delta t}}{\hat{P}^{\theta_{t}}_{t}}
   \big[\ln \hat{P}^{\theta_{t+\delta t}}_{t+\delta t}(\mathbf{s})
   - \ln(\mathbb{T}\hat{P}^{\theta_{t}}_{t})(\mathbf{s})\big] \\
\approx~& \mathbb{E}_{\mathbf{s}\sim\hat{P}^{\theta_{t}}_{t}}
   \left[1 + \dot{\theta}_{t}\!\cdot\!\nabla_{\theta_{t}}\ln\hat{P}^{\theta_{t}}_{t}\,\delta t\right]\Big\{ \ln \!\Big[ \hat{P}^{\theta_{t}}_{t} \!\Big(1 + \dot{\theta}_{t}\!\cdot\!\nabla_{\theta_{t}}
   \ln \hat{P}^{\theta_{t}}_{t}\,\delta t\Big)\Big]
   - \ln \!\Big[ \hat{P}^{\theta_{t}}_{t} \!\Big(1 + \partial_{t}\ln\hat{P}^{\theta_{t}}_{t}\,\delta t\Big)\Big] \Big\}
\label{eq:step1}\\
\approx~& \mathbb{E}_{\mathbf{s}\sim\hat{P}^{\theta_{t}}_{t}}
   \left[1 + \dot{\theta}_{t}\!\cdot\!\nabla_{\theta_{t}}\ln\hat{P}^{\theta_{t}}_{t}\,\delta t\right]
   \Big\{ \big[\dot{\theta}_{t}\!\cdot\!\nabla_{\theta_{t}}\ln\hat{P}^{\theta_{t}}_{t}
   - \partial_{t}\ln\hat{P}^{\theta_{t}}_{t}\big]\delta t
   - \frac{1}{2}\!\left[(\dot{\theta}_{t}\!\cdot\!\nabla_{\theta_{t}}\ln\hat{P}^{\theta_{t}}_{t})^{2}
   - (\partial_{t}\ln\hat{P}^{\theta_{t}}_{t})^{2}\right](\delta t)^{2} \Big\}
\label{eq:step2}\\
\approx~& \mathbb{E}_{\mathbf{s}\sim\hat{P}^{\theta_{t}}_{t}}
   \Big\{ \big[\dot{\theta}_{t}\!\cdot\!\nabla_{\theta_{t}}\ln\hat{P}^{\theta_{t}}_{t}
   - \partial_{t}\ln\hat{P}^{\theta_{t}}_{t}\big]\delta t
   + \frac{1}{2}\!\left(\dot{\theta}_{t}\!\cdot\!\nabla_{\theta_{t}}\ln\hat{P}^{\theta_{t}}_{t}
   - \partial_{t}\ln\hat{P}^{\theta_{t}}_{t}\right)^{2}(\delta t)^{2} \Big\}.
\label{eq:step3}
\end{align}

From Eq.~\eqref{eq:step1} to Eq.~\eqref{eq:step2}, we substitute the Taylor expansions of the logarithms and use the identity 
$\nabla_\theta \hat{P}^\theta = \hat{P}^\theta \nabla_\theta \ln \hat{P}^\theta$.  
From Eq.~\eqref{eq:step2} to Eq.~\eqref{eq:step3}, we apply the approximation 
$\ln(1+\epsilon) \approx \epsilon - \epsilon^2/2$ for $\epsilon = \mathcal{O}(\delta t)$, and expand the product up to order $\mathcal{O}(\delta t^2)$.  

Letting $a = \dot{\theta}_t \cdot \nabla_{\theta_t} \ln\hat{P}^{\theta_t}_t$ and $b = \partial_t \ln\hat{P}^{\theta_t}_t$, we multiply the series:
\begin{align}
\left(1 + a\,\delta t\right)\left[(a-b)\,\delta t - \tfrac{1}{2}(a^2 - b^2)\, \delta t^2\right] 
\nonumber=(a-b)\,\delta t + \left[-\tfrac{1}{2}(a^2 - b^2) + a(a-b)\right]\delta t^2 + \mathcal{O}(\delta t^3).
\end{align}
The second-order coefficient simplifies as
$
-\tfrac{1}{2}(a^2 - b^2) + a(a - b) = \tfrac{1}{2}(a - b)^2,
$
leading to the final approximation in Eq.~\eqref{eq:step3}.

Next, noting that the normalization condition $\sum_{\mathbf{s}} \hat{P}^{\theta_t}_t(\mathbf{s}) = 1$, we have 
$\mathbb{E}_{\mathbf{s}\sim\hat{P}^{\theta_t}_t}[\nabla_{\theta_t}\ln\hat{P}^{\theta_t}_t]
= \sum_{\mathbf{s}} \hat{P}^{\theta_t}_t(\mathbf{s})\,\nabla_{\theta_t}\ln\hat{P}^{\theta_t}_t(\mathbf{s})
= \sum_{\mathbf{s}} \nabla_{\theta_t} \hat{P}^{\theta_t}_t(\mathbf{s})
= \nabla_{\theta_t} \sum_{\mathbf{s}} \hat{P}^{\theta_t}_t(\mathbf{s}) = 0$,
so this term drops out.

This yields the variational condition:
\begin{equation}
\begin{split}
\dot{\theta}_t
= \operatorname*{arg\,min}_{\dot{\theta}}
\Bigl[
\tfrac12\,\dot{\theta_t}^{\top} S(\theta_t)\dot{\theta_t}
- \dot{\theta_t}^{\top}\,
\mathbb{E}_{\mathbf{s}\sim \hat{P}^{\theta_t}_t}
\bigl[\nabla_{\theta_t}\ln\hat{P}^{\theta_t}_t(\mathbf{s})
\cdot \partial_t \ln\hat{P}^{\theta_t}_t(\mathbf{s})\bigr]
\Bigr],
\end{split}
\label{eq:TDVP_condition}
\end{equation}
where $S(\theta_t) = \mathbb{E}_{\mathbf{s}\sim \hat{P}^{\theta_t}_t} [ (\nabla_{\theta_t} \ln\hat{P}^{\theta_t}_t) (\nabla_{\theta_t} \ln\hat{P}^{\theta_t}_t)^\top ]$ is the FIM (Eq.~\eqref{eq:Fisher}). The solution:
\begin{equation}
\dot{\theta}_t = S(\theta_t)^{-1} \cdot \mathbb{E}_{\mathbf{s}\sim \hat{P}^{\theta_t}_t} \left[ \nabla_{\theta_t} \ln\hat{P}^{\theta_t}_t(\mathbf{s}) \cdot \partial_t \ln\hat{P}^{\theta_t}_t(\mathbf{s}) \right],
\end{equation}
provides parameter updates respecting the system's intrinsic geometry. 
\end{widetext}

In practice, directly using the probability flow 
$\partial_t\!\ln\hat{P}^{\theta_t}_t(\mathbf{s})$ can be numerically unstable, 
as the flow may attain large values while the empirical Fisher matrix $S$ is 
often ill-conditioned. A more robust alternative is to incorporate the physical
step size and compute
\begin{equation}
\dot{\theta}_t
=
S^{-1}(\theta_t)\,
\mathbb{E}_{\mathbf{s}\sim\hat{P}^{\theta_t}_t}
\!\left[
\nabla_{\theta_t}\ln\hat{P}^{\theta_t}_t(\mathbf{s})\,
\bigl(\partial_t\!\ln\hat{P}^{\theta_t}_t(\mathbf{s})\,\delta t\bigr)
\right].
\end{equation}
Because $\partial_t\!\ln\hat{P}^{\theta_t}_t(\mathbf{s})\,\delta t$ is typically 
small, one may instead use the first-order approximation 
$x \approx \ln(1+x)$ for $x$ near zero, leading to
\begin{equation}
\dot{\theta}_t
=
S^{-1}(\theta_t)\,
\mathbb{E}_{\mathbf{s}\sim\hat{P}^{\theta_t}_t}
\!\left[
\nabla_{\theta_t}\ln\hat{P}^{\theta_t}_t(\mathbf{s})\,
\ln\bigl(1+\partial_t\!\ln\hat{P}^{\theta_t}_t(\mathbf{s})\,\delta t\bigr)
\right],
\end{equation}
which suppresses rare large flow values and improves numerical stability
while preserving first-order accuracy in~$\delta t$.  
The parameter update then follows as $\theta_{t+\delta t}=\theta_t+\dot{\theta}_t$.

\subsection{Optimization performance in the toggle switch}

\begin{figure}[ht]
\centering
\includegraphics[width=1\linewidth]{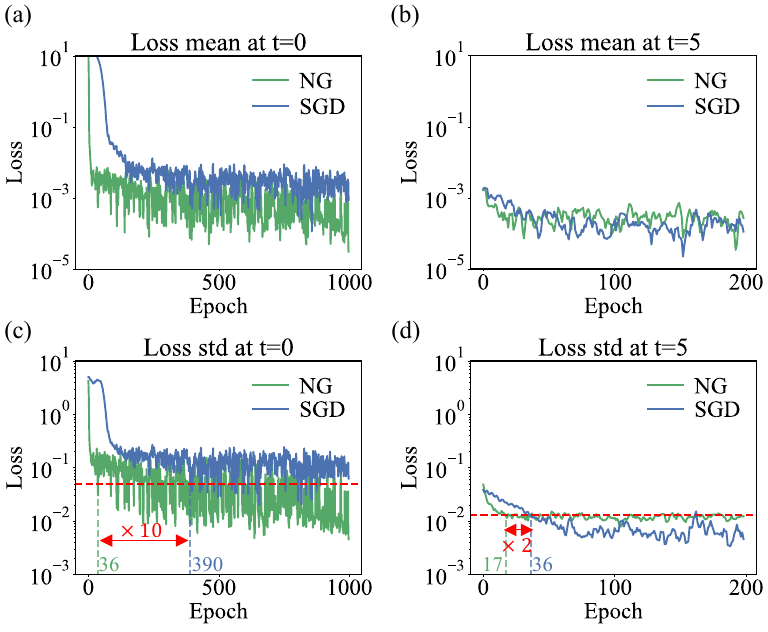}
\caption{Comparative analysis of training loss for NG and SGD. 
(a) Means of training loss at the initial time step $t=0$.
(b) Means of training loss at an early training stage $t=5$.
(c) Standard deviations of training loss at the initial time step $t=0$, the horizontal dotted lines indicate convergence threshold ($5\times10^{-2}$).
(d) Standard deviations of training loss at an early training stage $t=5$, the convergence threshold is set as $10^{-2}$. 
Red arrows quantify the epoch reduction ratio. All experiments were trained with reverse KL-divergence, the optimizer parameters: learning rate $\eta$ = 0.005, epoch = 100 at $t>0$,  epoch = 2000 at $t=0$ for SGD; $\eta$ = 0.5,  epoch = 5 for $t>0$,  epoch = 50 for $t=0$ for NG, in NADE architecture (width $16$, $N_b=2000$).}
\label{fig:SGD_NG}
\end{figure}

\begin{table}[!ht]
\centering
\begin{tabular}{l@{\quad}c}
\toprule
\text{Algorithm} & \text{Parameter update} \\
\midrule
Stochastic Gradient Descent & $\Delta \theta = -\eta \dfrac{\partial \mathcal{L}}{\partial \theta}$ \\
Natural Gradient & $\Delta \theta = -\eta S(\theta_t)^{-1} \dfrac{\partial \mathcal{L}}{\partial \theta}$\\
Time-Dependent Variational Principle & $\dot \theta =S(\theta_t)^{-1} F$\\
\bottomrule
\end{tabular}
\caption{Comparison of optimization algorithms and their update mechanisms.
SGD follows the negative gradient direction scaled by the learning rate $\eta$;
NG incorporates the FIM $S$ for preconditioning;
TDVP employs a force term $F$ derived from the system's Hamiltonian.
All methods iteratively minimize the reverse KL-divergence loss $\mathcal{L}$.}
\label{tab:al-sum2}
\end{table}

\begin{figure*}[ht]
\centering
\includegraphics[width=1\textwidth]{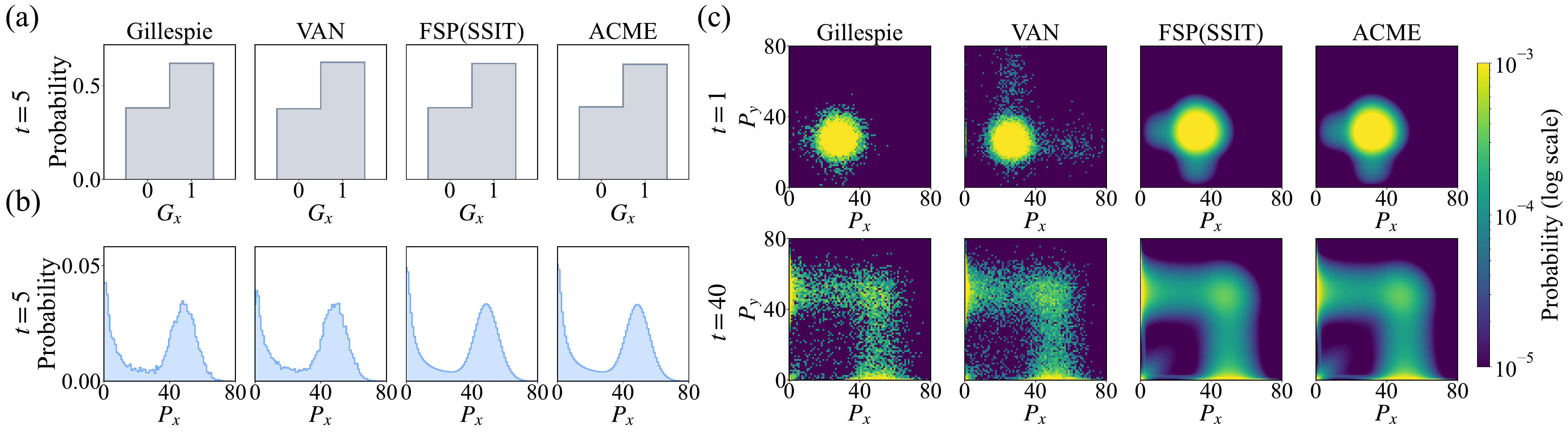}
\caption{Comparison of probability distributions for the genetic toggle-switch model obtained from Gillespie simulations, VAN, SSIT-FSP, and ACME.
(a) Distribution of the promoter state $G_x$ at $t=5$.
(b) Marginal distribution of protein $P_x$ at $t=5$.
(c) Joint distributions of $P_x$ and $P_y$ at $t=1$ and $t=40$, shown on a logarithmic probability scale.}
    \label{fig:appendix_toggle}
\end{figure*}

We evaluate the above three optimization algorithms for training VAN in the toggle switch system illustrated in \ref{subsec:Toggle}. Table~\ref{tab:al-sum2} specifies the mathematical framework for each algorithm, where $S(\theta_t)$ denotes the FIM and $F$ represents the Hamiltonian-derived force term in TDVP. The learning rate $\eta$ is algorithm-specific, with values empirically determined through hyperparameter tuning.

Comparative performance is evaluated for the three optimizers (Fig.~\ref{fig:optim-methods}). First, NG demonstrates a superior convergence speed. Fig.~\ref{fig:optim-methods}(a) illustrates the difference in the average loss between the NG and SGD algorithms under the same number of epochs, highlighting the convergence advantage of NG. Since the TDVP algorithm does not require iterative parameter updating, the comparison is not provided in this context. Second, all algorithms maintain Hellinger distances below $10^{-1}$ throughout the evolution, confirming the VAN framework's robustness. Third, variance comparisons reveal comparable accuracy across algorithms.

The training performance of the NG and SGD algorithms is meticulously compared in Fig.~\ref{fig:SGD_NG}, which underscores the superior training efficiency of the NG. Specifically, Fig.~\ref{fig:SGD_NG}(c) and Fig.~\ref{fig:SGD_NG}(d) reveal that NG exhibits a more rapid convergence rate compared to SGD. When the distributions are significantly different (t=0), NG's convergence advantage is pronounced, being approximately ten times faster than SGD. Even when the distributions are relatively similar (t=5), NG still maintains a notable edge, converging twice as fast as SGD. Moreover, Fig.~\ref{fig:SGD_NG}(a) and Fig.~\ref{fig:SGD_NG}(c) corroborate the precision of both gradient descent algorithms during the convergence process.

\section{Comparison with other methods}
\label{app:other_methods}

\begin{figure}[ht]
\centering
\includegraphics[width=1\linewidth]{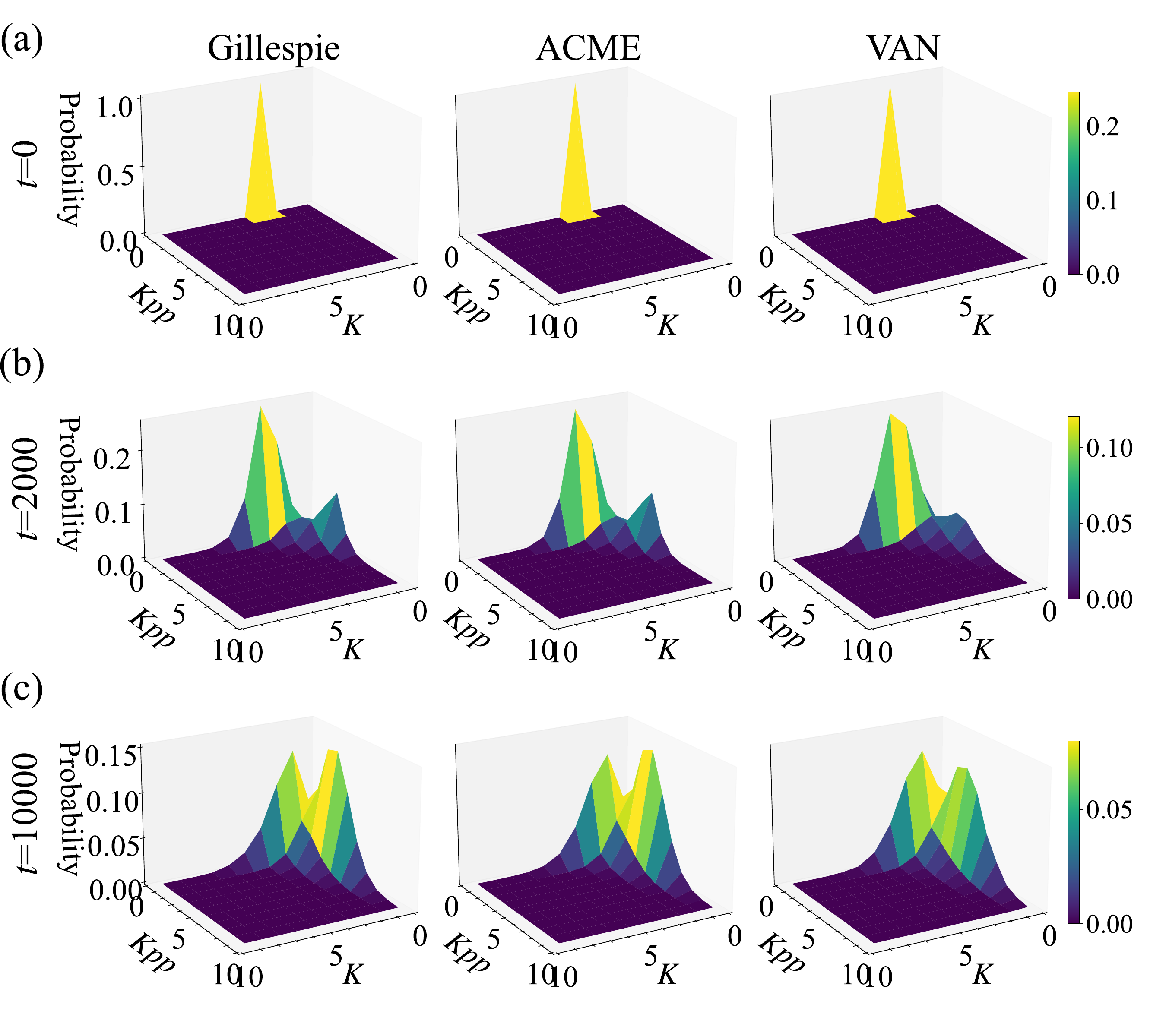}
\caption{Joint distributions of K and Kpp for the MAPK cascade at $t=0$, $2000$, and $10000$, obtained from Gillespie simulations, ACME, and VAN. The distribution evolves from the delta-initialized state to a bimodal long-time distribution, with close agreement among the three methods.}
    \label{fig:appendix_mapk}
\end{figure}

To further assess the accuracy of NNCME-2, we compare its results with Gillespie simulations and additional numerical methods. For the genetic toggle switch (Fig.~\ref{fig:appendix_toggle}), we include FSP using SSIT~\cite{popinga2026stochastic} and ACME~\cite{cao_accurate_2016}. For the MAPK cascade (Fig.~\ref{fig:appendix_mapk}), we compare the joint distribution obtained from NNCME-2 with Gillespie simulations and ACME at representative time points. Overall, NNCME-2 reproduces the marginal and joint distributions obtained from these reference methods.

\begin{figure*}[ht]
    \includegraphics[width=1\textwidth]{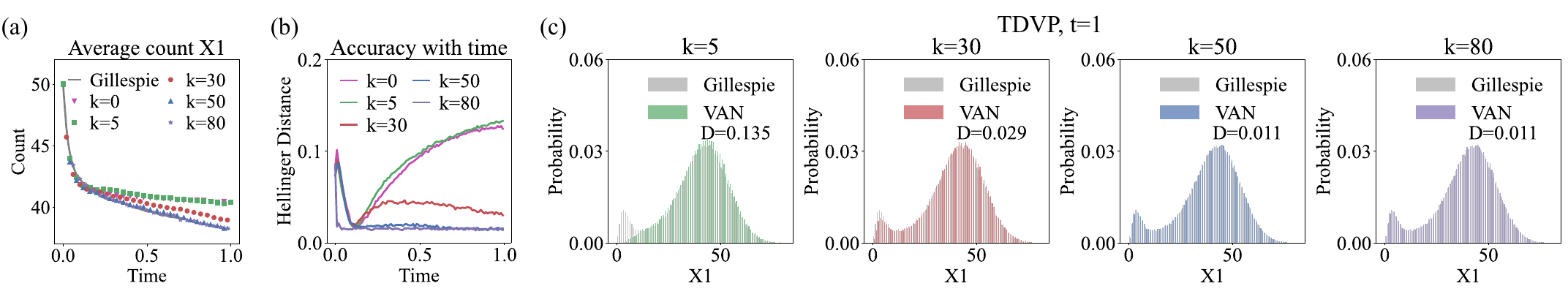}
    \caption{Effect of hyperparameter $k$ on diffusive ES for the Schl\"{o}gl (2 sites) model. 
    (a) The time evolution of the average count for $X_1$ with different $k$ from the VAN (dot) and the Gillespie simulation (gray line).
    (b) The accuracy of the VAN-generated distribution over time, quantified by the Hellinger distance from the benchmark $10^5$ Gillespie trajectories.
    (c) Marginal distribution of species $X_1$ from the VAN at $t=1$ for different $k$ values (colored) and compared with the Gillespie simulation (gray). 
    The learning parameters for the system are fixed at $T_{\text{step}}=1\times 10^5$ and $\delta t = 1\times 10^{-5}$. All experiments are performed using the NADE architecture with $\text{width}=16$ and $\text{depth}=1$. The Gillespie simulation has $10^5$ trajectories. 
    In these experiments, the VAN model employs NADE architecture ($\text{width}=16$, $\text{depth}=1$, and $N_b=2000$). Training utilizes reverse KL, with $\text{epoch}=50$, $\eta$ = 0.5 at $t=0$ using NG and $\text{epoch}=1$, $\eta$ =1 for $t>0$ using TDVP.
    }
    \label{fig:k_param_alpha_TDVP}
\end{figure*}

\section{Performance of enhanced-sampling strategies}
\label{app:enhanced_sampling}

In this section, we present a comparative evaluation of four sampling strategies applied to the Schl\"{o}gl (2 sites) system: \textit{Vanilla} (standard sampling from $P_\theta$), \textit{Mixture ES} (sampling from a mixture of model-generated and uniformly random configurations), \textit{Diffusive ES} (sampling from the convolved distribution), and $\alpha$ ES (sampling from a power-law distribution).

\subsection{Details of Diffusive ES}
\label{subapp:diffusive}
Here we provide the implementation details of the diffusive enhanced-sampling scheme introduced in the main text.
In this work, we adopt a \textit{uniform} diffusive kernel, which averages
$\hat{P}_t^{\theta_t}(n_i \mid \mathbf{n}_{<i})$ uniformly over the local
neighborhood ${\cal N}_k(n_i')=\{n_i : |n_i-n_i'|\le k\}$.
The kernel is given by
\begin{align}
\mathcal{K}_{U,k}(n_i \mid n_i')
=
\begin{cases}
\displaystyle \frac{1}{|{\cal N}_k(n_i')|}, & n_i \in {\cal N}_k(n_i'), \\[6pt]
0, & \text{otherwise},
\end{cases}
\end{align}
which produces an over-dispersed conditional distribution
$q_t^{(\mathrm{diff})}(n_i \mid \mathbf{n}_{<i})$ (Eq.~\eqref{eq:diffusive}) and enables smoother exploration across nearby values of $n_i$.

Samples are drawn from $q_t^{(\mathrm{diff})}$ (Eq.~\eqref{eq:diffusive}), while the reverse KL-divergence
loss~\eqref{loss2} is evaluated with the importance ratio
\begin{align}
\label{eq:diffusive_w_final}
w^{(\mathrm{diff})}(s)
=
\frac{\hat{P}_t^{\theta_t}(s)}
     {q_t^{(\mathrm{diff})}(s)}.
\end{align}
The corresponding diffusive estimator becomes
\begin{align}
\label{eq:diffusive_loss_final}
\mathcal{L}^{(\mathrm{diff})}
=
\mathbb{E}_{s\sim q_{t+\delta t}^{(\mathrm{diff})}}
\!\left[
w^{(\mathrm{diff})}(s)\,
\Bigl(
\ln \hat{P}_{t+\delta t}^{\theta_{t+\delta t}}(s)
- \ln (\mathbb{T}\hat{P}_{t}^{\theta_{t}})(s)
\Bigr)
\right],
\end{align}
which reduces to Eq.~\eqref{loss2} when the kernel collapses to
$\mathcal{K}_{U,k}(\mathbf{n}\mid \mathbf{n}')
= \delta_{\mathbf{n},\mathbf{n}'}$. Here, we also investigate the impact of the parameter: convolutional kernel size $k$ on the performance for exploring the rare region of the Schl\"{o}gl (2 sites) model, as illustrated in Fig.~\ref{fig:k_param_alpha_TDVP}. 

The results demonstrate a clear relationship between the choice of $k$ and the accuracy of the learned distribution. As shown in Fig.~\ref{fig:k_param_alpha_TDVP}, the models trained with larger $k$ values exhibit higher accuracy throughout the evolution. This observation is consistent with the hypothesis that sampling from a broader, over-dispersed distribution facilitates more efficient exploration of the state space, thereby yielding more accurate results. 

\subsection{\texorpdfstring{Details of $\alpha$ ES}{Details of alpha ES}}
\label{subapp:alpha}

\begin{figure*}[ht]
    \includegraphics[width=1\textwidth]{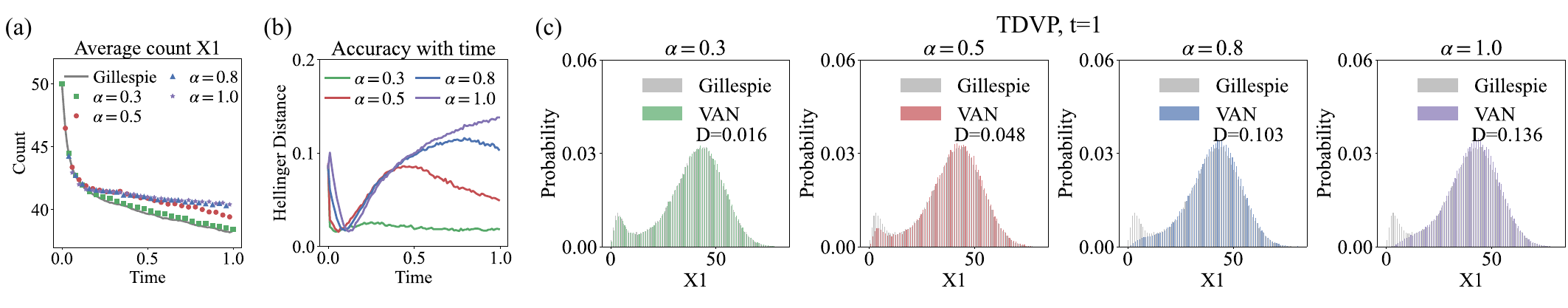}
    \caption{Effect of hyperparameter $\alpha$ on $\alpha$ ES for the Schl\"{o}gl (2 sites) model. Smaller values of $\alpha$ correspond to stronger exploration in the sampling process. The layout of the figure and the hyperparameters of VAN are the same as those of Fig.~\ref{fig:k_param_alpha_TDVP}.
    }
    \label{param_alpha_TDVP}
\end{figure*}

\begin{figure*}[ht]
    \centering
    \includegraphics[width=1\textwidth]{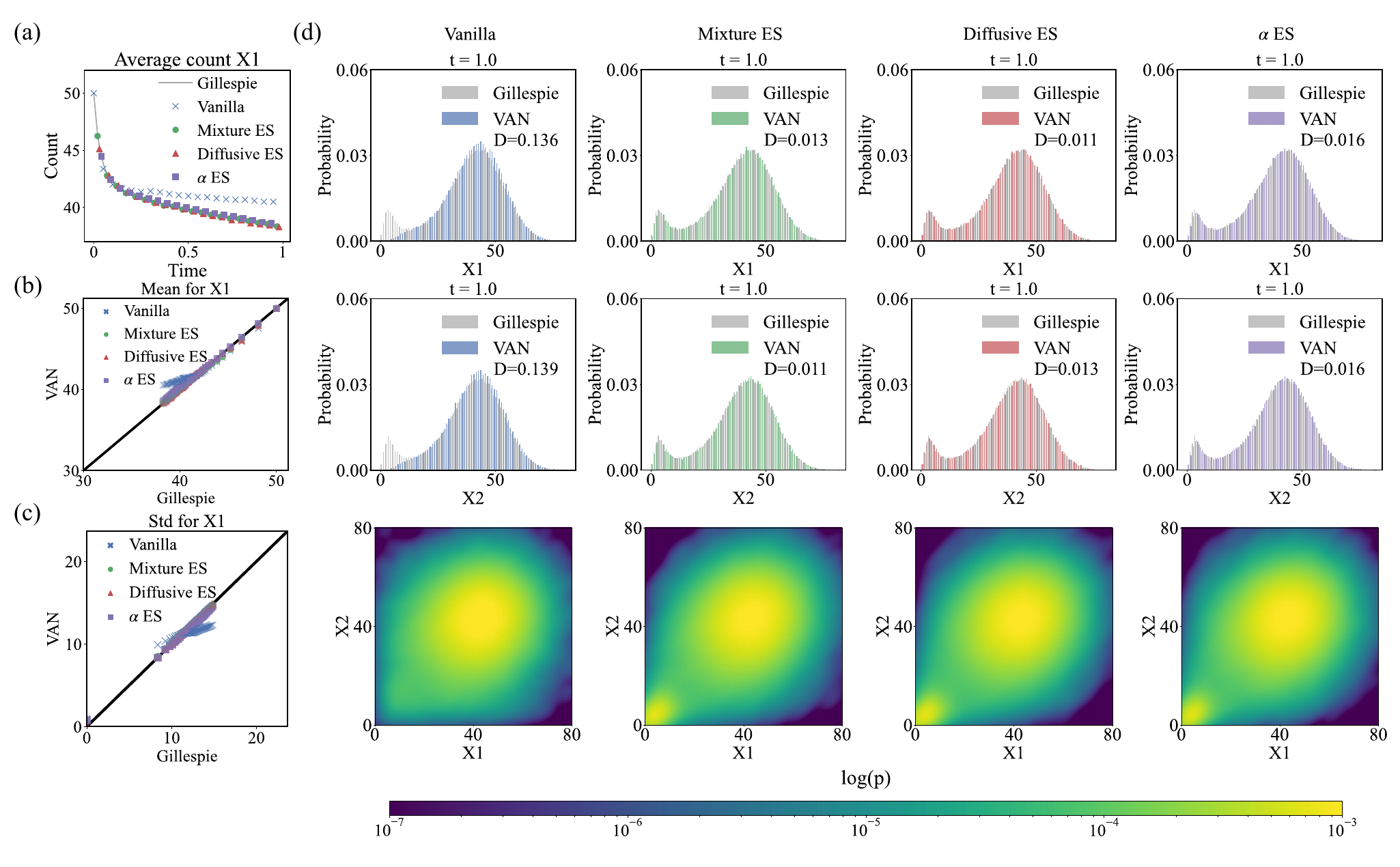}
\caption{Results of four ES strategies-\textit{Vanilla}, \textit{Mixture} ES, \textit{Diffusive} ES and \textit{$\alpha$} ES used in training VAN on the Schl\"{o}gl (2 sites) system.
(a) Time evolution of the means of species $X_1$ estimated by each method.
(b) Comparison of the means of $X_1$ between VAN-based estimates and Gillespie simulation.
(c) Comparison of the standard deviation of $X_1$ between VAN-based estimates and Gillespie simulation.
(d) Marginal distributions of species $X_1$ and $X_2$ at $t=1$ (first and second rows), and joint probability distribution heatmaps of $X_1$ and $X_2$ at $t=1$ (third row).
The Gillespie simulation has $10^5$ trajectories.
    In these experiments, the VAN model employs NADE architecture ($\text{width}=16$, $\text{depth}=1$, and $N_b=2000$). Training utilizes reverse KL, with $\text{epoch}=50$, $\eta$ = 0.5 at $t=0$ using NG and $\text{epoch}=1$, $\eta$ =1 for $t>0$ using TDVP.}
    \label{fig:Sample_methods}
\end{figure*}

We provide the technical details of the $\alpha$ ES used in the main text. Samples are drawn from $q_t^{(\alpha)}$ (Eq.~\eqref{eq:q_alpha}), and the reverse KL-divergence loss~\eqref{loss2} is evaluated using the importance ratio
\begin{equation}
\label{eq:w_alpha}
    w^{(\alpha)}(s) = 
    \frac{\hat{P}_t^{\theta_t}(s)}{q_t^{(\alpha)}(s)}.
\end{equation}
The reweighted estimator of the loss becomes
\begin{align}
\label{eq:alpha_loss}
\mathcal{L}^{(\alpha)}
= \mathbb{E}_{s\sim q_{t+\delta t}^{(\alpha)}}\!
\left[
w^{(\alpha)}(s)\,
\Bigl(
\ln \hat{P}_{t+\delta t}^{\theta_{t+\delta t}}(s)
- \ln (\mathbb{T}\hat{P}_{t}^{\theta_{t}})(s)
\Bigr)
\right],
\end{align}
which reduces to Eq.~\eqref{loss2} when $\alpha=1$.

For NG and TDVP, the FIM under the reweighted sampling measure is
\begin{equation}
S^{(\alpha)} =
\mathbb{E}_{s\sim q_t^{(\alpha)}}\!\left[
w_t^{(\alpha)}(s)\,
\nabla_{\theta_t}\!\ln \hat{P}_t^{\theta_t}(s)\,
\nabla_{\theta_t}\!\ln \hat{P}_t^{\theta_t}(s)^\top
\right].
\end{equation}
The corresponding TDVP update takes the form
\begin{equation}
\label{eq:alpha_update}
\dot{\theta}_t
= [S_t^{(\alpha)}]^{-1}
  \mathbb{E}_{s\sim q_t^{(\alpha)}}\!
  \left[
  w_t^{(\alpha)}(s)\,
  \nabla_{\theta_t}\ln \hat{P}_t^{\theta_t}(s)\,
  \partial_t\ln \hat{P}_t^{\theta_t}(s)
  \right],
\end{equation}
ensuring that the parameter updates respect the intrinsic geometry of the model distribution under the reweighted measure.

In Fig.~\ref{param_alpha_TDVP}, we present results obtained using the TDVP approach with different values of the parameter $\alpha \in \{0.3, 0.5, 0.8, 1.0\}$ in the Schl\"{o}gl (2 sites) system. The results indicate that sampling from a broader distribution, corresponding to smaller values of $\alpha$, leads to improved accuracy.

\subsection{Comparison of ES in VAN training}
\label{sec:Samlpe_Compare}

In this subsection, we conduct a comparative evaluation of four sampling strategies---\textit{Vanilla}, \textit{Mixture ES}, \textit{Diffusive ES} and $\alpha$ ES on the Schl\"{o}gl (2 sites) system. 

Fig.~\ref{fig:Sample_methods} (a) and (b) compare the means of $X_1$ learned by the VANs to those from Gillespie simulations, demonstrating that every ES strategy outperforms the Vanilla sampling. This superiority is also observed in the standard deviation depicted in Fig.~\ref{fig:Sample_methods} (c). 
Fig.~\ref{fig:Sample_methods}~(d) provides a multi-level analysis of the marginal and joint probability distributions at $t=1$. In particular, the \textit{Vanilla} method fails to capture the rare probability peak ($X \in [0,15]$), whereas other ES strategies succeed in tracking this rare event region. This suggests that the proposed ES improves the VAN\textquotesingle{}s ability to be aware of a broader distribution, which helps track rare probability regions.

\section{Performance evaluation of JAX vs PyTorch}
\label{sec_jax_vs_torch}
\begin{figure}[ht]
{\includegraphics[width=1\linewidth]{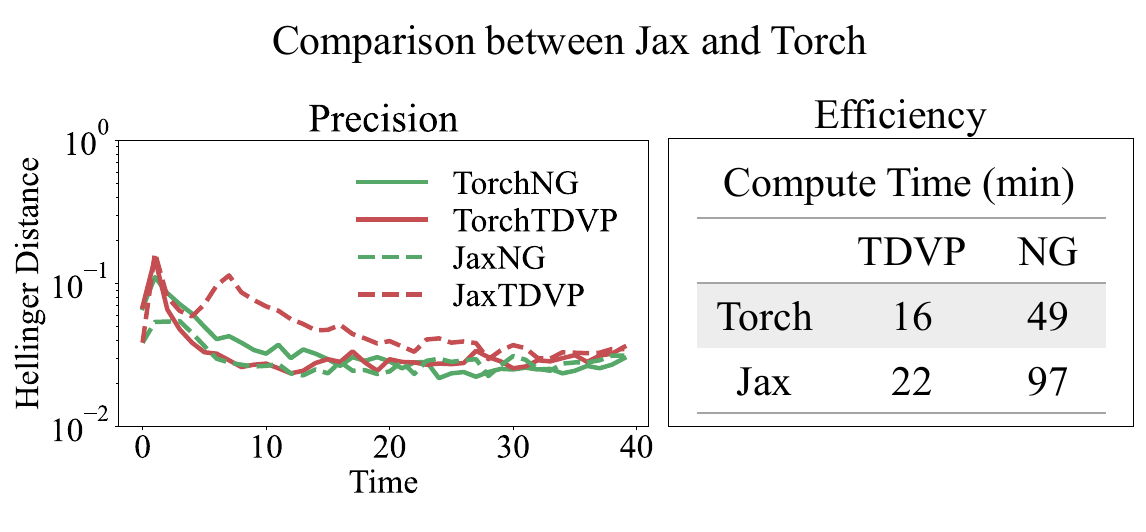}}
\caption{Comparison of computational efficiency and precision between PyTorch and JAX frameworks, framework implementations are distinguished by color (red: PyTorch, blue: JAX) and line style (solid: TDVP, dashed: NG algorithm). Experiments were conducted using the NADE net ($\text{width}=16, \text{depth}=1$). Performance metrics were evaluated through reverse KL-divergence measurements. Computational efficiency was evaluated through Hellinger distance measurements relative to $10^4$ Gillespie simulated trajectories. All experiments were executed on an NVIDIA Tesla V100 GPU under identical computational constraints.
}
\label{torchVSJax}
\end{figure}

Our implementation is primarily based on PyTorch~\cite{Ansel_PyTorch_2_Faster_2024}, but we also developed a JAX version~\cite{jax2018github} to examine differences in computational efficiency. Both are prominent open source libraries for machine learning and numerical computing. The notable distinction between JAX and PyTorch lies in their underlying design philosophies. JAX's compiler-driven approach allows for better optimization and efficiency during computations, while PyTorch's dynamic graph structure provides a more intuitive coding experience.
Under identical reaction parameters and model hyperparameters, we conducted benchmarking of NG and TDVP algorithms across both frameworks in the toggle switch system. As quantified in Figure \ref{torchVSJax}, the PyTorch implementation exhibited superior computational efficiency on a Tesla V100 GPU, completing TDVP simulations in 16 minutes compared to JAX's 22 minutes, with NG implementations showing analogous acceleration (PyTorch: 49.2 vs. JAX: 97.4 minutes).

\section{Details of the reaction network examples}
\label{sec:example}

\subsection{Genetic toggle switch}
\label{subsec:Toggle}
The genetic toggle switch~\cite{gardner_construction_2000,terebus_discrete_2019} has a multimodal distribution on the protein counts, depending on the genetic states of the two mutually inhibited genes. Thus, the system is suitable for testing the flexibility of the VAN to learn the multimodal distribution. The model has the chemical reactions:
\begingroup
\allowdisplaybreaks    
\begin{align*}
&G_{x}\stackrel{s_{x}}{\longrightarrow}G_{x}+P_{x},\quad
G_{y}\stackrel{s_{y}}{\longrightarrow}G_{y}+P_{y},\quad
\\&P_{x}\stackrel{d_{x}}{\longrightarrow}\emptyset,\quad
P_{y}\stackrel{d_{y}}{\longrightarrow}\emptyset, 
\\&2P_{x}+G_{y}\stackrel{b_{y}}{\longrightarrow}\bar{G}_{y},\quad
2P_{y}+G_{x}\stackrel{b_{x}}{\longrightarrow}\bar{G}_{x},\quad
\\&\bar{G}_{y}\stackrel{u_{y}}{\longrightarrow}2P_{x}+G_{y},\quad
\bar{G}_{x}\stackrel{u_{x}}{\longrightarrow}2P_{y}+G_{x},
\end{align*}
\endgroup
where $s_{x}$, $s_{y}$ are the synthesis rates of the proteins, and $p_{x}$, $p_{y}$ are the degradation rates of the proteins. 
The transition rate $b_{y}$ ($b_{x}$) is the binding rate of two copies of protein $P_{x}$ ($P_{y}$) to the $G_{y}$ ($G_{x}$), to form the complex $\bar{G}_{y}$ ($\bar{G}_{x}$). The unbinding of the complex $\bar{G}_{y}$ ($\bar{G}_{x}$) has the rate $u_{y}$ ($u_{x}$).   

The total count of the two promoter states for each gene is conserved, $G_{x}+\bar{G}_{x}=1$. This conservation imposes a constraint on the gene states, such that $G_{x}\in\{0,1\}$ and $G_{y}\in\{0,1\}$, and effectively reduces the number of independent variables by expressing $\bar{G}_{x}=1-G_{x}$ and $\bar{G}_{y}=1-G_{y}$. This constraint is explicitly imposed in the VAN representation. We consider a parameter regime with weak promoter binding ($s_x=s_y=50$, $d_x=d_y=1$, $b_x=b_y=10^{-4}$, $u_x=u_y=0.1$), in which the resulting joint probability distribution is multimodal with four distinct peaks, posing a challenge for accurately tracking the full distribution. In the VAN-based simulations, the temporal evolution is performed using a time step $\delta t = 0.005$ over $T_{\text{step}}=8000$ steps.

\subsection{MAPK cascade model}
\label{app:mapk}

The MAPK cascade is modeled following Ref.~\cite{cao_accurate_2016}, which describes a two-layer phosphorylation network involving extracellular signal-regulated kinase (ERK, denoted as $\mathrm{K}$), its upstream kinase MEK, and regulation by the phosphatase MKP3. The model includes both synthesis and degradation processes for ERK and MEK, as well as dual phosphorylation and dephosphorylation reactions at two conserved sites (T and Y). Under the standard assumptions that the count of MKP3 is fixed to unity and that phosphorylation does not protect ERK from degradation, the system comprises 16 molecular species and 35 reactions. All reactions obey mass-action kinetics, with rate constants taken from Ref.~\cite{cao_accurate_2016}.

Reaction rates $s_i$, $d_i$, $k_i$, and $h_i$ denote synthesis, degradation, phosphorylation, and dephosphorylation, respectively. Bidirectional arrows $\xrightleftharpoons[]{}$ represent reversible synthesis--degradation reactions, whereas single arrows correspond to irreversible phosphorylation or dephosphorylation steps. In our experiments, the temporal evolution is performed using a time step $\delta t=0.01$ over $T_{\mathrm{step}}=10^{6}$ time steps.

\begingroup
\label{tab:mapk_species}
\allowdisplaybreaks        
\small
\setlength{\jot}{1pt}     
\begin{align*}
R_1\!: & \emptyset \xrightleftharpoons[d_{2}]{s_{2}} \mathrm{MEK},\; s_{2} = 0.001,\; d_{2} = 0.15 \\[-2pt]
R_2\!: & \emptyset + \mathrm{Kpp} \xrightarrow{s_{3}} \mathrm{MEK} + \mathrm{Kpp},\; s_{3} = 0.005 \\[-2pt]
R_3\!: & \emptyset \xrightleftharpoons[d_{1}]{s_{1}} \mathrm{K},\; s_{1} = 0.00024,\; d_{1} = 0.0001 \\[-2pt]
R_4\!: & \mathrm{KpY}, \mathrm{KpT}, \mathrm{Kpp} \xrightarrow{d_{1}} \emptyset,\; d_{1} = 0.0001 \\[-2pt]
R_5\!: & \mathrm{K} + \mathrm{MEK} \xrightleftharpoons[k_{-1}]{k_{1}} \mathrm{K\_MEK\_Y},\; k_{1} = 0.375,\; k_{-1} = 1.0 \\[-2pt]
R_6\!: & \mathrm{K\_MEK\_Y} \xrightarrow{k_{2}} \mathrm{KpY} + \mathrm{MEK},\; k_{2} = 0.06 \\[-2pt]
R_7\!: & \mathrm{KpY} + \mathrm{MEK} \xrightleftharpoons[k_{-3}]{k_{3}} \mathrm{KpY\_MEK},\; k_{3} = 0.375,\; k_{-3} = 1.0 \\[-2pt]
R_8\!: & \mathrm{KpY\_MEK} \xrightarrow{k_{4}} \mathrm{Kpp} + \mathrm{MEK},\; k_{4} = 4.5 \\[-2pt]
R_9\!: & \mathrm{K} + \mathrm{MEK} \xrightleftharpoons[k_{-5}]{k_{5}} \mathrm{K\_MEK\_T},\; k_{5} = 0.375,\; k_{-5} = 1.0 \\[-2pt]
R_{10}\!: & \mathrm{K\_MEK\_T} \xrightarrow{k_{6}} \mathrm{KpT} + \mathrm{MEK},\; k_{6} = 0.06 \\[-2pt]
R_{11}\!: & \mathrm{KpT} + \mathrm{MEK} \xrightleftharpoons[k_{-7}]{k_{7}} \mathrm{KpT\_MEK},\; k_{7} = 0.375,\; k_{-7} = 1.0 \\[-2pt]
R_{12}\!: & \mathrm{KpT\_MEK} \xrightarrow{k_{8}} \mathrm{Kpp} + \mathrm{MEK},\; k_{8} = 4.5  \\[-2pt]
R_{13}\!: & \mathrm{Kpp} + \mathrm{MKP3} \xrightleftharpoons[h_{-1}]{h_{1}} \mathrm{Kpp\_MKP3},\; h_{1} = 0.015,\; h_{-1} = 1.0 \\[-2pt]
R_{14}\!: & \mathrm{Kpp\_MKP3} \xrightarrow{h_{2}} \mathrm{KpT\_MKP3\_Y},\; h_{2} = 0.032 \\[-2pt]
R_{15}\!: & \mathrm{KpT\_MKP3\_Y} \xrightleftharpoons[h_{-3}]{h_{3}} \mathrm{KpT} + \mathrm{MKP3},\; \\& h_{3} = 0.31,\; h_{-3} = 0.01 \\[-2pt]
R_{16}\!: & \mathrm{KpT} + \mathrm{MKP3} \xrightleftharpoons[h_{-4}]{h_{4}} \mathrm{KpT\_MKP3\_T},\;\\ & h_{4} = 0.01,\; h_{-4} = 1.0 \\[-2pt]
R_{17}\!: & \mathrm{KpT\_MKP3\_T} \xrightarrow{h_{5}} \mathrm{K\_MKP3\_T},\; h_{5} = 0.5 \\[-2pt]
R_{18}\!: & \mathrm{K\_MKP3\_T} \xrightleftharpoons[h_{-6}]{h_{6}} \mathrm{K} + \mathrm{MKP3},\; h_{6} = 0.086,\; h_{-6} = 0.0011 \\[-2pt]
R_{19}\!: & \mathrm{KpY} + \mathrm{MKP3} \xrightleftharpoons[h_{-7}]{h_{7}} \mathrm{KpY\_MKP3},\; h_{7} = 0.01,\; h_{-7} = 1.0 \\[-2pt]
R_{20}\!: & \mathrm{KpY\_MKP3} \xrightarrow{h_{8}} \mathrm{K\_MKP3\_Y},\; h_{8} = 0.47 \\[-2pt]
R_{21}\!: & \mathrm{K\_MKP3\_Y} \xrightleftharpoons[h_{-9}]{h_{9}} \mathrm{K} + \mathrm{MKP3},\; h_{9} = 0.14,\; h_{-9} = 0.0018.
\end{align*}
\endgroup

Such biochemical networks typically exhibit intricate connectivity, involving many interacting species and feedback pathways. The autoregressive neural networks can flexibly accommodate these arbitrary interaction patterns, whereas tensor-network approaches generally require adapting their architectures to reflect the underlying reaction graph~\cite{nicholson_quantifying_2023}, such as by modifying tensor connectivity or increasing the bond dimension to capture long-range correlations induced by the network topology.


\begin{thebibliography}{104}%
\makeatletter
\providecommand \@ifxundefined [1]{%
 \@ifx{#1\undefined}
}%
\providecommand \@ifnum [1]{%
 \ifnum #1\expandafter \@firstoftwo
 \else \expandafter \@secondoftwo
 \fi
}%
\providecommand \@ifx [1]{%
 \ifx #1\expandafter \@firstoftwo
 \else \expandafter \@secondoftwo
 \fi
}%
\providecommand \natexlab [1]{#1}%
\providecommand \enquote  [1]{``#1''}%
\providecommand \bibnamefont  [1]{#1}%
\providecommand \bibfnamefont [1]{#1}%
\providecommand \citenamefont [1]{#1}%
\providecommand \href@noop [0]{\@secondoftwo}%
\providecommand \href [0]{\begingroup \@sanitize@url \@href}%
\providecommand \@href[1]{\@@startlink{#1}\@@href}%
\providecommand \@@href[1]{\endgroup#1\@@endlink}%
\providecommand \@sanitize@url [0]{\catcode `\\12\catcode `\$12\catcode `\&12\catcode `\#12\catcode `\^12\catcode `\_12\catcode `\%12\relax}%
\providecommand \@@startlink[1]{}%
\providecommand \@@endlink[0]{}%
\providecommand \url  [0]{\begingroup\@sanitize@url \@url }%
\providecommand \@url [1]{\endgroup\@href {#1}{\urlprefix }}%
\providecommand \urlprefix  [0]{URL }%
\providecommand \Eprint [0]{\href }%
\providecommand \doibase [0]{https://doi.org/}%
\providecommand \selectlanguage [0]{\@gobble}%
\providecommand \bibinfo  [0]{\@secondoftwo}%
\providecommand \bibfield  [0]{\@secondoftwo}%
\providecommand \translation [1]{[#1]}%
\providecommand \BibitemOpen [0]{}%
\providecommand \bibitemStop [0]{}%
\providecommand \bibitemNoStop [0]{.\EOS\space}%
\providecommand \EOS [0]{\spacefactor3000\relax}%
\providecommand \BibitemShut  [1]{\csname bibitem#1\endcsname}%
\let\auto@bib@innerbib\@empty
\bibitem [{\citenamefont {Gillespie}(2007)}]{Gillespie_stochastic_2007}%
  \BibitemOpen
  \bibfield  {author} {\bibinfo {author} {\bibfnamefont {D.~T.}\ \bibnamefont {Gillespie}},\ }\bibfield  {title} {\bibinfo {title} {Stochastic simulation of chemical kinetics},\ }\href {https://www.annualreviews.org/doi/10.1146/annurev.physchem.58.032806.104637} {\bibfield  {journal} {\bibinfo  {journal} {Annu. Rev. Phys. Chem.}\ }\textbf {\bibinfo {volume} {58}},\ \bibinfo {pages} {35} (\bibinfo {year} {2007})}\BibitemShut {NoStop}%
\bibitem [{\citenamefont {van Kampen}(2007)}]{van2007stochastic}%
  \BibitemOpen
  \bibfield  {author} {\bibinfo {author} {\bibfnamefont {N.~G.}\ \bibnamefont {van Kampen}},\ }\href@noop {} {\emph {\bibinfo {title} {Stochastic Processes in Physics and Chemistry}}}\ (\bibinfo  {publisher} {Elsevier, New York},\ \bibinfo {year} {2007})\BibitemShut {NoStop}%
\bibitem [{\citenamefont {Weber}\ and\ \citenamefont {Frey}(2017)}]{weber2017master}%
  \BibitemOpen
  \bibfield  {author} {\bibinfo {author} {\bibfnamefont {M.~F.}\ \bibnamefont {Weber}}\ and\ \bibinfo {author} {\bibfnamefont {E.}~\bibnamefont {Frey}},\ }\bibfield  {title} {\bibinfo {title} {Master equations and the theory of stochastic path integrals},\ }\href {https://iopscience.iop.org/article/10.1088/1361-6633/aa5ae2} {\bibfield  {journal} {\bibinfo  {journal} {Rep. Prog. Phys.}\ }\textbf {\bibinfo {volume} {80}},\ \bibinfo {pages} {046601} (\bibinfo {year} {2017})}\BibitemShut {NoStop}%
\bibitem [{\citenamefont {Qian}\ and\ \citenamefont {Bishop}(2010)}]{qian_chemical_2010}%
  \BibitemOpen
  \bibfield  {author} {\bibinfo {author} {\bibfnamefont {H.}~\bibnamefont {Qian}}\ and\ \bibinfo {author} {\bibfnamefont {L.~M.}\ \bibnamefont {Bishop}},\ }\bibfield  {title} {\bibinfo {title} {The chemical master equation approach to nonequilibrium steady-state of open biochemical systems: {Linear} single-molecule enzyme kinetics and nonlinear biochemical reaction networks},\ }\href {https://www.mdpi.com/1422-0067/11/9/3472} {\bibfield  {journal} {\bibinfo  {journal} {IJMS}\ }\textbf {\bibinfo {volume} {11}},\ \bibinfo {pages} {3472} (\bibinfo {year} {2010})}\BibitemShut {NoStop}%
\bibitem [{\citenamefont {Zhang}\ and\ \citenamefont {Zhou}(2019)}]{zhang_markovian_2019}%
  \BibitemOpen
  \bibfield  {author} {\bibinfo {author} {\bibfnamefont {J.}~\bibnamefont {Zhang}}\ and\ \bibinfo {author} {\bibfnamefont {T.}~\bibnamefont {Zhou}},\ }\bibfield  {title} {\bibinfo {title} {Markovian approaches to modeling intracellular reaction processes with molecular memory},\ }\href {https://www.pnas.org/doi/10.1073/pnas.1913926116} {\bibfield  {journal} {\bibinfo  {journal} {Proc. Natl. Acad. Sci. U.S.A.}\ }\textbf {\bibinfo {volume} {116}},\ \bibinfo {pages} {23542 } (\bibinfo {year} {2019})}\BibitemShut {NoStop}%
\bibitem [{\citenamefont {Cao}\ \emph {et~al.}(2016)\citenamefont {Cao}, \citenamefont {Terebus},\ and\ \citenamefont {Liang}}]{cao_accurate_2016}%
  \BibitemOpen
  \bibfield  {author} {\bibinfo {author} {\bibfnamefont {Y.}~\bibnamefont {Cao}}, \bibinfo {author} {\bibfnamefont {A.}~\bibnamefont {Terebus}},\ and\ \bibinfo {author} {\bibfnamefont {J.}~\bibnamefont {Liang}},\ }\bibfield  {title} {\bibinfo {title} {Accurate chemical master equation solution using multi-finite buffers},\ }\href {https://epubs.siam.org/doi/10.1137/15M1034180} {\bibfield  {journal} {\bibinfo  {journal} {Multiscale Model. Sim.}\ }\textbf {\bibinfo {volume} {14}},\ \bibinfo {pages} {923} (\bibinfo {year} {2016})}\BibitemShut {NoStop}%
\bibitem [{\citenamefont {Cai}\ and\ \citenamefont {Lan}(2025)}]{cai2025revival}%
  \BibitemOpen
  \bibfield  {author} {\bibinfo {author} {\bibfnamefont {R.}~\bibnamefont {Cai}}\ and\ \bibinfo {author} {\bibfnamefont {Y.}~\bibnamefont {Lan}},\ }\bibfield  {title} {\bibinfo {title} {Revival of variational method in noisy cell signaling with fourier observer},\ }\href {https://doi.org/10.1088/1572-9494/adf38c} {\bibfield  {journal} {\bibinfo  {journal} {Commun. Theor. Phys.}\ }\textbf {\bibinfo {volume} {78}},\ \bibinfo {pages} {015601} (\bibinfo {year} {2025})}\BibitemShut {NoStop}%
\bibitem [{\citenamefont {Schlögl}(1972)}]{schlogl_chemical_1972}%
  \BibitemOpen
  \bibfield  {author} {\bibinfo {author} {\bibfnamefont {F.}~\bibnamefont {Schlögl}},\ }\bibfield  {title} {\bibinfo {title} {Chemical reaction models for non-equilibrium phase transitions},\ }\href {https://doi.org/10.1007/BF01379769} {\bibfield  {journal} {\bibinfo  {journal} {Z. Phys.}\ }\textbf {\bibinfo {volume} {253}},\ \bibinfo {pages} {147} (\bibinfo {year} {1972})}\BibitemShut {NoStop}%
\bibitem [{\citenamefont {Kim}\ \emph {et~al.}(2017)\citenamefont {Kim}, \citenamefont {Nonaka}, \citenamefont {Bell}, \citenamefont {Garcia},\ and\ \citenamefont {Donev}}]{kim_stochastic_2017}%
  \BibitemOpen
  \bibfield  {author} {\bibinfo {author} {\bibfnamefont {C.}~\bibnamefont {Kim}}, \bibinfo {author} {\bibfnamefont {A.}~\bibnamefont {Nonaka}}, \bibinfo {author} {\bibfnamefont {J.~B.}\ \bibnamefont {Bell}}, \bibinfo {author} {\bibfnamefont {A.~L.}\ \bibnamefont {Garcia}},\ and\ \bibinfo {author} {\bibfnamefont {A.}~\bibnamefont {Donev}},\ }\bibfield  {title} {\bibinfo {title} {Stochastic simulation of reaction-diffusion systems: {A} fluctuating-hydrodynamics approach},\ }\href {https://doi.org/10.1063/1.4978775} {\bibfield  {journal} {\bibinfo  {journal} {J. Chem. Phys.}\ }\textbf {\bibinfo {volume} {146}},\ \bibinfo {pages} {124110} (\bibinfo {year} {2017})}\BibitemShut {NoStop}%
\bibitem [{\citenamefont {Munsky}\ and\ \citenamefont {Khammash}(2006)}]{munsky_finite_2006}%
  \BibitemOpen
  \bibfield  {author} {\bibinfo {author} {\bibfnamefont {B.}~\bibnamefont {Munsky}}\ and\ \bibinfo {author} {\bibfnamefont {M.}~\bibnamefont {Khammash}},\ }\bibfield  {title} {\bibinfo {title} {The finite state projection algorithm for the solution of the chemical master equation},\ }\href {https://pubs.aip.org/jcp/article/124/4/044104/561868/The-finite-state-projection-algorithm-for-the} {\bibfield  {journal} {\bibinfo  {journal} {J. Chem. Phys.}\ }\textbf {\bibinfo {volume} {124}},\ \bibinfo {pages} {044104} (\bibinfo {year} {2006})}\BibitemShut {NoStop}%
\bibitem [{\citenamefont {Piho}\ and\ \citenamefont {Thomas}(2024)}]{FSP_cellular_sciadv}%
  \BibitemOpen
  \bibfield  {author} {\bibinfo {author} {\bibfnamefont {P.}~\bibnamefont {Piho}}\ and\ \bibinfo {author} {\bibfnamefont {P.}~\bibnamefont {Thomas}},\ }\bibfield  {title} {\bibinfo {title} {Feedback between stochastic gene networks and population dynamics enables cellular decision-making},\ }\href {https://doi.org/10.1126/sciadv.adl4895} {\bibfield  {journal} {\bibinfo  {journal} {Sci. Adv.}\ }\textbf {\bibinfo {volume} {10}},\ \bibinfo {pages} {eadl4895} (\bibinfo {year} {2024})}\BibitemShut {NoStop}%
\bibitem [{\citenamefont {Fang}\ \emph {et~al.}(2024)\citenamefont {Fang}, \citenamefont {Gupta}, \citenamefont {Kumar},\ and\ \citenamefont {Khammash}}]{fang_advanced_2024}%
  \BibitemOpen
  \bibfield  {author} {\bibinfo {author} {\bibfnamefont {Z.}~\bibnamefont {Fang}}, \bibinfo {author} {\bibfnamefont {A.}~\bibnamefont {Gupta}}, \bibinfo {author} {\bibfnamefont {S.}~\bibnamefont {Kumar}},\ and\ \bibinfo {author} {\bibfnamefont {M.}~\bibnamefont {Khammash}},\ }\bibfield  {title} {\bibinfo {title} {Advanced methods for gene network identification and noise decomposition from single-cell data},\ }\href {https://doi.org/10.1038/s41467-024-49177-1} {\bibfield  {journal} {\bibinfo  {journal} {Nat. Commun.}\ }\textbf {\bibinfo {volume} {15}},\ \bibinfo {pages} {4911} (\bibinfo {year} {2024})}\BibitemShut {NoStop}%
\bibitem [{\citenamefont {E}\ \emph {et~al.}(2019)\citenamefont {E}, \citenamefont {Li},\ and\ \citenamefont {Vanden-Eijnden}}]{e_applied_2019}%
  \BibitemOpen
  \bibfield  {author} {\bibinfo {author} {\bibfnamefont {W.}~\bibnamefont {E}}, \bibinfo {author} {\bibfnamefont {T.}~\bibnamefont {Li}},\ and\ \bibinfo {author} {\bibfnamefont {E.}~\bibnamefont {Vanden-Eijnden}},\ }\href@noop {} {\emph {\bibinfo {title} {Applied stochastic analysis}}},\ \bibinfo {series} {Graduate {Studies} in {Mathematics}}\ No.\ \bibinfo {number} {199}\ (\bibinfo  {publisher} {American Mathematical Society},\ \bibinfo {address} {Providence, Rhode Island},\ \bibinfo {year} {2019})\BibitemShut {NoStop}%
\bibitem [{\citenamefont {Liu}\ \emph {et~al.}(2025{\natexlab{a}})\citenamefont {Liu}, \citenamefont {Aden}, \citenamefont {Sarker},\ and\ \citenamefont {Song}}]{liu_dynamical_2025}%
  \BibitemOpen
  \bibfield  {author} {\bibinfo {author} {\bibfnamefont {J.}~\bibnamefont {Liu}}, \bibinfo {author} {\bibfnamefont {N.~M.}\ \bibnamefont {Aden}}, \bibinfo {author} {\bibfnamefont {D.}~\bibnamefont {Sarker}},\ and\ \bibinfo {author} {\bibfnamefont {C.}~\bibnamefont {Song}},\ }\bibfield  {title} {\bibinfo {title} {Dynamical phase transitions in nonequilibrium networks},\ }\href {https://link.aps.org/doi/10.1103/yls4-kdvj} {\bibfield  {journal} {\bibinfo  {journal} {Phys. Rev. Lett.}\ }\textbf {\bibinfo {volume} {135}},\ \bibinfo {pages} {167402} (\bibinfo {year} {2025}{\natexlab{a}})}\BibitemShut {NoStop}%
\bibitem [{\citenamefont {Wei}\ \emph {et~al.}(2025)\citenamefont {Wei}, \citenamefont {Zhu}, \citenamefont {Chu}, \citenamefont {Luo},\ and\ \citenamefont {Fu}}]{wei_non-kramers_2025}%
  \BibitemOpen
  \bibfield  {author} {\bibinfo {author} {\bibfnamefont {J.}~\bibnamefont {Wei}}, \bibinfo {author} {\bibfnamefont {J.}~\bibnamefont {Zhu}}, \bibinfo {author} {\bibfnamefont {P.}~\bibnamefont {Chu}}, \bibinfo {author} {\bibfnamefont {L.}~\bibnamefont {Luo}},\ and\ \bibinfo {author} {\bibfnamefont {X.}~\bibnamefont {Fu}},\ }\href@noop {} {\bibinfo {title} {Non-kramers state transitions in a synthetic toggle switch biosystem}} (\bibinfo {year} {2025}),\ \Eprint {https://arxiv.org/abs/2510.07797} {arXiv:2510.07797} \BibitemShut {NoStop}%
\bibitem [{\citenamefont {Jafarpour}\ \emph {et~al.}(2015)\citenamefont {Jafarpour}, \citenamefont {Biancalani},\ and\ \citenamefont {Goldenfeld}}]{earlylife2015PRL}%
  \BibitemOpen
  \bibfield  {author} {\bibinfo {author} {\bibfnamefont {F.}~\bibnamefont {Jafarpour}}, \bibinfo {author} {\bibfnamefont {T.}~\bibnamefont {Biancalani}},\ and\ \bibinfo {author} {\bibfnamefont {N.}~\bibnamefont {Goldenfeld}},\ }\bibfield  {title} {\bibinfo {title} {Noise-induced mechanism for biological homochirality of early life self-replicators},\ }\href {https://doi.org/10.1103/PhysRevLett.115.158101} {\bibfield  {journal} {\bibinfo  {journal} {Phys. Rev. Lett.}\ }\textbf {\bibinfo {volume} {115}},\ \bibinfo {pages} {158101} (\bibinfo {year} {2015})}\BibitemShut {NoStop}%
\bibitem [{\citenamefont {Halatek}\ and\ \citenamefont {Frey}(2018)}]{halatek2018patternformation}%
  \BibitemOpen
  \bibfield  {author} {\bibinfo {author} {\bibfnamefont {J.}~\bibnamefont {Halatek}}\ and\ \bibinfo {author} {\bibfnamefont {E.}~\bibnamefont {Frey}},\ }\bibfield  {title} {\bibinfo {title} {Rethinking pattern formation in reaction--diffusion systems},\ }\href@noop {} {\bibfield  {journal} {\bibinfo  {journal} {Nature Physics}\ }\textbf {\bibinfo {volume} {14}},\ \bibinfo {pages} {507} (\bibinfo {year} {2018})}\BibitemShut {NoStop}%
\bibitem [{\citenamefont {Gillespie}(1976)}]{gillespie_general_1976}%
  \BibitemOpen
  \bibfield  {author} {\bibinfo {author} {\bibfnamefont {D.~T.}\ \bibnamefont {Gillespie}},\ }\bibfield  {title} {\bibinfo {title} {A general method for numerically simulating the stochastic time evolution of coupled chemical reactions},\ }\href {https://www.sciencedirect.com/science/article/pii/0021999176900413} {\bibfield  {journal} {\bibinfo  {journal} {J. Comput. Phys.}\ }\textbf {\bibinfo {volume} {22}},\ \bibinfo {pages} {403} (\bibinfo {year} {1976})}\BibitemShut {NoStop}%
\bibitem [{\citenamefont {Pagare}\ \emph {et~al.}(2024)\citenamefont {Pagare}, \citenamefont {Zhang}, \citenamefont {Zheng},\ and\ \citenamefont {Lu}}]{pagare_stochastic_2024}%
  \BibitemOpen
  \bibfield  {author} {\bibinfo {author} {\bibfnamefont {A.}~\bibnamefont {Pagare}}, \bibinfo {author} {\bibfnamefont {Z.}~\bibnamefont {Zhang}}, \bibinfo {author} {\bibfnamefont {J.}~\bibnamefont {Zheng}},\ and\ \bibinfo {author} {\bibfnamefont {Z.}~\bibnamefont {Lu}},\ }\bibfield  {title} {\bibinfo {title} {Stochastic distinguishability of markovian trajectories},\ }\href {https://doi.org/10.1063/5.0203335} {\bibfield  {journal} {\bibinfo  {journal} {J. Chem. Phys.}\ }\textbf {\bibinfo {volume} {160}},\ \bibinfo {pages} {171101} (\bibinfo {year} {2024})}\BibitemShut {NoStop}%
\bibitem [{\citenamefont {Chittari}\ and\ \citenamefont {Lu}(2024)}]{10.1063/5.0217316}%
  \BibitemOpen
  \bibfield  {author} {\bibinfo {author} {\bibfnamefont {S.~S.}\ \bibnamefont {Chittari}}\ and\ \bibinfo {author} {\bibfnamefont {Z.}~\bibnamefont {Lu}},\ }\bibfield  {title} {\bibinfo {title} {Revisiting kinetic monte carlo algorithms for time-dependent processes: From open-loop control to feedback control},\ }\href {https://doi.org/10.1063/5.0217316} {\bibfield  {journal} {\bibinfo  {journal} {J. Chem. Phys.}\ }\textbf {\bibinfo {volume} {161}},\ \bibinfo {pages} {044104} (\bibinfo {year} {2024})}\BibitemShut {NoStop}%
\bibitem [{\citenamefont {Nicholson}\ and\ \citenamefont {Gingrich}(2023)}]{nicholson_quantifying_2023}%
  \BibitemOpen
  \bibfield  {author} {\bibinfo {author} {\bibfnamefont {S.~B.}\ \bibnamefont {Nicholson}}\ and\ \bibinfo {author} {\bibfnamefont {T.~R.}\ \bibnamefont {Gingrich}},\ }\bibfield  {title} {\bibinfo {title} {Quantifying rare events in stochastic reaction-diffusion dynamics using tensor networks},\ }\href {https://link.aps.org/doi/10.1103/PhysRevX.13.041006} {\bibfield  {journal} {\bibinfo  {journal} {Phys. Rev. X}\ }\textbf {\bibinfo {volume} {13}},\ \bibinfo {pages} {041006} (\bibinfo {year} {2023})}\BibitemShut {NoStop}%
\bibitem [{\citenamefont {Grima}(2012)}]{grima_study_2012}%
  \BibitemOpen
  \bibfield  {author} {\bibinfo {author} {\bibfnamefont {R.}~\bibnamefont {Grima}},\ }\bibfield  {title} {\bibinfo {title} {A study of the accuracy of moment-closure approximations for stochastic chemical kinetics.},\ }\href {https://doi.org/10.1063/1.3702848} {\bibfield  {journal} {\bibinfo  {journal} {J. Chem. Phys.}\ }\textbf {\bibinfo {volume} {136 15}},\ \bibinfo {pages} {154105} (\bibinfo {year} {2012})}\BibitemShut {NoStop}%
\bibitem [{\citenamefont {Pineda}\ and\ \citenamefont {Stamatakis}(2017)}]{pineda_beyond_2017}%
  \BibitemOpen
  \bibfield  {author} {\bibinfo {author} {\bibfnamefont {M.}~\bibnamefont {Pineda}}\ and\ \bibinfo {author} {\bibfnamefont {M.}~\bibnamefont {Stamatakis}},\ }\bibfield  {title} {\bibinfo {title} {Beyond mean-field approximations for accurate and computationally efficient models of on-lattice chemical kinetics.},\ }\href {https://doi.org/10.1063/1.4991690} {\bibfield  {journal} {\bibinfo  {journal} {J. Chem. Phys.}\ }\textbf {\bibinfo {volume} {147 2}},\ \bibinfo {pages} {024105} (\bibinfo {year} {2017})}\BibitemShut {NoStop}%
\bibitem [{\citenamefont {Gardner}\ \emph {et~al.}(2000)\citenamefont {Gardner}, \citenamefont {Cantor},\ and\ \citenamefont {Collins}}]{gardner_construction_2000}%
  \BibitemOpen
  \bibfield  {author} {\bibinfo {author} {\bibfnamefont {T.~S.}\ \bibnamefont {Gardner}}, \bibinfo {author} {\bibfnamefont {C.~R.}\ \bibnamefont {Cantor}},\ and\ \bibinfo {author} {\bibfnamefont {J.~J.}\ \bibnamefont {Collins}},\ }\bibfield  {title} {\bibinfo {title} {Construction of a genetic toggle switch in escherichia coli},\ }\href {https://www.nature.com/articles/35002131} {\bibfield  {journal} {\bibinfo  {journal} {Nature}\ }\textbf {\bibinfo {volume} {403}},\ \bibinfo {pages} {339} (\bibinfo {year} {2000})}\BibitemShut {NoStop}%
\bibitem [{\citenamefont {Terebus}\ \emph {et~al.}(2019)\citenamefont {Terebus}, \citenamefont {Liu},\ and\ \citenamefont {Liang}}]{terebus_discrete_2019}%
  \BibitemOpen
  \bibfield  {author} {\bibinfo {author} {\bibfnamefont {A.}~\bibnamefont {Terebus}}, \bibinfo {author} {\bibfnamefont {C.}~\bibnamefont {Liu}},\ and\ \bibinfo {author} {\bibfnamefont {J.}~\bibnamefont {Liang}},\ }\bibfield  {title} {\bibinfo {title} {Discrete and continuous models of probability flux of switching dynamics: {Uncovering} stochastic oscillations in a toggle-switch system},\ }\href {https://pubs.aip.org/jcp/article/151/18/185104/198040/Discrete-and-continuous-models-of-probability-flux} {\bibfield  {journal} {\bibinfo  {journal} {J. Chem. Phys.}\ }\textbf {\bibinfo {volume} {151}},\ \bibinfo {pages} {185104} (\bibinfo {year} {2019})}\BibitemShut {NoStop}%
\bibitem [{\citenamefont {Hartmann}\ and\ \citenamefont {Carleo}(2019)}]{PhysRevLett.122.250502}%
  \BibitemOpen
  \bibfield  {author} {\bibinfo {author} {\bibfnamefont {M.~J.}\ \bibnamefont {Hartmann}}\ and\ \bibinfo {author} {\bibfnamefont {G.}~\bibnamefont {Carleo}},\ }\bibfield  {title} {\bibinfo {title} {Neural-network approach to dissipative quantum many-body dynamics},\ }\href {https://doi.org/10.1103/PhysRevLett.122.250502} {\bibfield  {journal} {\bibinfo  {journal} {Phys. Rev. Lett.}\ }\textbf {\bibinfo {volume} {122}},\ \bibinfo {pages} {250502} (\bibinfo {year} {2019})}\BibitemShut {NoStop}%
\bibitem [{\citenamefont {Mehta}\ \emph {et~al.}(2019)\citenamefont {Mehta}, \citenamefont {Bukov}, \citenamefont {Wang}, \citenamefont {Day}, \citenamefont {Richardson}, \citenamefont {Fisher},\ and\ \citenamefont {Schwab}}]{mehta_high-bias_2019}%
  \BibitemOpen
  \bibfield  {author} {\bibinfo {author} {\bibfnamefont {P.}~\bibnamefont {Mehta}}, \bibinfo {author} {\bibfnamefont {M.}~\bibnamefont {Bukov}}, \bibinfo {author} {\bibfnamefont {C.-H.}\ \bibnamefont {Wang}}, \bibinfo {author} {\bibfnamefont {A.~G.}\ \bibnamefont {Day}}, \bibinfo {author} {\bibfnamefont {C.}~\bibnamefont {Richardson}}, \bibinfo {author} {\bibfnamefont {C.~K.}\ \bibnamefont {Fisher}},\ and\ \bibinfo {author} {\bibfnamefont {D.~J.}\ \bibnamefont {Schwab}},\ }\bibfield  {title} {\bibinfo {title} {A high-bias, low-variance introduction to machine learning for physicists},\ }\href {https://linkinghub.elsevier.com/retrieve/pii/S0370157319300766} {\bibfield  {journal} {\bibinfo  {journal} {Phys. Rep.}\ }\textbf {\bibinfo {volume} {810}},\ \bibinfo {pages} {1} (\bibinfo {year} {2019})}\BibitemShut {NoStop}%
\bibitem [{\citenamefont {Carleo}\ \emph {et~al.}(2019)\citenamefont {Carleo}, \citenamefont {Cirac}, \citenamefont {Cranmer}, \citenamefont {Daudet}, \citenamefont {Schuld}, \citenamefont {Tishby}, \citenamefont {Vogt-Maranto},\ and\ \citenamefont {Zdeborová}}]{carleo_machine_2019}%
  \BibitemOpen
  \bibfield  {author} {\bibinfo {author} {\bibfnamefont {G.}~\bibnamefont {Carleo}}, \bibinfo {author} {\bibfnamefont {I.}~\bibnamefont {Cirac}}, \bibinfo {author} {\bibfnamefont {K.}~\bibnamefont {Cranmer}}, \bibinfo {author} {\bibfnamefont {L.}~\bibnamefont {Daudet}}, \bibinfo {author} {\bibfnamefont {M.}~\bibnamefont {Schuld}}, \bibinfo {author} {\bibfnamefont {N.}~\bibnamefont {Tishby}}, \bibinfo {author} {\bibfnamefont {L.}~\bibnamefont {Vogt-Maranto}},\ and\ \bibinfo {author} {\bibfnamefont {L.}~\bibnamefont {Zdeborová}},\ }\bibfield  {title} {\bibinfo {title} {Machine learning and the physical sciences},\ }\href {https://link.aps.org/doi/10.1103/RevModPhys.91.045002} {\bibfield  {journal} {\bibinfo  {journal} {Rev. Mod. Phys.}\ }\textbf {\bibinfo {volume} {91}},\ \bibinfo {pages} {045002} (\bibinfo {year} {2019})}\BibitemShut {NoStop}%
\bibitem [{\citenamefont {Jiang}\ \emph {et~al.}(2021)\citenamefont {Jiang}, \citenamefont {Fu}, \citenamefont {Yan}, \citenamefont {Li}, \citenamefont {Du}, \citenamefont {Cao}, \citenamefont {Qian},\ and\ \citenamefont {Grima}}]{jiang_neural_2021}%
  \BibitemOpen
  \bibfield  {author} {\bibinfo {author} {\bibfnamefont {Q.}~\bibnamefont {Jiang}}, \bibinfo {author} {\bibfnamefont {X.}~\bibnamefont {Fu}}, \bibinfo {author} {\bibfnamefont {S.}~\bibnamefont {Yan}}, \bibinfo {author} {\bibfnamefont {R.}~\bibnamefont {Li}}, \bibinfo {author} {\bibfnamefont {W.}~\bibnamefont {Du}}, \bibinfo {author} {\bibfnamefont {Z.}~\bibnamefont {Cao}}, \bibinfo {author} {\bibfnamefont {F.}~\bibnamefont {Qian}},\ and\ \bibinfo {author} {\bibfnamefont {R.}~\bibnamefont {Grima}},\ }\bibfield  {title} {\bibinfo {title} {Neural network aided approximation and parameter inference of non-markovian models of gene expression},\ }\href {https://www.nature.com/articles/s41467-021-22919-1} {\bibfield  {journal} {\bibinfo  {journal} {Nat. Commun.}\ }\textbf {\bibinfo {volume} {12}},\ \bibinfo {pages} {2618} (\bibinfo {year} {2021})}\BibitemShut {NoStop}%
\bibitem [{\citenamefont {Sukys}\ \emph {et~al.}(2022)\citenamefont {Sukys}, \citenamefont {Öcal},\ and\ \citenamefont {Grima}}]{sukys_approximating_2022}%
  \BibitemOpen
  \bibfield  {author} {\bibinfo {author} {\bibfnamefont {A.}~\bibnamefont {Sukys}}, \bibinfo {author} {\bibfnamefont {K.}~\bibnamefont {Öcal}},\ and\ \bibinfo {author} {\bibfnamefont {R.}~\bibnamefont {Grima}},\ }\bibfield  {title} {\bibinfo {title} {Approximating solutions of the chemical master equation using neural networks},\ }\href {https://linkinghub.elsevier.com/retrieve/pii/S2589004222012822} {\bibfield  {journal} {\bibinfo  {journal} {iScience}\ }\textbf {\bibinfo {volume} {25}},\ \bibinfo {pages} {105010} (\bibinfo {year} {2022})}\BibitemShut {NoStop}%
\bibitem [{\citenamefont {Zhou}\ \emph {et~al.}(2025)\citenamefont {Zhou}, \citenamefont {Lu}, \citenamefont {Gao},\ and\ \citenamefont {Cao}}]{ZHOU2025121574}%
  \BibitemOpen
  \bibfield  {author} {\bibinfo {author} {\bibfnamefont {X.}~\bibnamefont {Zhou}}, \bibinfo {author} {\bibfnamefont {J.}~\bibnamefont {Lu}}, \bibinfo {author} {\bibfnamefont {F.}~\bibnamefont {Gao}},\ and\ \bibinfo {author} {\bibfnamefont {Z.}~\bibnamefont {Cao}},\ }\bibfield  {title} {\bibinfo {title} {Solving the chemical master equation for stochastic biochemical systems: A variational autoencoder approach with effective reactions},\ }\href {https://doi.org/https://doi.org/10.1016/j.ces.2025.121574} {\bibfield  {journal} {\bibinfo  {journal} {Chem. Eng. Sci.}\ }\textbf {\bibinfo {volume} {311}},\ \bibinfo {pages} {121574} (\bibinfo {year} {2025})}\BibitemShut {NoStop}%
\bibitem [{\citenamefont {Tang}\ and\ \citenamefont {Hoffmann}(2022)}]{tang_quantifying_2022}%
  \BibitemOpen
  \bibfield  {author} {\bibinfo {author} {\bibfnamefont {Y.}~\bibnamefont {Tang}}\ and\ \bibinfo {author} {\bibfnamefont {A.}~\bibnamefont {Hoffmann}},\ }\bibfield  {title} {\bibinfo {title} {Quantifying information of intracellular signaling: {Progress} with machine learning},\ }\href {https://iopscience.iop.org/article/10.1088/1361-6633/ac7a4a} {\bibfield  {journal} {\bibinfo  {journal} {Rep. Prog. Phys.}\ }\textbf {\bibinfo {volume} {85}},\ \bibinfo {pages} {086602} (\bibinfo {year} {2022})}\BibitemShut {NoStop}%
\bibitem [{\citenamefont {Anton}\ and\ \citenamefont {Stirnemann}(2026)}]{anton_computational_2025}%
  \BibitemOpen
  \bibfield  {author} {\bibinfo {author} {\bibfnamefont {O.}~\bibnamefont {Anton}}\ and\ \bibinfo {author} {\bibfnamefont {G.}~\bibnamefont {Stirnemann}},\ }\bibfield  {title} {\bibinfo {title} {Computational studies of prebiotic chemistry at the age of machine learning: {From} recent breakthroughs to future revolutions},\ }\href {https://doi.org/10.1002/syst.202500057} {\bibfield  {journal} {\bibinfo  {journal} {ChemSystemsChem}\ }\textbf {\bibinfo {volume} {8}},\ \bibinfo {pages} {e00057} (\bibinfo {year} {2026})}\BibitemShut {NoStop}%
\bibitem [{\citenamefont {Wu}\ \emph {et~al.}(2019)\citenamefont {Wu}, \citenamefont {Wang},\ and\ \citenamefont {Zhang}}]{wu_solving_2019}%
  \BibitemOpen
  \bibfield  {author} {\bibinfo {author} {\bibfnamefont {D.}~\bibnamefont {Wu}}, \bibinfo {author} {\bibfnamefont {L.}~\bibnamefont {Wang}},\ and\ \bibinfo {author} {\bibfnamefont {P.}~\bibnamefont {Zhang}},\ }\bibfield  {title} {\bibinfo {title} {Solving statistical mechanics using variational autoregressive networks},\ }\href {https://link.aps.org/doi/10.1103/PhysRevLett.122.080602} {\bibfield  {journal} {\bibinfo  {journal} {Phys. Rev. Lett.}\ }\textbf {\bibinfo {volume} {122}},\ \bibinfo {pages} {080602} (\bibinfo {year} {2019})}\BibitemShut {NoStop}%
\bibitem [{\citenamefont {Sharir}\ \emph {et~al.}(2020)\citenamefont {Sharir}, \citenamefont {Levine}, \citenamefont {Wies}, \citenamefont {Carleo},\ and\ \citenamefont {Shashua}}]{PhysRevLett.124.020503}%
  \BibitemOpen
  \bibfield  {author} {\bibinfo {author} {\bibfnamefont {O.}~\bibnamefont {Sharir}}, \bibinfo {author} {\bibfnamefont {Y.}~\bibnamefont {Levine}}, \bibinfo {author} {\bibfnamefont {N.}~\bibnamefont {Wies}}, \bibinfo {author} {\bibfnamefont {G.}~\bibnamefont {Carleo}},\ and\ \bibinfo {author} {\bibfnamefont {A.}~\bibnamefont {Shashua}},\ }\bibfield  {title} {\bibinfo {title} {Deep autoregressive models for the efficient variational simulation of many-body quantum systems},\ }\href {https://doi.org/10.1103/PhysRevLett.124.020503} {\bibfield  {journal} {\bibinfo  {journal} {Phys. Rev. Lett.}\ }\textbf {\bibinfo {volume} {124}},\ \bibinfo {pages} {020503} (\bibinfo {year} {2020})}\BibitemShut {NoStop}%
\bibitem [{\citenamefont {Barrett}\ \emph {et~al.}(2022)\citenamefont {Barrett}, \citenamefont {Malyshev},\ and\ \citenamefont {Lvovsky}}]{barrett2022autoregressive}%
  \BibitemOpen
  \bibfield  {author} {\bibinfo {author} {\bibfnamefont {T.~D.}\ \bibnamefont {Barrett}}, \bibinfo {author} {\bibfnamefont {A.}~\bibnamefont {Malyshev}},\ and\ \bibinfo {author} {\bibfnamefont {A.}~\bibnamefont {Lvovsky}},\ }\bibfield  {title} {\bibinfo {title} {Autoregressive neural-network wavefunctions for ab initio quantum chemistry},\ }\href {https://www.nature.com/articles/s42256-022-00461-z} {\bibfield  {journal} {\bibinfo  {journal} {Nat. Mach. Intell.}\ }\textbf {\bibinfo {volume} {4}},\ \bibinfo {pages} {351} (\bibinfo {year} {2022})}\BibitemShut {NoStop}%
\bibitem [{\citenamefont {Luo}\ \emph {et~al.}(2022)\citenamefont {Luo}, \citenamefont {Chen}, \citenamefont {Carrasquilla},\ and\ \citenamefont {Clark}}]{PhysRevLett.128.090501}%
  \BibitemOpen
  \bibfield  {author} {\bibinfo {author} {\bibfnamefont {D.}~\bibnamefont {Luo}}, \bibinfo {author} {\bibfnamefont {Z.}~\bibnamefont {Chen}}, \bibinfo {author} {\bibfnamefont {J.}~\bibnamefont {Carrasquilla}},\ and\ \bibinfo {author} {\bibfnamefont {B.~K.}\ \bibnamefont {Clark}},\ }\bibfield  {title} {\bibinfo {title} {Autoregressive neural network for simulating open quantum systems via a probabilistic formulation},\ }\href {https://doi.org/10.1103/PhysRevLett.128.090501} {\bibfield  {journal} {\bibinfo  {journal} {Phys. Rev. Lett.}\ }\textbf {\bibinfo {volume} {128}},\ \bibinfo {pages} {090501} (\bibinfo {year} {2022})}\BibitemShut {NoStop}%
\bibitem [{\citenamefont {Shin}\ \emph {et~al.}(2021)\citenamefont {Shin}, \citenamefont {Riesselman}, \citenamefont {Kollasch}, \citenamefont {McMahon}, \citenamefont {Simon}, \citenamefont {Sander}, \citenamefont {Manglik}, \citenamefont {Kruse},\ and\ \citenamefont {Marks}}]{shin2021protein}%
  \BibitemOpen
  \bibfield  {author} {\bibinfo {author} {\bibfnamefont {J.-E.}\ \bibnamefont {Shin}}, \bibinfo {author} {\bibfnamefont {A.~J.}\ \bibnamefont {Riesselman}}, \bibinfo {author} {\bibfnamefont {A.~W.}\ \bibnamefont {Kollasch}}, \bibinfo {author} {\bibfnamefont {C.}~\bibnamefont {McMahon}}, \bibinfo {author} {\bibfnamefont {E.}~\bibnamefont {Simon}}, \bibinfo {author} {\bibfnamefont {C.}~\bibnamefont {Sander}}, \bibinfo {author} {\bibfnamefont {A.}~\bibnamefont {Manglik}}, \bibinfo {author} {\bibfnamefont {A.~C.}\ \bibnamefont {Kruse}},\ and\ \bibinfo {author} {\bibfnamefont {D.~S.}\ \bibnamefont {Marks}},\ }\bibfield  {title} {\bibinfo {title} {Protein design and variant prediction using autoregressive generative models},\ }\href {https://www.nature.com/articles/s41467-021-22732-w} {\bibfield  {journal} {\bibinfo  {journal} {Nat. Commun.}\ }\textbf {\bibinfo {volume} {12}},\ \bibinfo {pages} {2403} (\bibinfo {year} {2021})}\BibitemShut {NoStop}%
\bibitem [{\citenamefont {Tang}\ \emph {et~al.}(2023)\citenamefont {Tang}, \citenamefont {Weng},\ and\ \citenamefont {Zhang}}]{tang_neural-network_2023}%
  \BibitemOpen
  \bibfield  {author} {\bibinfo {author} {\bibfnamefont {Y.}~\bibnamefont {Tang}}, \bibinfo {author} {\bibfnamefont {J.}~\bibnamefont {Weng}},\ and\ \bibinfo {author} {\bibfnamefont {P.}~\bibnamefont {Zhang}},\ }\bibfield  {title} {\bibinfo {title} {Neural-network solutions to stochastic reaction networks},\ }\href {https://doi.org/10.1038/s42256-023-00632-6} {\bibfield  {journal} {\bibinfo  {journal} {Nat. Mach. Intell.}\ }\textbf {\bibinfo {volume} {5}},\ \bibinfo {pages} {376} (\bibinfo {year} {2023})}\BibitemShut {NoStop}%
\bibitem [{\citenamefont {Alon}(2006)}]{alon2006introduction}%
  \BibitemOpen
  \bibfield  {author} {\bibinfo {author} {\bibfnamefont {U.}~\bibnamefont {Alon}},\ }\href@noop {} {\emph {\bibinfo {title} {An introduction to systems biology: design principles of biological circuits}}}\ (\bibinfo  {publisher} {CRC press, USA},\ \bibinfo {year} {2006})\BibitemShut {NoStop}%
\bibitem [{\citenamefont {de~Anna}\ \emph {et~al.}(2010)\citenamefont {de~Anna}, \citenamefont {Di~Patti}, \citenamefont {Fanelli}, \citenamefont {McKane},\ and\ \citenamefont {Dauxois}}]{PhysRevE.81.056110}%
  \BibitemOpen
  \bibfield  {author} {\bibinfo {author} {\bibfnamefont {P.}~\bibnamefont {de~Anna}}, \bibinfo {author} {\bibfnamefont {F.}~\bibnamefont {Di~Patti}}, \bibinfo {author} {\bibfnamefont {D.}~\bibnamefont {Fanelli}}, \bibinfo {author} {\bibfnamefont {A.~J.}\ \bibnamefont {McKane}},\ and\ \bibinfo {author} {\bibfnamefont {T.}~\bibnamefont {Dauxois}},\ }\bibfield  {title} {\bibinfo {title} {Spatial model of autocatalytic reactions},\ }\href {https://doi.org/10.1103/PhysRevE.81.056110} {\bibfield  {journal} {\bibinfo  {journal} {Phys. Rev. E}\ }\textbf {\bibinfo {volume} {81}},\ \bibinfo {pages} {056110} (\bibinfo {year} {2010})}\BibitemShut {NoStop}%
\bibitem [{\citenamefont {Fange}\ \emph {et~al.}(2010)\citenamefont {Fange}, \citenamefont {Berg}, \citenamefont {Sjöberg},\ and\ \citenamefont {Elf}}]{2010stochastic_reaction-diffusion}%
  \BibitemOpen
  \bibfield  {author} {\bibinfo {author} {\bibfnamefont {D.}~\bibnamefont {Fange}}, \bibinfo {author} {\bibfnamefont {O.~G.}\ \bibnamefont {Berg}}, \bibinfo {author} {\bibfnamefont {P.}~\bibnamefont {Sjöberg}},\ and\ \bibinfo {author} {\bibfnamefont {J.}~\bibnamefont {Elf}},\ }\bibfield  {title} {\bibinfo {title} {Stochastic reaction-diffusion kinetics in the microscopic limit},\ }\href {https://doi.org/10.1073/pnas.1006565107} {\bibfield  {journal} {\bibinfo  {journal} {Proc. Natl. Acad. Sci.}\ }\textbf {\bibinfo {volume} {107}},\ \bibinfo {pages} {19820} (\bibinfo {year} {2010})}\BibitemShut {NoStop}%
\bibitem [{\citenamefont {Isaacson}(2013)}]{2013CRDME}%
  \BibitemOpen
  \bibfield  {author} {\bibinfo {author} {\bibfnamefont {S.~A.}\ \bibnamefont {Isaacson}},\ }\bibfield  {title} {\bibinfo {title} {A convergent reaction-diffusion master equation},\ }\href {https://doi.org/10.1063/1.4816377} {\bibfield  {journal} {\bibinfo  {journal} {J. Chem. Phys.}\ }\textbf {\bibinfo {volume} {139}},\ \bibinfo {pages} {054101} (\bibinfo {year} {2013})}\BibitemShut {NoStop}%
\bibitem [{\citenamefont {Cirac}\ \emph {et~al.}(2021)\citenamefont {Cirac}, \citenamefont {P\'erez-Garc\'{\i}a}, \citenamefont {Schuch},\ and\ \citenamefont {Verstraete}}]{RevModPhys.93.045003}%
  \BibitemOpen
  \bibfield  {author} {\bibinfo {author} {\bibfnamefont {J.~I.}\ \bibnamefont {Cirac}}, \bibinfo {author} {\bibfnamefont {D.}~\bibnamefont {P\'erez-Garc\'{\i}a}}, \bibinfo {author} {\bibfnamefont {N.}~\bibnamefont {Schuch}},\ and\ \bibinfo {author} {\bibfnamefont {F.}~\bibnamefont {Verstraete}},\ }\bibfield  {title} {\bibinfo {title} {Matrix product states and projected entangled pair states: Concepts, symmetries, theorems},\ }\href {https://doi.org/10.1103/RevModPhys.93.045003} {\bibfield  {journal} {\bibinfo  {journal} {Rev. Mod. Phys.}\ }\textbf {\bibinfo {volume} {93}},\ \bibinfo {pages} {045003} (\bibinfo {year} {2021})}\BibitemShut {NoStop}%
\bibitem [{\citenamefont {Schuch}\ \emph {et~al.}(2007)\citenamefont {Schuch}, \citenamefont {Wolf}, \citenamefont {Verstraete},\ and\ \citenamefont {Cirac}}]{PhysRevLett.98.140506}%
  \BibitemOpen
  \bibfield  {author} {\bibinfo {author} {\bibfnamefont {N.}~\bibnamefont {Schuch}}, \bibinfo {author} {\bibfnamefont {M.~M.}\ \bibnamefont {Wolf}}, \bibinfo {author} {\bibfnamefont {F.}~\bibnamefont {Verstraete}},\ and\ \bibinfo {author} {\bibfnamefont {J.~I.}\ \bibnamefont {Cirac}},\ }\bibfield  {title} {\bibinfo {title} {Computational complexity of projected entangled pair states},\ }\href {https://doi.org/10.1103/PhysRevLett.98.140506} {\bibfield  {journal} {\bibinfo  {journal} {Phys. Rev. Lett.}\ }\textbf {\bibinfo {volume} {98}},\ \bibinfo {pages} {140506} (\bibinfo {year} {2007})}\BibitemShut {NoStop}%
\bibitem [{\citenamefont {Gonz\'alez-Garc\'{\i}a}\ \emph {et~al.}(2024)\citenamefont {Gonz\'alez-Garc\'{\i}a}, \citenamefont {Sang}, \citenamefont {Hsieh}, \citenamefont {Boixo}, \citenamefont {Vidal}, \citenamefont {Potter},\ and\ \citenamefont {Vasseur}}]{PhysRevB.109.235102}%
  \BibitemOpen
  \bibfield  {author} {\bibinfo {author} {\bibfnamefont {S.}~\bibnamefont {Gonz\'alez-Garc\'{\i}a}}, \bibinfo {author} {\bibfnamefont {S.}~\bibnamefont {Sang}}, \bibinfo {author} {\bibfnamefont {T.~H.}\ \bibnamefont {Hsieh}}, \bibinfo {author} {\bibfnamefont {S.}~\bibnamefont {Boixo}}, \bibinfo {author} {\bibfnamefont {G.}~\bibnamefont {Vidal}}, \bibinfo {author} {\bibfnamefont {A.~C.}\ \bibnamefont {Potter}},\ and\ \bibinfo {author} {\bibfnamefont {R.}~\bibnamefont {Vasseur}},\ }\bibfield  {title} {\bibinfo {title} {Random insights into the complexity of two-dimensional tensor network calculations},\ }\href {https://doi.org/10.1103/PhysRevB.109.235102} {\bibfield  {journal} {\bibinfo  {journal} {Phys. Rev. B}\ }\textbf {\bibinfo {volume} {109}},\ \bibinfo {pages} {235102} (\bibinfo {year} {2024})}\BibitemShut {NoStop}%
\bibitem [{\citenamefont {Xie}\ \emph {et~al.}(2014)\citenamefont {Xie}, \citenamefont {Chen}, \citenamefont {Yu}, \citenamefont {Kong}, \citenamefont {Normand},\ and\ \citenamefont {Xiang}}]{2014TensorRenormalization}%
  \BibitemOpen
  \bibfield  {author} {\bibinfo {author} {\bibfnamefont {Z.~Y.}\ \bibnamefont {Xie}}, \bibinfo {author} {\bibfnamefont {J.}~\bibnamefont {Chen}}, \bibinfo {author} {\bibfnamefont {J.~F.}\ \bibnamefont {Yu}}, \bibinfo {author} {\bibfnamefont {X.}~\bibnamefont {Kong}}, \bibinfo {author} {\bibfnamefont {B.}~\bibnamefont {Normand}},\ and\ \bibinfo {author} {\bibfnamefont {T.}~\bibnamefont {Xiang}},\ }\bibfield  {title} {\bibinfo {title} {Tensor renormalization of quantum many-body systems using projected entangled simplex states},\ }\href {https://doi.org/10.1103/PhysRevX.4.011025} {\bibfield  {journal} {\bibinfo  {journal} {Phys. Rev. X}\ }\textbf {\bibinfo {volume} {4}},\ \bibinfo {pages} {011025} (\bibinfo {year} {2014})}\BibitemShut {NoStop}%
\bibitem [{\citenamefont {Larochelle}\ and\ \citenamefont {Murray}(2011)}]{larochelle_neural_2011}%
  \BibitemOpen
  \bibfield  {author} {\bibinfo {author} {\bibfnamefont {H.}~\bibnamefont {Larochelle}}\ and\ \bibinfo {author} {\bibfnamefont {I.}~\bibnamefont {Murray}},\ }\bibfield  {title} {\bibinfo {title} {The neural autoregressive distribution estimator},\ }in\ \href@noop {} {\emph {\bibinfo {booktitle} {Proceedings of the {Fourteenth} {International} {Conference} on {Artificial} {Intelligence} and {Statistics}}}},\ Vol.~\bibinfo {volume} {15}\ (\bibinfo  {publisher} {PMLR},\ \bibinfo {address} {Fort Lauderdale, FL, USA},\ \bibinfo {year} {2011})\ pp.\ \bibinfo {pages} {29--37}\BibitemShut {NoStop}%
\bibitem [{\citenamefont {Uria}\ \emph {et~al.}(2016)\citenamefont {Uria}, \citenamefont {Côté}, \citenamefont {Gregor}, \citenamefont {Murray},\ and\ \citenamefont {Larochelle}}]{uria_neural_2016}%
  \BibitemOpen
  \bibfield  {author} {\bibinfo {author} {\bibfnamefont {B.}~\bibnamefont {Uria}}, \bibinfo {author} {\bibfnamefont {M.-A.}\ \bibnamefont {Côté}}, \bibinfo {author} {\bibfnamefont {K.}~\bibnamefont {Gregor}}, \bibinfo {author} {\bibfnamefont {I.}~\bibnamefont {Murray}},\ and\ \bibinfo {author} {\bibfnamefont {H.}~\bibnamefont {Larochelle}},\ }\bibfield  {title} {\bibinfo {title} {Neural autoregressive distribution estimation},\ }\href {http://jmlr.org/papers/v17/16-272.html} {\bibfield  {journal} {\bibinfo  {journal} {J. Mach. Learn. Res.}\ }\textbf {\bibinfo {volume} {17}},\ \bibinfo {pages} {1} (\bibinfo {year} {2016})}\BibitemShut {NoStop}%
\bibitem [{\citenamefont {Amari}(1998)}]{Amari1998naturalgradient}%
  \BibitemOpen
  \bibfield  {author} {\bibinfo {author} {\bibfnamefont {S.-i.}\ \bibnamefont {Amari}},\ }\bibfield  {title} {\bibinfo {title} {Natural gradient works efficiently in learning},\ }\href {https://doi.org/10.1162/089976698300017746} {\bibfield  {journal} {\bibinfo  {journal} {Neural Comput.}\ }\textbf {\bibinfo {volume} {10}},\ \bibinfo {pages} {251} (\bibinfo {year} {1998})}\BibitemShut {NoStop}%
\bibitem [{\citenamefont {Pascanu}\ and\ \citenamefont {Bengio}(2014)}]{pascanu2014natural}%
  \BibitemOpen
  \bibfield  {author} {\bibinfo {author} {\bibfnamefont {R.}~\bibnamefont {Pascanu}}\ and\ \bibinfo {author} {\bibfnamefont {Y.}~\bibnamefont {Bengio}},\ }\bibfield  {title} {\bibinfo {title} {Revisiting natural gradient for deep networks},\ }in\ \href@noop {} {\emph {\bibinfo {booktitle} {2nd International Conference on Learning Representations, {ICLR} 2014, Banff, AB, Canada, April 14-16, 2014, Conference Track Proceedings}}}\ (\bibinfo  {publisher} {arXiv},\ \bibinfo {year} {2014})\ \Eprint {https://arxiv.org/abs/1301.3584} {arXiv:1301.3584} \BibitemShut {NoStop}%
\bibitem [{\citenamefont {Liu}\ \emph {et~al.}(2025{\natexlab{b}})\citenamefont {Liu}, \citenamefont {Tang},\ and\ \citenamefont {Zhang}}]{liu_efficient_2025}%
  \BibitemOpen
  \bibfield  {author} {\bibinfo {author} {\bibfnamefont {J.}~\bibnamefont {Liu}}, \bibinfo {author} {\bibfnamefont {Y.}~\bibnamefont {Tang}},\ and\ \bibinfo {author} {\bibfnamefont {P.}~\bibnamefont {Zhang}},\ }\bibfield  {title} {\bibinfo {title} {Efficient optimization of variational autoregressive networks with natural gradient},\ }\href {https://link.aps.org/doi/10.1103/PhysRevE.111.025304} {\bibfield  {journal} {\bibinfo  {journal} {Phys. Rev. E}\ }\textbf {\bibinfo {volume} {111}},\ \bibinfo {pages} {025304} (\bibinfo {year} {2025}{\natexlab{b}})}\BibitemShut {NoStop}%
\bibitem [{\citenamefont {Reh}\ \emph {et~al.}(2021)\citenamefont {Reh}, \citenamefont {Schmitt},\ and\ \citenamefont {Gärttner}}]{reh_time-dependent_2021}%
  \BibitemOpen
  \bibfield  {author} {\bibinfo {author} {\bibfnamefont {M.}~\bibnamefont {Reh}}, \bibinfo {author} {\bibfnamefont {M.}~\bibnamefont {Schmitt}},\ and\ \bibinfo {author} {\bibfnamefont {M.}~\bibnamefont {Gärttner}},\ }\bibfield  {title} {\bibinfo {title} {Time-dependent variational principle for open quantum systems with artificial neural networks},\ }\href {https://link.aps.org/doi/10.1103/PhysRevLett.127.230501} {\bibfield  {journal} {\bibinfo  {journal} {Phys. Rev. Lett.}\ }\textbf {\bibinfo {volume} {127}},\ \bibinfo {pages} {230501} (\bibinfo {year} {2021})}\BibitemShut {NoStop}%
\bibitem [{\citenamefont {Chen}\ and\ \citenamefont {Heyl}(2024)}]{chen_empowering_2024}%
  \BibitemOpen
  \bibfield  {author} {\bibinfo {author} {\bibfnamefont {A.}~\bibnamefont {Chen}}\ and\ \bibinfo {author} {\bibfnamefont {M.}~\bibnamefont {Heyl}},\ }\bibfield  {title} {\bibinfo {title} {Empowering deep neural quantum states through efficient optimization},\ }\href {https://www.nature.com/articles/s41567-024-02566-1} {\bibfield  {journal} {\bibinfo  {journal} {Nat. Phys.}\ }\textbf {\bibinfo {volume} {20}},\ \bibinfo {pages} {1476} (\bibinfo {year} {2024})}\BibitemShut {NoStop}%
\bibitem [{\citenamefont {Rende}\ \emph {et~al.}(2024)\citenamefont {Rende}, \citenamefont {Viteritti}, \citenamefont {Bardone}, \citenamefont {Becca},\ and\ \citenamefont {Goldt}}]{rende_simple_2024}%
  \BibitemOpen
  \bibfield  {author} {\bibinfo {author} {\bibfnamefont {R.}~\bibnamefont {Rende}}, \bibinfo {author} {\bibfnamefont {L.~L.}\ \bibnamefont {Viteritti}}, \bibinfo {author} {\bibfnamefont {L.}~\bibnamefont {Bardone}}, \bibinfo {author} {\bibfnamefont {F.}~\bibnamefont {Becca}},\ and\ \bibinfo {author} {\bibfnamefont {S.}~\bibnamefont {Goldt}},\ }\bibfield  {title} {\bibinfo {title} {A simple linear algebra identity to optimize large-scale neural network quantum states},\ }\href {https://www.nature.com/articles/s42005-024-01732-4} {\bibfield  {journal} {\bibinfo  {journal} {Commun. Phys.}\ }\textbf {\bibinfo {volume} {7}},\ \bibinfo {pages} {260} (\bibinfo {year} {2024})}\BibitemShut {NoStop}%
\bibitem [{\citenamefont {Misery}\ \emph {et~al.}(2026)\citenamefont {Misery}, \citenamefont {Gravina}, \citenamefont {Santini},\ and\ \citenamefont {Vicentini}}]{misery_looking_2025}%
  \BibitemOpen
  \bibfield  {author} {\bibinfo {author} {\bibfnamefont {A.}~\bibnamefont {Misery}}, \bibinfo {author} {\bibfnamefont {L.}~\bibnamefont {Gravina}}, \bibinfo {author} {\bibfnamefont {A.}~\bibnamefont {Santini}},\ and\ \bibinfo {author} {\bibfnamefont {F.}~\bibnamefont {Vicentini}},\ }\bibfield  {title} {\bibinfo {title} {Looking elsewhere: improving variational {Monte} {Carlo} gradients by importance sampling},\ }\href {https://doi.org/10.1088/2632-2153/ae387f} {\bibfield  {journal} {\bibinfo  {journal} {Mach. Learn.: Sci. Technol.}\ }\textbf {\bibinfo {volume} {7}},\ \bibinfo {pages} {015035} (\bibinfo {year} {2026})}\BibitemShut {NoStop}%
\bibitem [{\citenamefont {Karan}\ and\ \citenamefont {Du}(2026)}]{karan2025reasoningsamplingbasemodel}%
  \BibitemOpen
  \bibfield  {author} {\bibinfo {author} {\bibfnamefont {A.}~\bibnamefont {Karan}}\ and\ \bibinfo {author} {\bibfnamefont {Y.}~\bibnamefont {Du}},\ }\bibfield  {title} {\bibinfo {title} {Reasoning with sampling: Your base model is smarter than you think},\ }in\ \href@noop {} {\emph {\bibinfo {booktitle} {International Conference on Learning Representations, {ICLR} 2026}}}\ (\bibinfo {address} {Rio de Janeiro, Brazil},\ \bibinfo {year} {2026})\ pp.\ \bibinfo {pages} {476--494}\BibitemShut {NoStop}%
\bibitem [{\citenamefont {Huang}\ and\ \citenamefont {Ferrell}(1996)}]{1996PNAS_MAPK}%
  \BibitemOpen
  \bibfield  {author} {\bibinfo {author} {\bibfnamefont {C.~Y.}\ \bibnamefont {Huang}}\ and\ \bibinfo {author} {\bibfnamefont {J.~E.}\ \bibnamefont {Ferrell}},\ }\bibfield  {title} {\bibinfo {title} {Ultrasensitivity in the mitogen-activated protein kinase cascade.},\ }\href {https://doi.org/10.1073/pnas.93.19.10078} {\bibfield  {journal} {\bibinfo  {journal} {Proc. Natl. Acad. Sci. U.S.A.}\ }\textbf {\bibinfo {volume} {93}},\ \bibinfo {pages} {10078} (\bibinfo {year} {1996})}\BibitemShut {NoStop}%
\bibitem [{\citenamefont {Johnson}\ and\ \citenamefont {Lapadat}(2002)}]{johnson_mitogen_2002}%
  \BibitemOpen
  \bibfield  {author} {\bibinfo {author} {\bibfnamefont {G.~L.}\ \bibnamefont {Johnson}}\ and\ \bibinfo {author} {\bibfnamefont {R.}~\bibnamefont {Lapadat}},\ }\bibfield  {title} {\bibinfo {title} {Mitogen-activated protein kinase pathways mediated by {ERK}, {JNK}, and p38 protein kinases},\ }\href {https://doi.org/10.1126/science.1072682} {\bibfield  {journal} {\bibinfo  {journal} {Science}\ }\textbf {\bibinfo {volume} {298}},\ \bibinfo {pages} {1911} (\bibinfo {year} {2002})}\BibitemShut {NoStop}%
\bibitem [{\citenamefont {Vellela}\ and\ \citenamefont {Qian}(2009)}]{vellela_stochastic_2009}%
  \BibitemOpen
  \bibfield  {author} {\bibinfo {author} {\bibfnamefont {M.}~\bibnamefont {Vellela}}\ and\ \bibinfo {author} {\bibfnamefont {H.}~\bibnamefont {Qian}},\ }\bibfield  {title} {\bibinfo {title} {Stochastic dynamics and non-equilibrium thermodynamics of a bistable chemical system: the {Schl\"{o}gl} model revisited},\ }\href {https://doi.org/10.1098/rsif.2008.0476} {\bibfield  {journal} {\bibinfo  {journal} {J. R. Soc. Interface}\ }\textbf {\bibinfo {volume} {6}},\ \bibinfo {pages} {925} (\bibinfo {year} {2009})}\BibitemShut {NoStop}%
\bibitem [{\citenamefont {Ge}\ \emph {et~al.}(2012)\citenamefont {Ge}, \citenamefont {Qian},\ and\ \citenamefont {Qian}}]{ge2012stochastic}%
  \BibitemOpen
  \bibfield  {author} {\bibinfo {author} {\bibfnamefont {H.}~\bibnamefont {Ge}}, \bibinfo {author} {\bibfnamefont {M.}~\bibnamefont {Qian}},\ and\ \bibinfo {author} {\bibfnamefont {H.}~\bibnamefont {Qian}},\ }\bibfield  {title} {\bibinfo {title} {Stochastic theory of nonequilibrium steady states. part ii: Applications in chemical biophysics},\ }\href {http://www.sciencedirect.com/science/article/pii/S0370157311002419} {\bibfield  {journal} {\bibinfo  {journal} {Phys. Rep.}\ }\textbf {\bibinfo {volume} {510}},\ \bibinfo {pages} {87} (\bibinfo {year} {2012})}\BibitemShut {NoStop}%
\bibitem [{\citenamefont {Hibat-Allah}\ \emph {et~al.}(2020)\citenamefont {Hibat-Allah}, \citenamefont {Ganahl}, \citenamefont {Hayward}, \citenamefont {Melko},\ and\ \citenamefont {Carrasquilla}}]{hibat-allah_recurrent_2020}%
  \BibitemOpen
  \bibfield  {author} {\bibinfo {author} {\bibfnamefont {M.}~\bibnamefont {Hibat-Allah}}, \bibinfo {author} {\bibfnamefont {M.}~\bibnamefont {Ganahl}}, \bibinfo {author} {\bibfnamefont {L.~E.}\ \bibnamefont {Hayward}}, \bibinfo {author} {\bibfnamefont {R.~G.}\ \bibnamefont {Melko}},\ and\ \bibinfo {author} {\bibfnamefont {J.}~\bibnamefont {Carrasquilla}},\ }\bibfield  {title} {\bibinfo {title} {Recurrent neural network wave functions},\ }\href {https://link.aps.org/doi/10.1103/PhysRevResearch.2.023358} {\bibfield  {journal} {\bibinfo  {journal} {Phys. Rev. Research}\ }\textbf {\bibinfo {volume} {2}},\ \bibinfo {pages} {023358} (\bibinfo {year} {2020})}\BibitemShut {NoStop}%
\bibitem [{\citenamefont {Germain}\ \emph {et~al.}(2015)\citenamefont {Germain}, \citenamefont {Gregor}, \citenamefont {Murray},\ and\ \citenamefont {Larochelle}}]{germain_made_2015}%
  \BibitemOpen
  \bibfield  {author} {\bibinfo {author} {\bibfnamefont {M.}~\bibnamefont {Germain}}, \bibinfo {author} {\bibfnamefont {K.}~\bibnamefont {Gregor}}, \bibinfo {author} {\bibfnamefont {I.}~\bibnamefont {Murray}},\ and\ \bibinfo {author} {\bibfnamefont {H.}~\bibnamefont {Larochelle}},\ }\bibfield  {title} {\bibinfo {title} {{MADE}: {Masked} autoencoder for distribution estimation},\ }in\ \href@noop {} {\emph {\bibinfo {booktitle} {Proceedings of the 32nd {International} {Conference} on {Machine} {Learning}}}},\ Vol.~\bibinfo {volume} {37}\ (\bibinfo  {publisher} {PMLR},\ \bibinfo {address} {Lille, France},\ \bibinfo {year} {2015})\ pp.\ \bibinfo {pages} {881--889}\BibitemShut {NoStop}%
\bibitem [{\citenamefont {Elman}(1990)}]{Elman1990}%
  \BibitemOpen
  \bibfield  {author} {\bibinfo {author} {\bibfnamefont {J.~L.}\ \bibnamefont {Elman}},\ }\bibfield  {title} {\bibinfo {title} {Finding structure in time},\ }\href@noop {} {\bibfield  {journal} {\bibinfo  {journal} {Cognitive Science}\ }\textbf {\bibinfo {volume} {14}},\ \bibinfo {pages} {179} (\bibinfo {year} {1990})}\BibitemShut {NoStop}%
\bibitem [{\citenamefont {Hochreiter}\ and\ \citenamefont {Schmidhuber}(1997)}]{LSTM_1997}%
  \BibitemOpen
  \bibfield  {author} {\bibinfo {author} {\bibfnamefont {S.}~\bibnamefont {Hochreiter}}\ and\ \bibinfo {author} {\bibfnamefont {J.}~\bibnamefont {Schmidhuber}},\ }\bibfield  {title} {\bibinfo {title} {Long short-term memory},\ }\href {https://doi.org/10.1162/neco.1997.9.8.1735} {\bibfield  {journal} {\bibinfo  {journal} {Neural Computation}\ }\textbf {\bibinfo {volume} {9}},\ \bibinfo {pages} {1735} (\bibinfo {year} {1997})}\BibitemShut {NoStop}%
\bibitem [{\citenamefont {Cho}\ \emph {et~al.}(2014)\citenamefont {Cho}, \citenamefont {Van~Merrienboer}, \citenamefont {Gulcehre}, \citenamefont {Bahdanau}, \citenamefont {Bougares}, \citenamefont {Schwenk},\ and\ \citenamefont {Bengio}}]{cho_learning_2014}%
  \BibitemOpen
  \bibfield  {author} {\bibinfo {author} {\bibfnamefont {K.}~\bibnamefont {Cho}}, \bibinfo {author} {\bibfnamefont {B.}~\bibnamefont {Van~Merrienboer}}, \bibinfo {author} {\bibfnamefont {C.}~\bibnamefont {Gulcehre}}, \bibinfo {author} {\bibfnamefont {D.}~\bibnamefont {Bahdanau}}, \bibinfo {author} {\bibfnamefont {F.}~\bibnamefont {Bougares}}, \bibinfo {author} {\bibfnamefont {H.}~\bibnamefont {Schwenk}},\ and\ \bibinfo {author} {\bibfnamefont {Y.}~\bibnamefont {Bengio}},\ }\bibfield  {title} {\bibinfo {title} {Learning phrase representations using {RNN} encoder–decoder for statistical machine translation},\ }in\ \href@noop {} {\emph {\bibinfo {booktitle} {Proceedings of the 2014 {Conference} on {Empirical} {Methods} in {Natural} {Language} {Processing} ({EMNLP})}}}\ (\bibinfo  {publisher} {Association for Computational Linguistics},\ \bibinfo {address} {Doha, Qatar},\ \bibinfo {year} {2014})\ pp.\ \bibinfo {pages} {1724--1734},\ \Eprint {https://arxiv.org/abs/1406.1078} {arXiv:1406.1078} \BibitemShut {NoStop}%
\bibitem [{\citenamefont {Sutskever}\ \emph {et~al.}(2014)\citenamefont {Sutskever}, \citenamefont {Vinyals},\ and\ \citenamefont {Le}}]{sutskever_sequence_2014}%
  \BibitemOpen
  \bibfield  {author} {\bibinfo {author} {\bibfnamefont {I.}~\bibnamefont {Sutskever}}, \bibinfo {author} {\bibfnamefont {O.}~\bibnamefont {Vinyals}},\ and\ \bibinfo {author} {\bibfnamefont {Q.~V.}\ \bibnamefont {Le}},\ }\bibfield  {title} {\bibinfo {title} {Sequence to sequence learning with neural networks},\ }in\ \href@noop {} {\emph {\bibinfo {booktitle} {Advances in {Neural} {Information} {Processing} {Systems}}}}\ (\bibinfo  {publisher} {Curran Associates, Inc.},\ \bibinfo {address} {Red Hook, NY, USA},\ \bibinfo {year} {2014})\ \Eprint {https://arxiv.org/abs/1409.3215} {1409.3215} \BibitemShut {NoStop}%
\bibitem [{\citenamefont {Vaswani}\ \emph {et~al.}(2017)\citenamefont {Vaswani}, \citenamefont {Shazeer}, \citenamefont {Parmar}, \citenamefont {Uszkoreit}, \citenamefont {Jones}, \citenamefont {Gomez}, \citenamefont {Kaiser},\ and\ \citenamefont {Polosukhin}}]{vaswani_attention_2017}%
  \BibitemOpen
  \bibfield  {author} {\bibinfo {author} {\bibfnamefont {A.}~\bibnamefont {Vaswani}}, \bibinfo {author} {\bibfnamefont {N.}~\bibnamefont {Shazeer}}, \bibinfo {author} {\bibfnamefont {N.}~\bibnamefont {Parmar}}, \bibinfo {author} {\bibfnamefont {J.}~\bibnamefont {Uszkoreit}}, \bibinfo {author} {\bibfnamefont {L.}~\bibnamefont {Jones}}, \bibinfo {author} {\bibfnamefont {A.~N.}\ \bibnamefont {Gomez}}, \bibinfo {author} {\bibfnamefont {{\L}.}~\bibnamefont {Kaiser}},\ and\ \bibinfo {author} {\bibfnamefont {I.}~\bibnamefont {Polosukhin}},\ }\bibfield  {title} {\bibinfo {title} {Attention is all you need},\ }in\ \href@noop {} {\emph {\bibinfo {booktitle} {Proceedings of the 31st {International} {Conference} on {Neural} {Information} {Processing} {Systems}}}},\ \bibinfo {series and number} {{NIPS}'17}\ (\bibinfo  {publisher} {Curran Associates Inc.},\ \bibinfo {address} {Red Hook, NY, USA},\ \bibinfo {year} {2017})\ pp.\ \bibinfo {pages} {6000--6010},\ \Eprint {https://arxiv.org/abs/1706.03762} {arXiv:1706.03762} \BibitemShut {NoStop}%
\bibitem [{\citenamefont {He}\ \emph {et~al.}(2025)\citenamefont {He}, \citenamefont {Chen}, \citenamefont {Zhang}, \citenamefont {Barber},\ and\ \citenamefont {Hern{\'a}ndez-Lobato}}]{pmlr-diffusiveKL}%
  \BibitemOpen
  \bibfield  {author} {\bibinfo {author} {\bibfnamefont {J.}~\bibnamefont {He}}, \bibinfo {author} {\bibfnamefont {W.}~\bibnamefont {Chen}}, \bibinfo {author} {\bibfnamefont {M.}~\bibnamefont {Zhang}}, \bibinfo {author} {\bibfnamefont {D.}~\bibnamefont {Barber}},\ and\ \bibinfo {author} {\bibfnamefont {J.~M.}\ \bibnamefont {Hern{\'a}ndez-Lobato}},\ }\bibfield  {title} {\bibinfo {title} {Training neural samplers with reverse diffusive kl divergence},\ }in\ \href@noop {} {\emph {\bibinfo {booktitle} {Proceedings of The 28th International Conference on Artificial Intelligence and Statistics}}},\ \bibinfo {series} {Proceedings of Machine Learning Research}, Vol.\ \bibinfo {volume} {258}\ (\bibinfo  {publisher} {PMLR},\ \bibinfo {address} {Mai Khao, Thailand},\ \bibinfo {year} {2025})\ pp.\ \bibinfo {pages} {5167--5175},\ \Eprint {https://arxiv.org/abs/2410.12456} {arXiv:2410.12456} \BibitemShut {NoStop}%
\bibitem [{\citenamefont {Pauloski}\ \emph {et~al.}(2020)\citenamefont {Pauloski}, \citenamefont {Zhang}, \citenamefont {Huang}, \citenamefont {Xu},\ and\ \citenamefont {Foster}}]{pauloski_convolutional_2020}%
  \BibitemOpen
  \bibfield  {author} {\bibinfo {author} {\bibfnamefont {J.~G.}\ \bibnamefont {Pauloski}}, \bibinfo {author} {\bibfnamefont {Z.}~\bibnamefont {Zhang}}, \bibinfo {author} {\bibfnamefont {L.}~\bibnamefont {Huang}}, \bibinfo {author} {\bibfnamefont {W.}~\bibnamefont {Xu}},\ and\ \bibinfo {author} {\bibfnamefont {I.~T.}\ \bibnamefont {Foster}},\ }\bibfield  {title} {\bibinfo {title} {Convolutional neural network training with distributed {K}-{FAC}},\ }in\ \href@noop {} {\emph {\bibinfo {booktitle} {Proceedings of the {International} {Conference} for {High} {Performance} {Computing}, {Networking}, {Storage} and {Analysis}}}},\ \bibinfo {series and number} {{SC} '20}\ (\bibinfo  {publisher} {IEEE Press},\ \bibinfo {address} {Atlanta, Georgia},\ \bibinfo {year} {2020})\ pp.\ \bibinfo {pages} {1--12},\ \Eprint {https://arxiv.org/abs/2007.00784} {arXiv:2007.00784} \BibitemShut {NoStop}%
\bibitem [{\citenamefont {Kaul}\ and\ \citenamefont {Lall}(2023)}]{kaul_projective_2023}%
  \BibitemOpen
  \bibfield  {author} {\bibinfo {author} {\bibfnamefont {P.}~\bibnamefont {Kaul}}\ and\ \bibinfo {author} {\bibfnamefont {B.}~\bibnamefont {Lall}},\ }\bibfield  {title} {\bibinfo {title} {Projective fisher information for natural gradient descent},\ }\href {https://ieeexplore.ieee.org/document/9721554/} {\bibfield  {journal} {\bibinfo  {journal} {IEEE Trans. Artif. Intell.}\ }\textbf {\bibinfo {volume} {4}},\ \bibinfo {pages} {304} (\bibinfo {year} {2023})}\BibitemShut {NoStop}%
\bibitem [{\citenamefont {Zhang}\ \emph {et~al.}(2023)\citenamefont {Zhang}, \citenamefont {Yao}, \citenamefont {Shi},\ and\ \citenamefont {Gu}}]{zhang_kronecker-factored_2023}%
  \BibitemOpen
  \bibfield  {author} {\bibinfo {author} {\bibfnamefont {C.}~\bibnamefont {Zhang}}, \bibinfo {author} {\bibfnamefont {X.}~\bibnamefont {Yao}}, \bibinfo {author} {\bibfnamefont {C.}~\bibnamefont {Shi}},\ and\ \bibinfo {author} {\bibfnamefont {M.}~\bibnamefont {Gu}},\ }\bibfield  {title} {\bibinfo {title} {Kronecker-factored approximate curvature with adaptive learning rate for optimizing model-agnostic meta-learning},\ }\href {https://link.springer.com/10.1007/s00530-023-01159-x} {\bibfield  {journal} {\bibinfo  {journal} {Multimed. Syst.}\ }\textbf {\bibinfo {volume} {29}},\ \bibinfo {pages} {3169} (\bibinfo {year} {2023})}\BibitemShut {NoStop}%
\bibitem [{\citenamefont {Grosse}\ and\ \citenamefont {Salakhutdinov}(2015)}]{grosse_scaling_2015}%
  \BibitemOpen
  \bibfield  {author} {\bibinfo {author} {\bibfnamefont {R.~B.}\ \bibnamefont {Grosse}}\ and\ \bibinfo {author} {\bibfnamefont {R.}~\bibnamefont {Salakhutdinov}},\ }\bibfield  {title} {\bibinfo {title} {Scaling up natural gradient by sparsely factorizing the inverse fisher matrix},\ }in\ \href@noop {} {\emph {\bibinfo {booktitle} {Proceedings of the 32nd International Conference on International Conference on Machine Learning - Volume 37}}},\ \bibinfo {series and number} {ICML'15}\ (\bibinfo  {publisher} {JMLR.org},\ \bibinfo {address} {Lille, France},\ \bibinfo {year} {2015})\ p.\ \bibinfo {pages} {2304–2313},\ \Eprint {https://arxiv.org/abs/2502.08696} {2502.08696} \BibitemShut {NoStop}%
\bibitem [{\citenamefont {Kolotouros}\ and\ \citenamefont {Wallden}(2024)}]{kolotouros_random_2024}%
  \BibitemOpen
  \bibfield  {author} {\bibinfo {author} {\bibfnamefont {I.}~\bibnamefont {Kolotouros}}\ and\ \bibinfo {author} {\bibfnamefont {P.}~\bibnamefont {Wallden}},\ }\bibfield  {title} {\bibinfo {title} {Random natural gradient},\ }\href {https://quantum-journal.org/papers/q-2024-10-22-1503/} {\bibfield  {journal} {\bibinfo  {journal} {Quantum}\ }\textbf {\bibinfo {volume} {8}},\ \bibinfo {pages} {1503} (\bibinfo {year} {2024})}\BibitemShut {NoStop}%
\bibitem [{\citenamefont {Sen}\ \emph {et~al.}(2024)\citenamefont {Sen}, \citenamefont {Qin}, \citenamefont {C}, \citenamefont {N}, \citenamefont {Chen},\ and\ \citenamefont {Raman}}]{sen_sofim_2024}%
  \BibitemOpen
  \bibfield  {author} {\bibinfo {author} {\bibfnamefont {M.}~\bibnamefont {Sen}}, \bibinfo {author} {\bibfnamefont {A.~K.}\ \bibnamefont {Qin}}, \bibinfo {author} {\bibfnamefont {G.}~\bibnamefont {C}}, \bibinfo {author} {\bibfnamefont {R.~K.}\ \bibnamefont {N}}, \bibinfo {author} {\bibfnamefont {Y.-W.}\ \bibnamefont {Chen}},\ and\ \bibinfo {author} {\bibfnamefont {B.}~\bibnamefont {Raman}},\ }\bibfield  {title} {\bibinfo {title} {{SOFIM}: {Stochastic} optimization using regularized fisher information matrix},\ }in\ \href@noop {} {\emph {\bibinfo {booktitle} {2024 {International} {Joint} {Conference} on {Neural} {Networks} ({IJCNN})}}}\ (\bibinfo  {publisher} {IEEE},\ \bibinfo {address} {Yokohama, Japan},\ \bibinfo {year} {2024})\ \Eprint {https://arxiv.org/abs/2403.02833} {2403.02833} \BibitemShut {NoStop}%
\bibitem [{\citenamefont {Wu}\ \emph {et~al.}(2024)\citenamefont {Wu}, \citenamefont {Hu}, \citenamefont {Zhang},\ and\ \citenamefont {Wen}}]{wu_convergence_2024}%
  \BibitemOpen
  \bibfield  {author} {\bibinfo {author} {\bibfnamefont {J.}~\bibnamefont {Wu}}, \bibinfo {author} {\bibfnamefont {J.}~\bibnamefont {Hu}}, \bibinfo {author} {\bibfnamefont {H.}~\bibnamefont {Zhang}},\ and\ \bibinfo {author} {\bibfnamefont {Z.}~\bibnamefont {Wen}},\ }\bibfield  {title} {\bibinfo {title} {Convergence analysis of an adaptively regularized natural gradient method},\ }\href {https://doi.org/10.1109/TSP.2024.3398496} {\bibfield  {journal} {\bibinfo  {journal} {IEEE Trans. Signal Process.}\ }\textbf {\bibinfo {volume} {72}},\ \bibinfo {pages} {2527} (\bibinfo {year} {2024})}\BibitemShut {NoStop}%
\bibitem [{\citenamefont {Haegeman}\ \emph {et~al.}(2011)\citenamefont {Haegeman}, \citenamefont {Cirac}, \citenamefont {Osborne}, \citenamefont {Pižorn}, \citenamefont {Verschelde},\ and\ \citenamefont {Verstraete}}]{haegeman_time-dependent_2011}%
  \BibitemOpen
  \bibfield  {author} {\bibinfo {author} {\bibfnamefont {J.}~\bibnamefont {Haegeman}}, \bibinfo {author} {\bibfnamefont {J.~I.}\ \bibnamefont {Cirac}}, \bibinfo {author} {\bibfnamefont {T.~J.}\ \bibnamefont {Osborne}}, \bibinfo {author} {\bibfnamefont {I.}~\bibnamefont {Pižorn}}, \bibinfo {author} {\bibfnamefont {H.}~\bibnamefont {Verschelde}},\ and\ \bibinfo {author} {\bibfnamefont {F.}~\bibnamefont {Verstraete}},\ }\bibfield  {title} {\bibinfo {title} {Time-dependent variational principle for quantum lattices},\ }\href {https://link.aps.org/doi/10.1103/PhysRevLett.107.070601} {\bibfield  {journal} {\bibinfo  {journal} {Phys. Rev. Lett.}\ }\textbf {\bibinfo {volume} {107}},\ \bibinfo {pages} {070601} (\bibinfo {year} {2011})}\BibitemShut {NoStop}%
\bibitem [{\citenamefont {Tang}\ \emph {et~al.}(2024)\citenamefont {Tang}, \citenamefont {Liu}, \citenamefont {Zhang},\ and\ \citenamefont {Zhang}}]{tang_learning_2024}%
  \BibitemOpen
  \bibfield  {author} {\bibinfo {author} {\bibfnamefont {Y.}~\bibnamefont {Tang}}, \bibinfo {author} {\bibfnamefont {J.}~\bibnamefont {Liu}}, \bibinfo {author} {\bibfnamefont {J.}~\bibnamefont {Zhang}},\ and\ \bibinfo {author} {\bibfnamefont {P.}~\bibnamefont {Zhang}},\ }\bibfield  {title} {\bibinfo {title} {Learning nonequilibrium statistical mechanics and dynamical phase transitions},\ }\href {https://www.nature.com/articles/s41467-024-45172-8} {\bibfield  {journal} {\bibinfo  {journal} {Nat. Commun.}\ }\textbf {\bibinfo {volume} {15}},\ \bibinfo {pages} {1117} (\bibinfo {year} {2024})}\BibitemShut {NoStop}%
\bibitem [{\citenamefont {Guo}\ \emph {et~al.}(2026)\citenamefont {Guo}, \citenamefont {Tao},\ and\ \citenamefont {Chen}}]{guo2025complexityanalysisnormalizingconstant}%
  \BibitemOpen
  \bibfield  {author} {\bibinfo {author} {\bibfnamefont {W.}~\bibnamefont {Guo}}, \bibinfo {author} {\bibfnamefont {M.}~\bibnamefont {Tao}},\ and\ \bibinfo {author} {\bibfnamefont {Y.}~\bibnamefont {Chen}},\ }\bibfield  {title} {\bibinfo {title} {Complexity analysis of normalizing constant estimation: from jarzynski equality to annealed importance sampling and beyond},\ }in\ \href@noop {} {\emph {\bibinfo {booktitle} {International Conference on Learning Representations, {ICLR} 2026}}}\ (\bibinfo {address} {Rio de Janeiro, Brazil},\ \bibinfo {year} {2026})\ pp.\ \bibinfo {pages} {117994--118041}\BibitemShut {NoStop}%
\bibitem [{\citenamefont {Chehab}\ \emph {et~al.}(2025)\citenamefont {Chehab}, \citenamefont {Korba}, \citenamefont {Stromme},\ and\ \citenamefont {Vacher}}]{chehab2025provableconvergencelimitationsgeometric}%
  \BibitemOpen
  \bibfield  {author} {\bibinfo {author} {\bibfnamefont {O.}~\bibnamefont {Chehab}}, \bibinfo {author} {\bibfnamefont {A.}~\bibnamefont {Korba}}, \bibinfo {author} {\bibfnamefont {A.}~\bibnamefont {Stromme}},\ and\ \bibinfo {author} {\bibfnamefont {A.}~\bibnamefont {Vacher}},\ }\bibfield  {title} {\bibinfo {title} {Provable convergence and limitations of geometric tempering for langevin dynamics},\ }in\ \href@noop {} {\emph {\bibinfo {booktitle} {International Conference on Learning Representations, {ICLR} 2025}}}\ (\bibinfo {address} {Singapore},\ \bibinfo {year} {2025})\ pp.\ \bibinfo {pages} {5451--5493}\BibitemShut {NoStop}%
\bibitem [{\citenamefont {Zhang}\ \emph {et~al.}(2026)\citenamefont {Zhang}, \citenamefont {Webber}, \citenamefont {Lindsey}, \citenamefont {Berkelbach},\ and\ \citenamefont {Weare}}]{zhang2025weightedVMC}%
  \BibitemOpen
  \bibfield  {author} {\bibinfo {author} {\bibfnamefont {H.}~\bibnamefont {Zhang}}, \bibinfo {author} {\bibfnamefont {R.~J.}\ \bibnamefont {Webber}}, \bibinfo {author} {\bibfnamefont {M.}~\bibnamefont {Lindsey}}, \bibinfo {author} {\bibfnamefont {T.~C.}\ \bibnamefont {Berkelbach}},\ and\ \bibinfo {author} {\bibfnamefont {J.}~\bibnamefont {Weare}},\ }\bibfield  {title} {\bibinfo {title} {Improved energies and local energies with weighted variational monte carlo},\ }\href {https://doi.org/10.1103/6b3v-38jp} {\bibfield  {journal} {\bibinfo  {journal} {Phys. Rev. Res.}\ }\textbf {\bibinfo {volume} {8}},\ \bibinfo {pages} {013213} (\bibinfo {year} {2026})}\BibitemShut {NoStop}%
\bibitem [{\citenamefont {Popinga}\ \emph {et~al.}(2026)\citenamefont {Popinga}, \citenamefont {Forman}, \citenamefont {Svetlov}, \citenamefont {Vo},\ and\ \citenamefont {Munsky}}]{popinga2026stochastic}%
  \BibitemOpen
  \bibfield  {author} {\bibinfo {author} {\bibfnamefont {A.~N.}\ \bibnamefont {Popinga}}, \bibinfo {author} {\bibfnamefont {J.}~\bibnamefont {Forman}}, \bibinfo {author} {\bibfnamefont {D.}~\bibnamefont {Svetlov}}, \bibinfo {author} {\bibfnamefont {H.}~\bibnamefont {Vo}},\ and\ \bibinfo {author} {\bibfnamefont {B.}~\bibnamefont {Munsky}},\ }\bibfield  {title} {\bibinfo {title} {The stochastic system identification toolkit (ssit) to model, fit, predict, and design experiments},\ }\href {https://doi.org/10.64898/2026.02.20.707039} {\bibfield  {journal} {\bibinfo  {journal} {bioRxiv}\ ,\ \bibinfo {pages} {2026.02.20.707039}} (\bibinfo {year} {2026})}\BibitemShut {NoStop}%
\bibitem [{\citenamefont {Farhat}\ \emph {et~al.}(2026)\citenamefont {Farhat}, \citenamefont {Rodrigues},\ and\ \citenamefont {Hendrik}}]{bioacme}%
  \BibitemOpen
  \bibfield  {author} {\bibinfo {author} {\bibfnamefont {A.}~\bibnamefont {Farhat}}, \bibinfo {author} {\bibfnamefont {J.~C.~J.}\ \bibnamefont {Rodrigues}},\ and\ \bibinfo {author} {\bibfnamefont {H.}~\bibnamefont {Hendrik}},\ }\href {https://github.com/alifarhat40/bioacme} {\bibinfo {title} {{BIOACME}}},\ \bibinfo {howpublished} {GitHub repository} (\bibinfo {year} {2026}),\ \bibinfo {note} {\url{https://github.com/alifarhat40/bioacme}}\BibitemShut {NoStop}%
\bibitem [{\citenamefont {Chen}\ \emph {et~al.}(2026)\citenamefont {Chen}, \citenamefont {Liu}, \citenamefont {Deng},\ and\ \citenamefont {Zhang}}]{chen2025tensor}%
  \BibitemOpen
  \bibfield  {author} {\bibinfo {author} {\bibfnamefont {T.}~\bibnamefont {Chen}}, \bibinfo {author} {\bibfnamefont {J.}~\bibnamefont {Liu}}, \bibinfo {author} {\bibfnamefont {Y.}~\bibnamefont {Deng}},\ and\ \bibinfo {author} {\bibfnamefont {P.}~\bibnamefont {Zhang}},\ }\bibfield  {title} {\bibinfo {title} {Tensor-network markov-chain monte carlo: Efficient sampling of three-dimensional spin glasses},\ }\href {https://doi.org/10.1103/96lr-z7tb} {\bibfield  {journal} {\bibinfo  {journal} {Phys. Rev. Lett.}\ }\textbf {\bibinfo {volume} {137}},\ \bibinfo {pages} {117101} (\bibinfo {year} {2026})}\BibitemShut {NoStop}%
\bibitem [{\citenamefont {Gupta}\ \emph {et~al.}(2021)\citenamefont {Gupta}, \citenamefont {Schwab},\ and\ \citenamefont {Khammash}}]{gupta_deepcme_2021}%
  \BibitemOpen
  \bibfield  {author} {\bibinfo {author} {\bibfnamefont {A.}~\bibnamefont {Gupta}}, \bibinfo {author} {\bibfnamefont {C.}~\bibnamefont {Schwab}},\ and\ \bibinfo {author} {\bibfnamefont {M.}~\bibnamefont {Khammash}},\ }\bibfield  {title} {\bibinfo {title} {{DeepCME}: {A} deep learning framework for computing solution statistics of the chemical master equation},\ }\href {https://dx.plos.org/10.1371/journal.pcbi.1009623} {\bibfield  {journal} {\bibinfo  {journal} {PLoS Comput. Biol.}\ }\textbf {\bibinfo {volume} {17}},\ \bibinfo {pages} {e1009623} (\bibinfo {year} {2021})}\BibitemShut {NoStop}%
\bibitem [{\citenamefont {Gupta}\ and\ \citenamefont {Khammash}(2025)}]{gupta_spectral_2025}%
  \BibitemOpen
  \bibfield  {author} {\bibinfo {author} {\bibfnamefont {A.}~\bibnamefont {Gupta}}\ and\ \bibinfo {author} {\bibfnamefont {M.}~\bibnamefont {Khammash}},\ }\href@noop {} {\bibinfo {title} {A spectral koopman approximation framework for stochastic reaction networks}} (\bibinfo {year} {2025}),\ \Eprint {https://arxiv.org/abs/2511.23114} {arXiv:2511.23114} \BibitemShut {NoStop}%
\bibitem [{\citenamefont {Badolle}\ \emph {et~al.}(2025)\citenamefont {Badolle}, \citenamefont {Theuer}, \citenamefont {Fang}, \citenamefont {Gupta},\ and\ \citenamefont {Khammash}}]{badolle_interpretable_2025}%
  \BibitemOpen
  \bibfield  {author} {\bibinfo {author} {\bibfnamefont {Q.}~\bibnamefont {Badolle}}, \bibinfo {author} {\bibfnamefont {A.}~\bibnamefont {Theuer}}, \bibinfo {author} {\bibfnamefont {Z.}~\bibnamefont {Fang}}, \bibinfo {author} {\bibfnamefont {A.}~\bibnamefont {Gupta}},\ and\ \bibinfo {author} {\bibfnamefont {M.}~\bibnamefont {Khammash}},\ }\href@noop {} {\bibinfo {title} {Interpretable neural approximation of stochastic reaction dynamics with guaranteed reliability}} (\bibinfo {year} {2025}),\ \Eprint {https://arxiv.org/abs/2512.06294} {arXiv:2512.06294} \BibitemShut {NoStop}%
\bibitem [{\citenamefont {Wang}\ \emph {et~al.}(2025)\citenamefont {Wang}, \citenamefont {Shi}, \citenamefont {Heess}, \citenamefont {Gretton},\ and\ \citenamefont {Titsias}}]{LO_ARM}%
  \BibitemOpen
  \bibfield  {author} {\bibinfo {author} {\bibfnamefont {Z.}~\bibnamefont {Wang}}, \bibinfo {author} {\bibfnamefont {J.}~\bibnamefont {Shi}}, \bibinfo {author} {\bibfnamefont {N.}~\bibnamefont {Heess}}, \bibinfo {author} {\bibfnamefont {A.}~\bibnamefont {Gretton}},\ and\ \bibinfo {author} {\bibfnamefont {M.}~\bibnamefont {Titsias}},\ }\bibfield  {title} {\bibinfo {title} {Learning-order autoregressive models with application to molecular graph generation},\ }in\ \href@noop {} {\emph {\bibinfo {booktitle} {Proceedings of the 42nd International Conference on Machine Learning}}},\ \bibinfo {series} {Proceedings of Machine Learning Research}, Vol.\ \bibinfo {volume} {267}\ (\bibinfo  {publisher} {PMLR},\ \bibinfo {year} {2025})\ pp.\ \bibinfo {pages} {64282--64301},\ \Eprint {https://arxiv.org/abs/2503.05979} {arXiv:2503.05979} \BibitemShut {NoStop}%
\bibitem [{\citenamefont {Cross}\ and\ \citenamefont {Hohenberg}(1993)}]{cross_pattern_1993}%
  \BibitemOpen
  \bibfield  {author} {\bibinfo {author} {\bibfnamefont {M.~C.}\ \bibnamefont {Cross}}\ and\ \bibinfo {author} {\bibfnamefont {P.~C.}\ \bibnamefont {Hohenberg}},\ }\bibfield  {title} {\bibinfo {title} {Pattern formation outside of equilibrium},\ }\href {https://doi.org/10.1103/RevModPhys.65.851} {\bibfield  {journal} {\bibinfo  {journal} {Rev. Mod. Phys.}\ }\textbf {\bibinfo {volume} {65}},\ \bibinfo {pages} {851} (\bibinfo {year} {1993})}\BibitemShut {NoStop}%
\bibitem [{\citenamefont {Kamenev}(2011)}]{kamenev_field_2011}%
  \BibitemOpen
  \bibfield  {author} {\bibinfo {author} {\bibfnamefont {A.}~\bibnamefont {Kamenev}},\ }\href@noop {} {\emph {\bibinfo {title} {Field theory of non-equilibrium systems}}},\ \bibinfo {edition} {1st}\ ed.\ (\bibinfo  {publisher} {Cambridge University Press},\ \bibinfo {address} {Cambridge, UK},\ \bibinfo {year} {2011})\BibitemShut {NoStop}%
\bibitem [{\citenamefont {Nagaosa}\ and\ \citenamefont {Tokura}(2013)}]{nagaosa_topological_2013}%
  \BibitemOpen
  \bibfield  {author} {\bibinfo {author} {\bibfnamefont {N.}~\bibnamefont {Nagaosa}}\ and\ \bibinfo {author} {\bibfnamefont {Y.}~\bibnamefont {Tokura}},\ }\bibfield  {title} {\bibinfo {title} {Topological properties and dynamics of magnetic skyrmions},\ }\href {https://doi.org/10.1038/nnano.2013.243} {\bibfield  {journal} {\bibinfo  {journal} {Nature Nanotech.}\ }\textbf {\bibinfo {volume} {8}},\ \bibinfo {pages} {899} (\bibinfo {year} {2013})}\BibitemShut {NoStop}%
\bibitem [{\citenamefont {Fert}\ \emph {et~al.}(2017)\citenamefont {Fert}, \citenamefont {Reyren},\ and\ \citenamefont {Cros}}]{fert_magnetic_2017}%
  \BibitemOpen
  \bibfield  {author} {\bibinfo {author} {\bibfnamefont {A.}~\bibnamefont {Fert}}, \bibinfo {author} {\bibfnamefont {N.}~\bibnamefont {Reyren}},\ and\ \bibinfo {author} {\bibfnamefont {V.}~\bibnamefont {Cros}},\ }\bibfield  {title} {\bibinfo {title} {Magnetic skyrmions: {Advances} in physics and potential applications},\ }\href {https://doi.org/10.1038/natrevmats.2017.31} {\bibfield  {journal} {\bibinfo  {journal} {Nat. Rev. Mater.}\ }\textbf {\bibinfo {volume} {2}},\ \bibinfo {pages} {17031} (\bibinfo {year} {2017})}\BibitemShut {NoStop}%
\bibitem [{\citenamefont {Zhu}\ \emph {et~al.}(2026)\citenamefont {Zhu}, \citenamefont {Guan}, \citenamefont {Sun}, \citenamefont {Zhang},\ and\ \citenamefont {Song}}]{zhu2025universaltwostagedynamicsphase}%
  \BibitemOpen
  \bibfield  {author} {\bibinfo {author} {\bibfnamefont {S.}~\bibnamefont {Zhu}}, \bibinfo {author} {\bibfnamefont {X.}~\bibnamefont {Guan}}, \bibinfo {author} {\bibfnamefont {Z.}~\bibnamefont {Sun}}, \bibinfo {author} {\bibfnamefont {Q.}~\bibnamefont {Zhang}},\ and\ \bibinfo {author} {\bibfnamefont {C.}~\bibnamefont {Song}},\ }\bibfield  {title} {\bibinfo {title} {Two-stage dynamics and phase control of skyrmion formation in chiral magnets},\ }\href {https://doi.org/10.1103/z6k3-1zvy} {\bibfield  {journal} {\bibinfo  {journal} {Phys. Rev. B}\ }\textbf {\bibinfo {volume} {113}},\ \bibinfo {pages} {L100403} (\bibinfo {year} {2026})}\BibitemShut {NoStop}%
\bibitem [{\citenamefont {Xiong}\ \emph {et~al.}(2025)\citenamefont {Xiong}, \citenamefont {Zhou}, \citenamefont {Hu}, \citenamefont {Nian},\ and\ \citenamefont {Zheng}}]{PhysRevB.111.184415}%
  \BibitemOpen
  \bibfield  {author} {\bibinfo {author} {\bibfnamefont {L.}~\bibnamefont {Xiong}}, \bibinfo {author} {\bibfnamefont {N.-J.}\ \bibnamefont {Zhou}}, \bibinfo {author} {\bibfnamefont {S.-Q.}\ \bibnamefont {Hu}}, \bibinfo {author} {\bibfnamefont {L.-L.}\ \bibnamefont {Nian}},\ and\ \bibinfo {author} {\bibfnamefont {B.}~\bibnamefont {Zheng}},\ }\bibfield  {title} {\bibinfo {title} {Capturing the dynamics of the phase transition of skyrmions with a nonstationary machine learning approach},\ }\href {https://doi.org/10.1103/PhysRevB.111.184415} {\bibfield  {journal} {\bibinfo  {journal} {Phys. Rev. B}\ }\textbf {\bibinfo {volume} {111}},\ \bibinfo {pages} {184415} (\bibinfo {year} {2025})}\BibitemShut {NoStop}%
\bibitem [{\citenamefont {Li}\ \emph {et~al.}(2026)\citenamefont {Li}, \citenamefont {Liu}, \citenamefont {Wang}, \citenamefont {Liao},\ and\ \citenamefont {Li}}]{li2025weightflow}%
  \BibitemOpen
  \bibfield  {author} {\bibinfo {author} {\bibfnamefont {R.}~\bibnamefont {Li}}, \bibinfo {author} {\bibfnamefont {J.}~\bibnamefont {Liu}}, \bibinfo {author} {\bibfnamefont {H.}~\bibnamefont {Wang}}, \bibinfo {author} {\bibfnamefont {Q.}~\bibnamefont {Liao}},\ and\ \bibinfo {author} {\bibfnamefont {Y.}~\bibnamefont {Li}},\ }\bibfield  {title} {\bibinfo {title} {Weightflow: Learning stochastic dynamics via evolving weight of neural network},\ }in\ \href {https://doi.org/10.1609/aaai.v40i1.37029} {\emph {\bibinfo {booktitle} {Proceedings of the AAAI Conference on Artificial Intelligence}}},\ Vol.~\bibinfo {volume} {40}\ (\bibinfo {year} {2026})\ pp.\ \bibinfo {pages} {641--649}\BibitemShut {NoStop}%
\bibitem [{\citenamefont {Weng}\ \emph {et~al.}(2025)\citenamefont {Weng}, \citenamefont {Zhu}, \citenamefont {Liu}, \citenamefont {L{\"u}}, \citenamefont {Zhang},\ and\ \citenamefont {Tang}}]{NNCME2_2025}%
  \BibitemOpen
  \bibfield  {author} {\bibinfo {author} {\bibfnamefont {J.}~\bibnamefont {Weng}}, \bibinfo {author} {\bibfnamefont {X.}~\bibnamefont {Zhu}}, \bibinfo {author} {\bibfnamefont {J.}~\bibnamefont {Liu}}, \bibinfo {author} {\bibfnamefont {L.}~\bibnamefont {L{\"u}}}, \bibinfo {author} {\bibfnamefont {P.}~\bibnamefont {Zhang}},\ and\ \bibinfo {author} {\bibfnamefont {Y.}~\bibnamefont {Tang}},\ }\href {https://github.com/Machine-learning-and-complex-systems/NNCME} {\bibinfo {title} {{NNCME}}},\ \bibinfo {howpublished} {GitHub repository} (\bibinfo {year} {2025}),\ \bibinfo {note} {\url{https://github.com/Machine-learning-and-complex-systems/NNCME}}\BibitemShut {NoStop}%
\bibitem [{\citenamefont {Kullback}\ and\ \citenamefont {Leibler}(1951)}]{kullback_information_1951}%
  \BibitemOpen
  \bibfield  {author} {\bibinfo {author} {\bibfnamefont {S.}~\bibnamefont {Kullback}}\ and\ \bibinfo {author} {\bibfnamefont {R.~A.}\ \bibnamefont {Leibler}},\ }\bibfield  {title} {\bibinfo {title} {On information and sufficiency},\ }\href {http://projecteuclid.org/euclid.aoms/1177729694} {\bibfield  {journal} {\bibinfo  {journal} {Ann. Math. Statist.}\ }\textbf {\bibinfo {volume} {22}},\ \bibinfo {pages} {79} (\bibinfo {year} {1951})}\BibitemShut {NoStop}%
\bibitem [{\citenamefont {Minka}(2005)}]{minka_divergence_2005}%
  \BibitemOpen
  \bibfield  {author} {\bibinfo {author} {\bibfnamefont {T.~P.}\ \bibnamefont {Minka}},\ }\href@noop {} {\emph {\bibinfo {title} {Divergence measures and message passing}}},\ \bibinfo {type} {Technical report}\ (\bibinfo  {institution} {Microsoft Research},\ \bibinfo {year} {2005})\BibitemShut {NoStop}%
\bibitem [{\citenamefont {Naesseth}\ \emph {et~al.}(2020)\citenamefont {Naesseth}, \citenamefont {Lindsten},\ and\ \citenamefont {Blei}}]{naesseth_markovian_2020}%
  \BibitemOpen
  \bibfield  {author} {\bibinfo {author} {\bibfnamefont {C.~A.}\ \bibnamefont {Naesseth}}, \bibinfo {author} {\bibfnamefont {F.}~\bibnamefont {Lindsten}},\ and\ \bibinfo {author} {\bibfnamefont {D.}~\bibnamefont {Blei}},\ }\bibfield  {title} {\bibinfo {title} {Markovian score climbing: {Variational} inference with {KL}(p{\textbar}{\textbar}q)},\ }in\ \href@noop {} {\emph {\bibinfo {booktitle} {Proceedings of the 34th {International} {Conference} on {Neural} {Information} {Processing} {Systems}}}},\ \bibinfo {series and number} {{NIPS} '20}\ (\bibinfo  {publisher} {Curran Associates Inc.},\ \bibinfo {address} {Red Hook, NY, USA},\ \bibinfo {year} {2020})\ pp.\ \bibinfo {pages} {15499--15510},\ \Eprint {https://arxiv.org/abs/2003.10374} {arXiv:2003.10374} \BibitemShut {NoStop}%
\bibitem [{\citenamefont {Gu}\ \emph {et~al.}(2015)\citenamefont {Gu}, \citenamefont {Ghahramani},\ and\ \citenamefont {Turner}}]{gu_neural_2015}%
  \BibitemOpen
  \bibfield  {author} {\bibinfo {author} {\bibfnamefont {S.}~\bibnamefont {Gu}}, \bibinfo {author} {\bibfnamefont {Z.}~\bibnamefont {Ghahramani}},\ and\ \bibinfo {author} {\bibfnamefont {R.~E.}\ \bibnamefont {Turner}},\ }\bibfield  {title} {\bibinfo {title} {Neural adaptive sequential monte carlo},\ }in\ \href@noop {} {\emph {\bibinfo {booktitle} {Proceedings of the 29th International Conference on Neural Information Processing Systems - Volume 2}}},\ \bibinfo {series and number} {NIPS'15},\ \bibinfo {editor} {edited by\ \bibinfo {editor} {\bibfnamefont {C.}~\bibnamefont {Cortes}}, \bibinfo {editor} {\bibfnamefont {N.~D.}\ \bibnamefont {Lawrence}}, \bibinfo {editor} {\bibfnamefont {D.~D.}\ \bibnamefont {Lee}}, \bibinfo {editor} {\bibfnamefont {M.}~\bibnamefont {Sugiyama}},\ and\ \bibinfo {editor} {\bibfnamefont {R.}~\bibnamefont {Garnett}}}\ (\bibinfo  {publisher} {MIT Press},\ \bibinfo {address} {Cambridge, MA, USA},\ \bibinfo {year} {2015})\ pp.\ \bibinfo {pages} {2629--2637}\BibitemShut {NoStop}%
\bibitem [{\citenamefont {Schulman}\ \emph {et~al.}(2017)\citenamefont {Schulman}, \citenamefont {Wolski}, \citenamefont {Dhariwal}, \citenamefont {Radford},\ and\ \citenamefont {Klimov}}]{schulman_proximal_2017}%
  \BibitemOpen
  \bibfield  {author} {\bibinfo {author} {\bibfnamefont {J.}~\bibnamefont {Schulman}}, \bibinfo {author} {\bibfnamefont {F.}~\bibnamefont {Wolski}}, \bibinfo {author} {\bibfnamefont {P.}~\bibnamefont {Dhariwal}}, \bibinfo {author} {\bibfnamefont {A.}~\bibnamefont {Radford}},\ and\ \bibinfo {author} {\bibfnamefont {O.}~\bibnamefont {Klimov}},\ }\href@noop {} {\bibinfo {title} {Proximal policy optimization algorithms}} (\bibinfo {year} {2017}),\ \Eprint {https://arxiv.org/abs/1707.06347} {arXiv:1707.06347} \BibitemShut {NoStop}%
\bibitem [{\citenamefont {Hellinger}(1907)}]{hellinger_orthogonalinvarianten_1907}%
  \BibitemOpen
  \bibfield  {author} {\bibinfo {author} {\bibfnamefont {E.}~\bibnamefont {Hellinger}},\ }\href@noop {} {\emph {\bibinfo {title} {Die orthogonalinvarianten quadratischer formen von unendlichvielen variabelen}}},\ \bibinfo {number} {84 p.}\ (\bibinfo  {publisher} {W. Fr. Kaestner},\ \bibinfo {address} {Göttingen},\ \bibinfo {year} {1907})\BibitemShut {NoStop}%
\bibitem [{\citenamefont {Ansel}\ \emph {et~al.}(2024)\citenamefont {Ansel}, \citenamefont {Yang}, \citenamefont {He}, \citenamefont {Gimelshein}, \citenamefont {Jain}, \citenamefont {Voznesensky}, \citenamefont {Bao}, \citenamefont {Bell}, \citenamefont {Berard}, \citenamefont {Burovski}, \citenamefont {Chauhan}, \citenamefont {Chourdia}, \citenamefont {Constable}, \citenamefont {Desmaison}, \citenamefont {DeVito}, \citenamefont {Ellison}, \citenamefont {Feng}, \citenamefont {Gong}, \citenamefont {Gschwind}, \citenamefont {Hirsh},\ and\ \citenamefont {et~al.}}]{Ansel_PyTorch_2_Faster_2024}%
  \BibitemOpen
  \bibfield  {author} {\bibinfo {author} {\bibfnamefont {J.}~\bibnamefont {Ansel}}, \bibinfo {author} {\bibfnamefont {E.}~\bibnamefont {Yang}}, \bibinfo {author} {\bibfnamefont {H.}~\bibnamefont {He}}, \bibinfo {author} {\bibfnamefont {N.}~\bibnamefont {Gimelshein}}, \bibinfo {author} {\bibfnamefont {A.}~\bibnamefont {Jain}}, \bibinfo {author} {\bibfnamefont {M.}~\bibnamefont {Voznesensky}}, \bibinfo {author} {\bibfnamefont {B.}~\bibnamefont {Bao}}, \bibinfo {author} {\bibfnamefont {P.}~\bibnamefont {Bell}}, \bibinfo {author} {\bibfnamefont {D.}~\bibnamefont {Berard}}, \bibinfo {author} {\bibfnamefont {E.}~\bibnamefont {Burovski}}, \bibinfo {author} {\bibfnamefont {G.}~\bibnamefont {Chauhan}}, \bibinfo {author} {\bibfnamefont {A.}~\bibnamefont {Chourdia}}, \bibinfo {author} {\bibfnamefont {W.}~\bibnamefont {Constable}}, \bibinfo {author} {\bibfnamefont {A.}~\bibnamefont {Desmaison}}, \bibinfo {author} {\bibfnamefont {Z.}~\bibnamefont {DeVito}}, \bibinfo {author} {\bibfnamefont {E.}~\bibnamefont {Ellison}}, \bibinfo {author} {\bibfnamefont {W.}~\bibnamefont {Feng}}, \bibinfo {author} {\bibfnamefont {J.}~\bibnamefont {Gong}}, \bibinfo {author} {\bibfnamefont {M.}~\bibnamefont {Gschwind}}, \bibinfo {author} {\bibfnamefont {B.}~\bibnamefont {Hirsh}},\ and\ \bibinfo {author} {\bibnamefont {et~al.}},\ }\bibfield  {title} {\bibinfo {title} {{PyTorch 2: Faster Machine Learning Through Dynamic Python Bytecode Transformation and Graph Compilation}},\ }in\ \href@noop {} {\emph {\bibinfo {booktitle} {29th ACM International Conference on Architectural Support for Programming Languages and Operating Systems, Volume 2 (ASPLOS '24)}}}\ (\bibinfo  {publisher} {ACM},\ \bibinfo {address} {New York, NY, USA},\ \bibinfo {year} {2024})\BibitemShut {NoStop}%
\bibitem [{\citenamefont {Bradbury}\ \emph {et~al.}(2018)\citenamefont {Bradbury}, \citenamefont {Frostig}, \citenamefont {Hawkins}, \citenamefont {Johnson}, \citenamefont {Leary}, \citenamefont {Maclaurin}, \citenamefont {Necula}, \citenamefont {Paszke}, \citenamefont {Vander{P}las}, \citenamefont {Wanderman-{M}ilne},\ and\ \citenamefont {Zhang}}]{jax2018github}%
  \BibitemOpen
  \bibfield  {author} {\bibinfo {author} {\bibfnamefont {J.}~\bibnamefont {Bradbury}}, \bibinfo {author} {\bibfnamefont {R.}~\bibnamefont {Frostig}}, \bibinfo {author} {\bibfnamefont {P.}~\bibnamefont {Hawkins}}, \bibinfo {author} {\bibfnamefont {M.~J.}\ \bibnamefont {Johnson}}, \bibinfo {author} {\bibfnamefont {C.}~\bibnamefont {Leary}}, \bibinfo {author} {\bibfnamefont {D.}~\bibnamefont {Maclaurin}}, \bibinfo {author} {\bibfnamefont {G.}~\bibnamefont {Necula}}, \bibinfo {author} {\bibfnamefont {A.}~\bibnamefont {Paszke}}, \bibinfo {author} {\bibfnamefont {J.}~\bibnamefont {Vander{P}las}}, \bibinfo {author} {\bibfnamefont {S.}~\bibnamefont {Wanderman-{M}ilne}},\ and\ \bibinfo {author} {\bibfnamefont {Q.}~\bibnamefont {Zhang}},\ }\href {http://github.com/jax-ml/jax} {\bibinfo {title} {{JAX}: composable transformations of {P}ython+{N}um{P}y programs}} (\bibinfo {year} {2018})\BibitemShut {NoStop}%
\end{thebibliography}
\end{document}